June 2009

# Astrophysical Sources of Cosmic Rays and Related Measurements with the Pierre Auger Observatory

Presentations for the
31st International Cosmic Ray Conference, Łódź , Poland, July 2009

# PIERRE AUGER COLLABORATION

**J. Abraham[8], P. Abreu[71], M. Aglietta[54], C. Aguirre[12], E.J. Ahn[87], D. Allard[31], I. Allekotte[1], J. Allen[90], J. Alvarez-Muñiz[78], M. Ambrosio[48], L. Anchordoqui[104], S. Andringa[71], A. Anzalone[53], C. Aramo[48], E. Arganda[75], S. Argirò[51], K. Arisaka[95], F. Arneodo[55], F. Arqueros[75], T. Asch[38], H. Asorey[1], P. Assis[71], J. Aublin[33], M. Ave[96], G. Avila[10], T. Bäcker[42], D. Badagnani[6], K.B. Barber[11], A.F. Barbosa[14], S.L.C. Barroso[20], B. Baughman[92], P. Bauleo[85], J.J. Beatty[92], T. Beau[31], B.R. Becker[101], K.H. Becker[36], A. Bellétoile[34], J.A. Bellido[11, 93], S. BenZvi[103], C. Berat[34], P. Bernardini[47], X. Bertou[1], P.L. Biermann[39], P. Billoir[33], O. Blanch-Bigas[33], F. Blanco[75], C. Bleve[47], H. Blümer[41, 37], M. Boháčová[96, 27], D. Boncioli[49], C. Bonifazi[33], R. Bonino[54], N. Borodai[69], J. Brack[85], P. Brogueira[71], W.C. Brown[86], R. Bruijn[81], P. Buchholz[42], A. Bueno[77], R.E. Burton[83], N.G. Busca[31], K.S. Caballero-Mora[41], L. Caramete[39], R. Caruso[50], W. Carvalho[17], A. Castellina[54], O. Catalano[53], L. Cazon[96], R. Cester[51], J. Chauvin[34], A. Chiavassa[54], J.A. Chinellato[18], A. Chou[87, 90], J. Chudoba[27], J. Chye[89d], R.W. Clay[11], E. Colombo[2], R. Conceição[71], B. Connolly[102], F. Contreras[9], J. Coppens[65, 67], A. Cordier[32], U. Cotti[63], S. Coutu[93], C.E. Covault[83], A. Creusot[73], A. Criss[93], J. Cronin[96], A. Curutiu[39], S. Dagoret-Campagne[32], R. Dallier[35], K. Daumiller[37], B.R. Dawson[11], R.M. de Almeida[18], M. De Domenico[50], C. De Donato[46], S.J. de Jong[65], G. De La Vega[8], W.J.M. de Mello Junior[18], J.R.T. de Mello Neto[23], I. De Mitri[47], V. de Souza[16], K.D. de Vries[66], G. Decerprit[31], L. del Peral[76], O. Deligny[30], A. Della Selva[48], C. Delle Fratte[49], H. Dembinski[40], C. Di Giulio[49], J.C. Diaz[89], P.N. Diep[105], C. Dobrigkeit [18], J.C. D'Olivo[64], P.N. Dong[105], A. Dorofeev[88], J.C. dos Anjos[14], M.T. Dova[6], D. D'Urso[48], I. Dutan[39], M.A. DuVernois[98], R. Engel[37], M. Erdmann[40], C.O. Escobar[18], A. Etchegoyen[2], P. Facal San Luis[96, 78], H. Falcke[65, 68], G. Farrar[90], A.C. Fauth[18], N. Fazzini[87], F. Ferrer[83], A. Ferrero[2], B. Fick[89], A. Filevich[2], A. Filipčič[72, 73], I. Fleck[42], S. Fliescher[40], C.E. Fracchiolla[85], E.D. Fraenkel[66], W. Fulgione[54], R.F. Gamarra[2], S. Gambetta[44], B. García[8], D. García Gámez[77], D. Garcia-Pinto[75], X. Garrido[37, 32], G. Gelmini[95], H. Gemmeke[38], P.L. Ghia[30, 54], U. Giaccari[47], M. Giller[70], H. Glass[87], L.M. Goggin[104], M.S. Gold[101], G. Golup[1], F. Gomez Albarracin[6], M. Gómez Berisso[1], P. Gonçalves[71], M. Gonçalves do Amaral[24], D. Gonzalez[41], J.G. Gonzalez[77, 88], D. Góra[41, 69], A. Gorgi[54], P. Gouffon[17], S.R. Gozzini[81], E. Grashorn[92], S. Grebe[65], M. Grigat[40], A.F. Grillo[55], Y. Guardincerri[4], F. Guarino[48], G.P. Guedes[19], J. Gutiérrez[76], J.D. Hague[101], V. Halenka[28], P. Hansen[6], D. Harari[1], S. Harmsma[66, 67], J.L. Harton[85], A. Haungs[37], M.D. Healy[95], T. Hebbeker[40], G. Hebrero[76], D. Heck[37], V.C. Holmes[11], P. Homola[69], J.R. Hörandel[65], A. Horneffer[65], M. Hrabovský[28, 27], T. Huege[37], M. Hussain[73], M. Iarlori[45], A. Insolia[50], F. Ionita[96], A. Italiano[50], S. Jiraskova[65], M. Kaducak[87], K.H. Kampert[36], T. Karova[27], P. Kasper[87], B. Kégl[32], B. Keilhauer[37], E. Kemp[18], R.M. Kieckhafer[89], H.O. Klages[37], M. Kleifges[38], J. Kleinfeller[37], R. Knapik[85], J. Knapp[81], D.-H. Koang[34], A. Krieger[2], O. Krömer[38], D. Kruppke-Hansen[36], F. Kuehn[87], D. Kuempel[36], N. Kunka[38], A. Kusenko[95], G. La Rosa[53], C. Lachaud[31], B.L. Lago[23], P. Lautridou[35], M.S.A.B. Leão[22], D. Lebrun[34], P. Lebrun[87], J. Lee[95], M.A. Leigui de Oliveira[22], A. Lemiere[30], A. Letessier-Selvon[33], M. Leuthold[40], I. Lhenry-Yvon[30], R. López[59], A. Lopez Agüera[78], K. Louedec[32], J. Lozano Bahilo[77], A. Lucero[54], H. Lyberis[30], M.C. Maccarone[53], C. Macolino[45], S. Maldera[54], D. Mandat[27], P. Mantsch[87], A.G. Mariazzi[6], I.C. Maris[41], H.R. Marquez Falcon[63], D. Martello[47], O. Martínez Bravo[59], H.J. Mathes[37], J. Matthews[88, 94], J.A.J. Matthews[101], G. Matthiae[49], D. Maurizio[51], P.O. Mazur[87], M. McEwen[76], R.R. McNeil[88], G. Medina-Tanco[64], M. Melissas[41], D. Melo[51], E. Menichetti[51], A. Menshikov[38], R. Meyhandan[14], M.I. Micheletti[2], G. Miele[48], W. Miller[101], L. Miramonti[46], S. Mollerach[1], M. Monasor[75], D. Monnier Ragaigne[32], F. Montanet[34], B. Morales[64], C. Morello[54], J.C. Moreno[6], C. Morris[92], M. Mostafá[85], C.A. Moura[48], S. Mueller[37], M.A. Muller[18], R. Mussa[51], G. Navarra[54], J.L. Navarro[77], S. Navas[77], P. Necesal[27], L. Nellen[64], C. Newman-Holmes[87], D. Newton[81], P.T. Nhung[105], N. Nierstenhoefer[36], D. Nitz[89], D. Nosek[26], L. Nožka[27], M. Nyklicek[27], J. Oehlschläger[37], A. Olinto[96], P. Oliva[36], V.M. Olmos-Gilbaja[78], M. Ortiz[75], N. Pacheco[76], D. Pakk Selmi-Dei[18], M. Palatka[27], J. Pallotta[3], G. Parente[78], E. Parizot[31], S. Parlati[55], S. Pastor[74], M. Patel[81], T. Paul[91], V. Pavlidou[96c], K. Payet[34], M. Pech[27], J. Pękala[69], I.M. Pepe[21], L. Perrone[52], R. Pesce[44], E. Petermann[100], S. Petrera[45], P. Petrinca[49], A. Petrolini[44], Y. Petrov[85], J. Petrovic[67], C. Pfendner[103], R. Piegaia[4], T. Pierog[37], M. Pimenta[71], T. Pinto[74], V. Pirronello[50], O. Pisanti[48], M. Platino[2], J. Pochon[1], V.H. Ponce[1], M. Pontz[42], P. Privitera[96], M. Prouza[27], E.J. Quel[3], J. Rautenberg[36], O. Ravel[35], D. Ravignani[2],**

**A. Redondo[76], B. Revenu[35], F.A.S. Rezende[14], J. Ridky[27], S. Riggi[50], M. Risse[36], C. Rivière[34], V. Rizi[45], C. Robledo[59], G. Rodriguez[49], J. Rodriguez Martino[50], J. Rodriguez Rojo[9], I. Rodriguez-Cabo[78], M.D. Rodríguez-Frías[76], G. Ros[75, 76], J. Rosado[75], T. Rossler[28], M. Roth[37], B. Rouillé-d'Orfeuil[31], E. Roulet[1], A.C. Rovero[7], F. Salamida[45], H. Salazar[59b], G. Salina[49], F. Sánchez[64], M. Santander[9], C.E. Santo[71], E.M. Santos[23], F. Sarazin[84], S. Sarkar[79], R. Sato[9], N. Scharf[40], V. Scherini[36], H. Schieler[37], P. Schiffer[40], A. Schmidt[38], F. Schmidt[96], T. Schmidt[41], O. Scholten[66], H. Schoorlemmer[65], J. Schovancova[27], P. Schovánek[27], F. Schroeder[37], S. Schulte[40], F. Schüssler[37], D. Schuster[84], S.J. Sciutto[6], M. Scuderi[50], A. Segreto[53], D. Semikoz[31], M. Settimo[47], R.C. Shellard[14, 15], I. Sidelnik[2], B.B. Siffert[23], A. Śmiałkowski[70], R. Šmída[27], B.E. Smith[81], G.R. Snow[100], P. Sommers[93], J. Sorokin[11], H. Spinka[82, 87], R. Squartini[9], E. Strazzeri[32], A. Stutz[34], F. Suarez[2], T. Suomijärvi[30], A.D. Supanitsky[64], M.S. Sutherland[92], J. Swain[91], Z. Szadkowski[70], A. Tamashiro[7], A. Tamburro[41], T. Tarutina[6], O. Taşcău[36], R. Tcaciuc[42], D. Tcherniakhovski[38], D. Tegolo[58], N.T. Thao[105], D. Thomas[85], R. Ticona[13], J. Tiffenberg[4], C. Timmermans[67, 65], W. Tkaczyk[70], C.J. Todero Peixoto[22], B. Tomé[71], A. Tonachini[51], I. Torres[59], P. Travnicek[27], D.B. Tridapalli[17], G. Tristram[31], E. Trovato[50], M. Tueros[6], R. Ulrich[37], M. Unger[37], M. Urban[32], J.F. Valdés Galicia[64], I. Valiño[37], L. Valore[48], A.M. van den Berg[66], J.R. Vázquez[75], R.A. Vázquez[78], D. Veberič[73, 72], A. Velarde[13], T. Venters[96], V. Verzi[49], M. Videla[8], L. Villaseñor[63], S. Vorobiov[73], L. Voyvodic[87‡], H. Wahlberg[6], P. Wahrlich[11], O. Wainberg[2], D. Warner[85], A.A. Watson[81], S. Westerhoff[103], B.J. Whelan[11], G. Wieczorek[70], L. Wiencke[84], B. Wilczyńska[69], H. Wilczyński[69], C. Wileman[81], M.G. Winnick[11], H. Wu[32], B. Wundheiler[2], T. Yamamoto[96a], P. Younk[85], G. Yuan[88], A. Yushkov[48], E. Zas[78], D. Zavrtanik[73, 72], M. Zavrtanik[72, 73], I. Zaw[90], A. Zepeda[60b], M. Ziolkowski[42]**

*[1] Centro Atómico Bariloche and Instituto Balseiro (CNEA-UNCuyo-CONICET), San Carlos de Bariloche, Argentina*

*[2] Centro Atómico Constituyentes (Comisión Nacional de Energía Atómica/CONICET/UTN- FRBA), Buenos Aires, Argentina*

*[3] Centro de Investigaciones en Láseres y Aplicaciones, CITEFA and CONICET, Argentina*

*[4] Departamento de Física, FCEyN, Universidad de Buenos Aires y CONICET, Argentina*

*[6] IFLP, Universidad Nacional de La Plata and CONICET, La Plata, Argentina*

*[7] Instituto de Astronomía y Física del Espacio (CONICET), Buenos Aires, Argentina*

*[8] National Technological University, Faculty Mendoza (CONICET/CNEA), Mendoza, Argentina*

*[9] Pierre Auger Southern Observatory, Malargüe, Argentina*

*[10] Pierre Auger Southern Observatory and Comisión Nacional de Energía Atómica, Malargüe, Argentina*

*[11] University of Adelaide, Adelaide, S.A., Australia*

*[12] Universidad Catolica de Bolivia, La Paz, Bolivia*

*[13] Universidad Mayor de San Andrés, Bolivia*

*[14] Centro Brasileiro de Pesquisas Fisicas, Rio de Janeiro, RJ, Brazil*

*[15] Pontifícia Universidade Católica, Rio de Janeiro, RJ, Brazil*

*[16] Universidade de São Paulo, Instituto de Física, São Carlos, SP, Brazil*

*[17] Universidade de São Paulo, Instituto de Física, São Paulo, SP, Brazil*

*[18] Universidade Estadual de Campinas, IFGW, Campinas, SP, Brazil*

*[19] Universidade Estadual de Feira de Santana, Brazil*

*[20] Universidade Estadual do Sudoeste da Bahia, Vitoria da Conquista, BA, Brazil*

*[21] Universidade Federal da Bahia, Salvador, BA, Brazil*

*[22] Universidade Federal do ABC, Santo André, SP, Brazil*

*[23] Universidade Federal do Rio de Janeiro, Instituto de Física, Rio de Janeiro, RJ, Brazil*

*[24] Universidade Federal Fluminense, Instituto de Fisica, Niterói, RJ, Brazil*

*[26] Charles University, Faculty of Mathematics and Physics, Institute of Particle and Nuclear Physics, Prague, Czech Republic*

*[27] Institute of Physics of the Academy of Sciences of the Czech Republic, Prague, Czech Republic*

*[28] Palacký University, Olomouc, Czech Republic*

*[30] Institut de Physique Nucléaire d'Orsay (IPNO), Université Paris 11, CNRS-IN2P3, Orsay, France*

*[31] Laboratoire AstroParticule et Cosmologie (APC), Université Paris 7, CNRS-IN2P3, Paris, France*

*[32] Laboratoire de l'Accélérateur Linéaire (LAL), Université Paris 11, CNRS-IN2P3, Orsay, France*

*[33] Laboratoire de Physique Nucléaire et de Hautes Energies (LPNHE), Universités Paris 6 et Paris 7, Paris Cedex 05, France*

[34] Laboratoire de Physique Subatomique et de Cosmologie (LPSC), Université Joseph Fourier, INPG, CNRS-IN2P3, Grenoble, France
[35] SUBATECH, Nantes, France
[36] Bergische Universität Wuppertal, Wuppertal, Germany
[37] Forschungszentrum Karlsruhe, Institut für Kernphysik, Karlsruhe, Germany
[38] Forschungszentrum Karlsruhe, Institut für Prozessdatenverarbeitung und Elektronik, Karlsruhe, Germany
[39] Max-Planck-Institut für Radioastronomie, Bonn, Germany
[40] RWTH Aachen University, III. Physikalisches Institut A, Aachen, Germany
[41] Universität Karlsruhe (TH), Institut für Experimentelle Kernphysik (IEKP), Karlsruhe, Germany
[42] Universität Siegen, Siegen, Germany
[44] Dipartimento di Fisica dell'Università and INFN, Genova, Italy
[45] Università dell'Aquila and INFN, L'Aquila, Italy
[46] Università di Milano and Sezione INFN, Milan, Italy
[47] Dipartimento di Fisica dell'Università del Salento and Sezione INFN, Lecce, Italy
[48] Università di Napoli "Federico II" and Sezione INFN, Napoli, Italy
[49] Università di Roma II "Tor Vergata" and Sezione INFN, Roma, Italy
[50] Università di Catania and Sezione INFN, Catania, Italy
[51] Università di Torino and Sezione INFN, Torino, Italy
[52] Dipartimento di Ingegneria dell'Innovazione dell'Università del Salento and Sezione INFN, Lecce, Italy
[53] Istituto di Astrofisica Spaziale e Fisica Cosmica di Palermo (INAF), Palermo, Italy
[54] Istituto di Fisica dello Spazio Interplanetario (INAF), Università di Torino and Sezione INFN, Torino, Italy
[55] INFN, Laboratori Nazionali del Gran Sasso, Assergi (L'Aquila), Italy
[58] Università di Palermo and Sezione INFN, Catania, Italy
[59] Benemérita Universidad Autónoma de Puebla, Puebla, Mexico
[60] Centro de Investigación y de Estudios Avanzados del IPN (CINVESTAV), México, D.F., Mexico
[61] Instituto Nacional de Astrofisica, Optica y Electronica, Tonantzintla, Puebla, Mexico
[63] Universidad Michoacana de San Nicolas de Hidalgo, Morelia, Michoacan, Mexico
[64] Universidad Nacional Autonoma de Mexico, Mexico, D.F., Mexico
[65] IMAPP, Radboud University, Nijmegen, Netherlands
[66] Kernfysisch Versneller Instituut, University of Groningen, Groningen, Netherlands
[67] NIKHEF, Amsterdam, Netherlands
[68] ASTRON, Dwingeloo, Netherlands
[69] Institute of Nuclear Physics PAN, Krakow, Poland
[70] University of Łódź, Łódź, Poland
[71] LIP and Instituto Superior Técnico, Lisboa, Portugal
[72] J. Stefan Institute, Ljubljana, Slovenia
[73] Laboratory for Astroparticle Physics, University of Nova Gorica, Slovenia
[74] Instituto de Física Corpuscular, CSIC-Universitat de València, Valencia, Spain
[75] Universidad Complutense de Madrid, Madrid, Spain
[76] Universidad de Alcalá, Alcalá de Henares (Madrid), Spain
[77] Universidad de Granada & C.A.F.P.E., Granada, Spain
[78] Universidad de Santiago de Compostela, Spain
[79] Rudolf Peierls Centre for Theoretical Physics, University of Oxford, Oxford, United Kingdom
[81] School of Physics and Astronomy, University of Leeds, United Kingdom
[82] Argonne National Laboratory, Argonne, IL, USA
[83] Case Western Reserve University, Cleveland, OH, USA
[84] Colorado School of Mines, Golden, CO, USA
[85] Colorado State University, Fort Collins, CO, USA
[86] Colorado State University, Pueblo, CO, USA
[87] Fermilab, Batavia, IL, USA
[88] Louisiana State University, Baton Rouge, LA, USA
[89] Michigan Technological University, Houghton, MI, USA
[90] New York University, New York, NY, USA
[91] Northeastern University, Boston, MA, USA
[92] Ohio State University, Columbus, OH, USA
[93] Pennsylvania State University, University Park, PA, USA
[94] Southern University, Baton Rouge, LA, USA
[95] University of California, Los Angeles, CA, USA

*[96] University of Chicago, Enrico Fermi Institute, Chicago, IL, USA*
*[98] University of Hawaii, Honolulu, HI, USA*
*[100] University of Nebraska, Lincoln, NE, USA*
*[101] University of New Mexico, Albuquerque, NM, USA*
*[102] University of Pennsylvania, Philadelphia, PA, USA*
*[103] University of Wisconsin, Madison, WI, USA*
*[104] University of Wisconsin, Milwaukee, WI, USA*
*[105] Institute for Nuclear Science and Technology (INST), Hanoi, Vietnam*
*[‡] Deceased*
*[a] at Konan University, Kobe, Japan*
*[b] On leave of absence at the Instituto Nacional de Astrofisica, Optica y Electronica*
*[c] at Caltech, Pasadena, USA*
*[d] at Hawaii Pacific University*

*Note added: An additional author, C. Hojvat, Fermilab, Batavia, IL, USA, should be added to papers 3,4,5,7,8,10 in this collection*

# Correlation of the Highest Energy Cosmic Rays with Nearby Extragalactic Objects in Pierre Auger Observatory Data

**J. D. Hague* for The Pierre Auger Collaboration†**

*University of New Mexico, Albuquerque, New Mexico USA
†Observatorio Pierre Auger, Av. San Martín Norte 304, (5613) Malargüe, Mendoza, Argentina

***Abstract*. We update the analysis of correlation between the arrival directions of the highest energy cosmic rays observed by the Pierre Auger Observatory and the positions of nearby active galaxies.**

## I. Introduction

Using data collected between 1 January, 2004 and 31 August, 2007, the Pierre Auger Observatory has reported [1] evidence of anisotropy in the arrival directions of cosmic rays (CR) with energies exceeding $\sim 60$ EeV (1 EeV is $10^{18}$ eV). The arrival directions were correlated with the positions of nearby objects from the 12th edition of the catalog of quasars and active galactic nuclei (AGN) by Véron-Cetty and Véron [2] (VCV catalog). This catalog is not an unbiased statistical sample, since it is neither homogeneous nor statistically complete. This is not an obstacle to demonstrating the existence of anisotropy if CR arrive preferentially close to the positions of nearby objects in this sample. The nature of the catalog, however, limits the ability of the correlation method to identify the actual sources of cosmic rays. The observed correlation identifies neither individual sources nor a specific class of astrophysical sites of origin. It provides clues to the extragalactic origin of the CR with the highest energies and suggests that the suppression of the flux (see [3] and [4]) is due to interaction with the cosmic background radiation.

In this article we update the analysis of correlation with AGN in the VCV catalog by including data collected through 31 March, 2009. We also analyse the distribution of arrival directions with respect to the location of the Centaurus cluster and the radio source Cen A. Alternative tests that may discriminate among different populations of source candidates are presented in a separate paper at this conference [5].

## II. Data

The data set analyzed here consists of events observed by the Pierre Auger Observatory prior to 31 March, 2009. We consider events with zenith angles smaller than 60°. The event selection implemented in the present analysis requires that at least five active nearest-neighbors surround the station with the highest signal when the event was recorded, and that the reconstructed shower core be inside an active equilateral triangle of detectors. The integrated exposure for this event selection amounts to 17040 km$^2$ sr yr ($\pm 3\%$), nearly twice the exposure used in [1].

In [1] we published the list of 27 events with $E >$ 57 EeV. Since then, the reconstruction algorithms and calibration procedures of the Pierre Auger Observatory have been updated. The lowest energy among these same 27 events is 55 EeV according to the latest reconstruction. Reconstructed values for the arrival directions of these events differ by less than $0.1^\circ$ from their previous determination. There are now 31 additional events above the energy threshold of 55 EeV. The systematic uncertainty of the observed energy for events used here is $\sim 22\%$ and the energy resolution is $\sim 17\%$ [6], [7]. The angular resolution of the arrival directions for events with energy above this threshold is better than $0.9^\circ$ [8].

## III. Update of the correlation with AGN

To avoid the negative impact of trial factors in *a posteriori* analyses, the statistical significance of the anisotropy reported in [1] was established through a test with independent data. The parameters of the test were chosen by an exploratory scan using events observed prior to 27 May, 2006. The scan searched for a correlation of CR with objects in the VCV catalog with redshift less than $z_{\rm max}$ at an angular scale $\psi_{\rm max}$ and energy threshold $E_{\rm th}$. The scan was implemented to find a minimum of the probability $P$ that $k$ or more out of a total of $N$ events from an isotropic flux are correlated by chance with the selected objects at the chosen angular scale, given by

$$P = \sum_{j=k}^{N} \left( \begin{array}{c} N \\ j \end{array} \right) {p_{\rm iso}}^j (1 - p_{\rm iso})^{N-j} \ . \qquad (1)$$

We take $p_{\rm iso}$ to be the exposure-weighted fraction of the sky accessible to the Pierre Auger Observatory that is within $\psi_{\rm max}$ degrees of the selected potential sources. The minimum value of $P$ was found for the parameters $\psi_{\rm max} = 3.1^\circ$, $z_{\rm max} = 0.018$ and $E_{\rm th} = 55$ EeV (in the present energy calibration). The probability that an individual event from an isotropic flux arrives within the fraction of the sky prescribed by these parameters by chance is $p_{\rm iso} = 0.21$.

Of the 27 events observed prior to 31 August, 2007, 13 were observed after the exploratory phase. Nine of these arrival directions were within the prescribed area of the sky, where 2.7 are expected on average if the

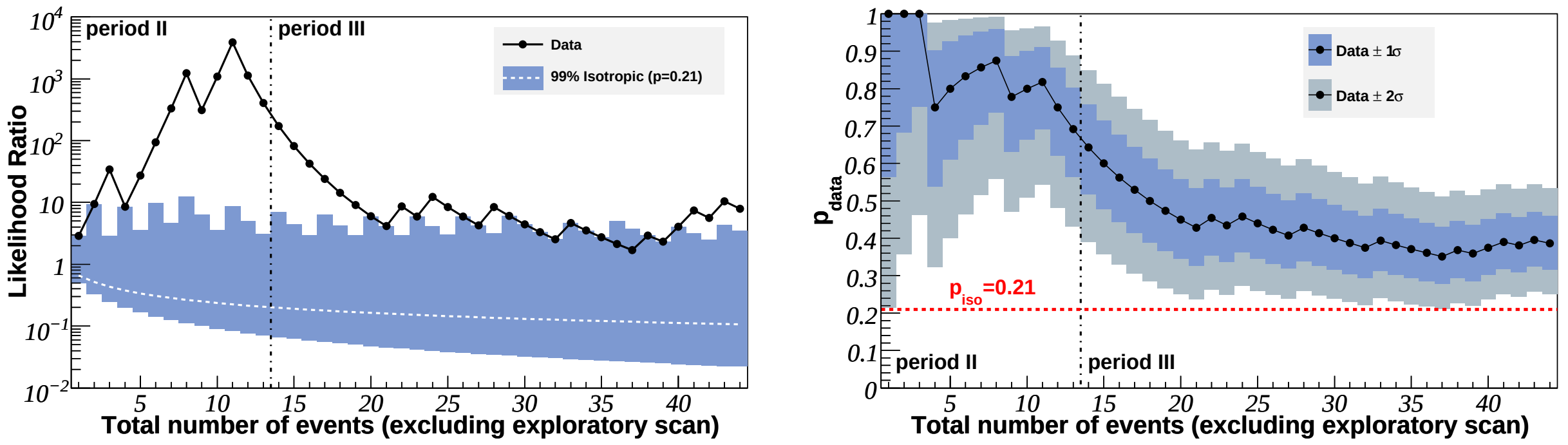

Fig. 1. Monitoring the correlation signal. *Left:* The sequential analysis of cosmic rays with energy greater than 55 EeV arriving after 27 May, 2006. The likelihood ratio $\log_{10} R$ (see Eqn (2)) for the data is plotted in black circles. Events that arrive within $\psi_{\text{max}} = 3.1^\circ$ of an AGN with maximum redshift $z_{\text{max}} = 0.018$ result in an up-tick of this line. Values above the area shaded in blue have less than 1% chance probability to arise from an isotropic distribution ($p_{\text{iso}} = 0.21$). *Right:* The most likely value of the binomial parameter $p_{\text{data}} = k/N$ is plotted with black circles as a function of time. The $1\sigma$ and $2\sigma$ uncertainties in the observed value are shaded. The horizontal dashed line shows the isotropic value $p_{\text{iso}} = 0.21$. The current estimate of the signal is $0.38 \pm 0.07$. In both plots events to the left of the dashed vertical line correspond to period II of Table I and those to the right, collected after [1], correspond to period III.

TABLE I

A NUMERICAL SUMMARY OF RESULTS FOR EVENTS WITH $E \geq 55$ EEV. SEE THE TEXT FOR A DESCRIPTION OF THE ENTRIES.

| Period | Exposure | GP | $N$ | $k$ | $k_{\text{iso}}$ | $P$ |
|---|---|---|---|---|---|---|
| I | 4390 | unmasked | 14 | 9 | 2.9 | |
| | | masked | 10 | 8 | 2.5 | |
| II | 4500 | unmasked | 13 | 9 | 2.7 | $2 \times 10^{-4}$ |
| | | masked | 11 | 9 | 2.8 | $1 \times 10^{-4}$ |
| III | 8150 | unmasked | 31 | 8 | 6.5 | 0.33 |
| | | masked | 24 | 8 | 6.0 | 0.22 |
| II+III | 12650 | **unmasked** | **44** | **17** | **9.2** | $\mathbf{6 \times 10^{-3}}$ |
| | | masked | 35 | 17 | 8.8 | $2 \times 10^{-3}$ |
| I+II | 8890 | unmasked | 27 | 18 | 5.7 | |
| | | masked | 21 | 17 | 5.3 | |
| I+II+III | 17040 | unmasked | 58 | 26 | 12.2 | |
| | | masked | 45 | 25 | 11.3 | |

flux were isotropic. This degree of correlation provided a 99% significance level for rejecting the hypothesis that the distribution of arrival directions is isotropic.

The left panel of Fig. 1 displays the likelihood ratio of correlation as a function of the total number of time-ordered events observed since 27 May, 2006, i.e. excluding the data used in the exploratory scan that lead to the choice of parameters. The likelihood ratio $R$ is defined as (see [9] and [10])

$$R = \frac{\int_{p_{\text{iso}}}^{1} p^k (1-p)^{N-k} \, dp}{{p_{\text{iso}}}^k (1 - p_{\text{iso}})^{N-k+1}} \,. \quad (2)$$

This quantity is the ratio between the binomial probability of correlation – marginalized over its range of possible values and assuming a flat prior – and the binomial probability in the isotropic case ($p_{\text{iso}} = 0.21$). A sequential test rejects the isotropic hypothesis at the 99% significance level (and with less than 5% chance of incorrectly accepting the null hypothesis) if $R > 95$. The likelihood ratio test indicated a 99% significance level for the anisotropy of the arrival directions using the independent data reported in [1]. Subsequent data neither strengthen the case for anisotropy, nor do they contradict the earlier result. The departure from isotropy remains at the 1% level as measured by the cumulative binomial probability ($P = 0.006$), with 17 out of 44 events in correlation.

In the right panel of Fig. 1 we plot the degree of correlation ($p_{\text{data}}$) with objects in the VCV catalog as a function of the total number of time-ordered events observed since 27 May, 2006. For each new event the best estimate of $p_{\text{data}}$ is $k/N$. The $1\sigma$ and $2\sigma$ uncertainties in this value are determined such that the area under the posterior distribution function is equal to 68% and 95%, respectively. The current estimate, with 17 out of 44 events that correlate in the independent data, is $p_{\text{data}} = 0.38$, or more than two standard deviations from the value expected from a purely isotropic distribution of events. More data are needed to accurately constrain this parameter.

The correlations between events with $E \geq 55$ EeV and AGN in the VCV catalog during the pre- and post-exploratory periods of data collection are summarized in Table I. The left most column shows the period in which the data was collected. Period I is the exploratory period from 1 January, 2004 through 26 May, 2006. The data collected during this period was scanned to establish the parameters which maximize the correlation. Period II is from 27 May, 2006 through 31 August, 2007 and period III includes data collected after [1], from 1 September, 2007 through 31 March, 2009. The numbers in bold correspond to period II+III and give the results for the post-exploratory data (see Fig. 1). The exposure for each period is listed in units of $\mathrm{km}^2$ sr yr and has an uncertainty of $3\%$. If the region of the sky within $12°$ of the galactic plane (GP) is included in the analysis then the third column is marked "unmasked" (and $p_{\mathrm{iso}} = 0.21$), if not then it is marked "masked" (and $p_{\mathrm{iso}} = 0.25$). The average number of events from an isotropic flux expected to correlate is listed as $k_{\mathrm{iso}} = Np_{\mathrm{iso}}$, where $N$ is the total number of events observed during each period. $k$ is the number of events that arrive within $3.1°$ of an AGN with a redshift of $0.018$. The cumulative binomial probability (see Eqn (1)) is shown in the right most column. We do not include this value for any row containing period I because this period was used to determine the correlation parameters for the rest of the table and cannot, therefore, be interpreted as a statistical significance.

Note that during period I+II (reported in [1]), 18 out of 27 events arrive within $3.1°$ of an AGN in the VCV catalog with redshift less than $0.018$. [1] There are 31 additional events (during period III) above the specified energy threshold, 8 of which have arrival directions within the prescribed area of the sky, not significantly more than the 6.5 events that are expected to arrive on average if the flux were isotropic.

While the degree of correlation with objects in the VCV catalog has decreased with the accumulation of new data, a re-scan of the complete data set shows that the values of $\psi_{\mathrm{max}}$, $z_{\mathrm{max}}$ and $E_{\mathrm{th}}$ that characterise the correlation have not changed appreciably from the values reported in [1].

## IV. A POSTERIORI ANALYSES

In this section we further analyze the complete set of 58 events with energy larger than 55 EeV collected before 31 March, 2009.

To complement the information given in Table I over different angular scales, we plot in Fig. 2 the distribution of angular separations between the arrival directions of the 58 events with $E > 55$ EeV and the position of the closest object in the VCV catalog within redshift $z_{\mathrm{max}} \leq 0.018$. The cumulative distribution is plotted in the left panel and the differential distribution is plotted in the right. The average distribution expected for 58 events drawn from an isotropic flux is also shown. In the right panel the 13 events with galactic latitudes $|b| < 12°$ have been shaded. Note that only 1 of these 13 events is within $3°$ of a selected AGN. Incompleteness of the VCV catalog due to obscuration by the Milky Way or larger magnetic bending of CR trajectories along the galactic disk are potential causes for smaller correlation of arrival directions at small galactic latitudes.

An excess of events as compared to isotropic expectations is observed from a region of the sky close to the location of the radio source Cen A ($(l, b) = (-50.5°, 19.4°)$ [11]). In Fig. 3 we plot the distribution of events as a function of angular distance from Cen A. In a Kolmogorov-Smirnov [12] test 2% of isotropic realizations have maximum departure from the isotropic expectation greater than or equal to the maximum departure for the observed events. The excess of events in circular windows around Cen A with the smallest isotropic chance probability corresponds to a radius of $18°$, which contains 12 events where 2.7 are expected on average if the flux were isotropic. The (differential) histogram of angular distances from Cen A is in the right panel of Fig. 3.

By contrast, the region around the Virgo cluster is densely populated with galaxies but does not have an excess of events above isotropic expectations. In particular, a circle of radius $20°$ centred at the location of M87 ($(l, b) = (76.2°, 74.5°)$ [11]) does not contain any of the 58 events with energy $E > 55$ EeV. This is a region of relatively low exposure for the Pierre Auger Observatory and only 1.2 event is expected on average with the current statistics if the flux were isotropic.

## V. DISCUSSION

With data collected by the Pierre Auger Observatory between 1 January, 2004 and 31 March, 2009, we have updated the analysis reported in [1] of correlation between the arrival directions of the highest energy cosmic rays and the positions of nearby objects from the 12th edition of the VCV catalog of quasars and active galactic nuclei. The total number of events above 55 EeV is 58. A subset of 44 events are independent of those used to determine the parameters ($\psi_{\mathrm{max}} = 3.1°$, $z_{\mathrm{max}} = 0.018$ and $E_{\mathrm{th}} = 55$ EeV) with which we monitor the correlation signal (see Table I for more details). 17 of these 44 events correlate under these parameters. This correlation has a less than 1% probability to occur by chance if the arrival directions are isotropically distributed. The evidence for anisotropy has not strengthened since the analysis reported in [1]. The degree of correlation with objects in the VCV catalog appears to be weaker than suggested by the earliest data.

We note that there is an excess of events in the present data set close to the direction of the radio source Cen A, a region dense in potential sources. This excess is based on *a posteriori* data but suggests that the region of the sky near Cen A warrants further study.

Additional data are needed to make further progress in the quest to identify the sites of ultra high energy CR origin. Alternative tests that may discriminate among different populations of source candidates are presented in a separate paper at this conference [5].

[1] Two additional events correlate within a slightly larger angular distance, as reported in [1]. Here we restrict the analysis to the parameters chosen to monitor the correlation signal.

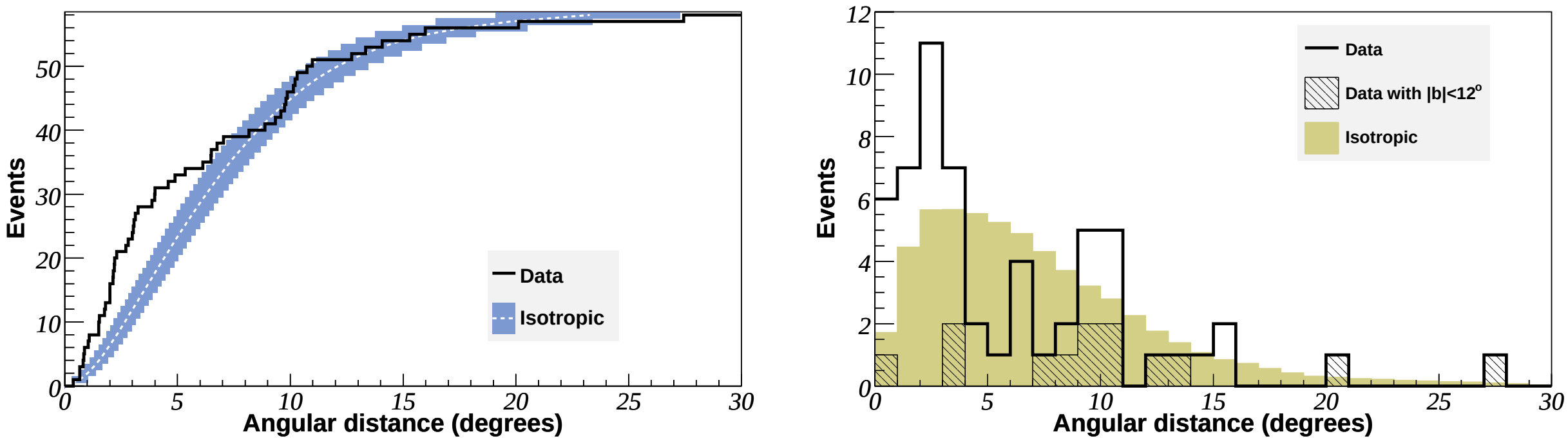

Fig. 2. The distribution of angular separations between the 58 events with $E > 55$ EeV and the closest AGN in the VCV catalog within 75 Mpc. *Left:* The cumulative number of events as a function of angular distance. The 68% the confidence intervals for the isotropic expectation is shaded blue. *Right:* The histogram of events as a function of angular distance. The 13 events with galactic latitudes $|b| < 12°$ are shown with hatching. The average isotropic expectation is shaded brown.

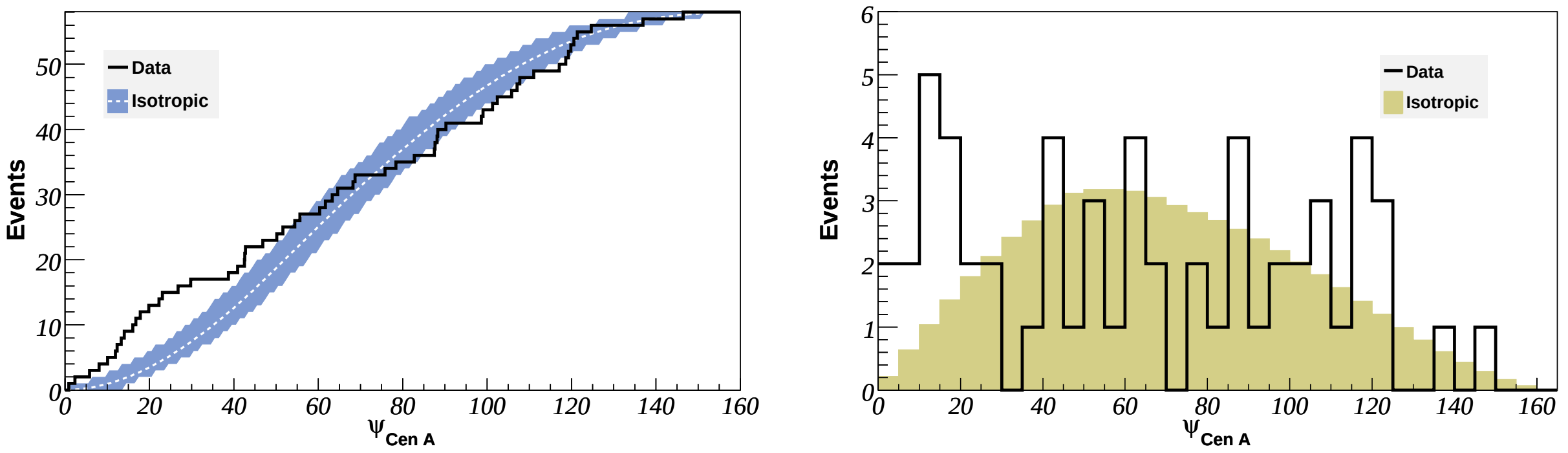

Fig. 3. *Left:* The cumulative number of events with $E \geq 55$ EeV as a function of angular distance from Cen A. The average isotropic expectation with approximate 68% confidence intervals is shaded blue. *Right:* The histogram of events as a function of angular distance from Cen A. The average isotropic expectation is shaded brown.

# Discriminating potential astrophysical sources of the highest energy cosmic rays with the Pierre Auger Observatory

**Julien Aublin***, **for the Pierre Auger Collaboration**†

* *LPNHE, Université Paris 6, 4 place Jussieu, 75252 Paris Cedex 05, France*
† *Pierre Auger Observatory, av. San Martín Norte 304, (5613) Malargüe, Argentina.*

***Abstract.*** **We compare the distribution of arrival directions of the highest energy cosmic rays detected by the Pierre Auger Observatory from 1 January 2004 to 31 March 2009 with that of populations of potential astrophysical sources. For this purpose, we use several complementary statistical tests allowing one to describe and quantify the degree of compatibility between data and a given catalogue of sources. We applied these tests to active galactic nuclei detected in X-rays by SWIFT-BAT and to galaxies found in the HI Parkes and in the 2 Micron All-Sky Surveys.**

## I. Introduction

The origin and nature of the ultra high energy cosmic rays (UHECRs) are still unknown after more than half a century since their discovery. The deflections encountered by UHECRs during their propagation through galactic and extra galactic magnetic fields make a direct identification of their sources difficult.

Recently, the Pierre Auger Collaboration reported a correlation between the arrival directions of the highest energy events observed ($E \geq 6 \times 10^{19}$ eV) and the directions of known active galaxies closer than 100 Mpc [1], [2]. An update of this analysis [3] with data collected up to 31 March 2009 shows that the evidence for anisotropy remains at the 99% confidence level, although the correlation has not strenghtened. In any case, this result does not imply that AGN are indeed the actual sources of UHECRs, since many other source scenarios could in principle reproduce the observed data.

In the present study, we compare the arrival directions of data from the Pierre Auger Observatory with the position of potential astrophysical sources. First, we compute the standard cross-correlation function between the observed arrival directions and a volume-selected sample of galaxies from the 2MRS [4] catalogue, under the simple assumption of an equal contribution of each source to the cosmic ray flux. Then we compare our data with three different catalogues (X-ray AGNs detected by SWIFT [5], galaxies in the HI-Parkes [6], [7] and 2MRS surveys) taking into account the intrinsic luminosity and the distance of the sources. For this comparison, we use two complementary methods: a likelihood test and a test based on the scalar product of functions on the sphere.

## II. Data set

The data set consists of 58 events recorded by the Pierre Auger Observatory from 1 January 2004 to 31 March 2009, with energies reconstructed above $55$ EeV and zenith angles smaller than $60°$. The energy resolution is $17\%$ , with a systematic uncertainty of 22%[8]. The angular resolution, defined as the angular radius that would contain 68% of the reconstructed events is $\leq 0.9°$. We use the energy threshold that maximizes the departure from isotropy through the correlation with AGN [1]. This particular value corresponds to the region where the energy spectrum of UHECRs [8] presents a significant deviation from the power-law extrapolated from lower energy. This supports the idea of a sharp reduction of the cosmic rays horizon due to the GZK effect [9], [10], at energies greater than $\simeq 50-60$ EeV, limiting drastically the number of contributing sources.

## III. Astrophysical catalogues

Recent analysis comparing Auger data with the SWIFT-BAT and HIPASS catalogues can be found in [11], [12], [13]. The 22 months SWIFT-BAT [5] catalogue provides the most uniform all-sky hard X-ray survey to date, it contains a total of 261 Seyfert galaxies and AGN. The HIPASS [6], [7] galaxy catalogue contains a large number of extragalactic HI sources that could host preferentially GRBs and magnetars producing UHECRs. Here, we adopt the flux limit $S_{int} > 9.4$ Jy km s$^{-1}$ following [13], leading to a total number of 3058 galaxies. As in [12], we also consider a sub-sample of the HIPASS catalogue, that we call HIPASS HL in the following text, that contains the 765 most luminous galaxies. We use the compilation (2MRS) provided by Huchra et al. [4] of the redshifts of the $K_{\rm mag} < 11.25$ brightest galaxies from the 2MASS[14] catalogue. The catalogue, containing $\simeq$ 23000 sources, provides an excellent image of the distribution of local matter. For the cross-correlation analysis, we use a volume-selected sample of galaxies from the 2MRS catalogue (2MRS VS thereafter) to prevent a bias toward the faint galaxies at small distances. We select galaxies with 10 Mpc $<$ $d <$ 200 Mpc and absolute magnitudes $M_k < -25.25$, leading to 1940 objects in this sub-sample. For 2MRS, we exclude from the analysis UHECRs events (and sources for 2MRS VS) that have $|b| < 10°$ to avoid a bias due to incompleteness in the galactic plane region.

## IV. Data analysis

### A. Cross-correlation

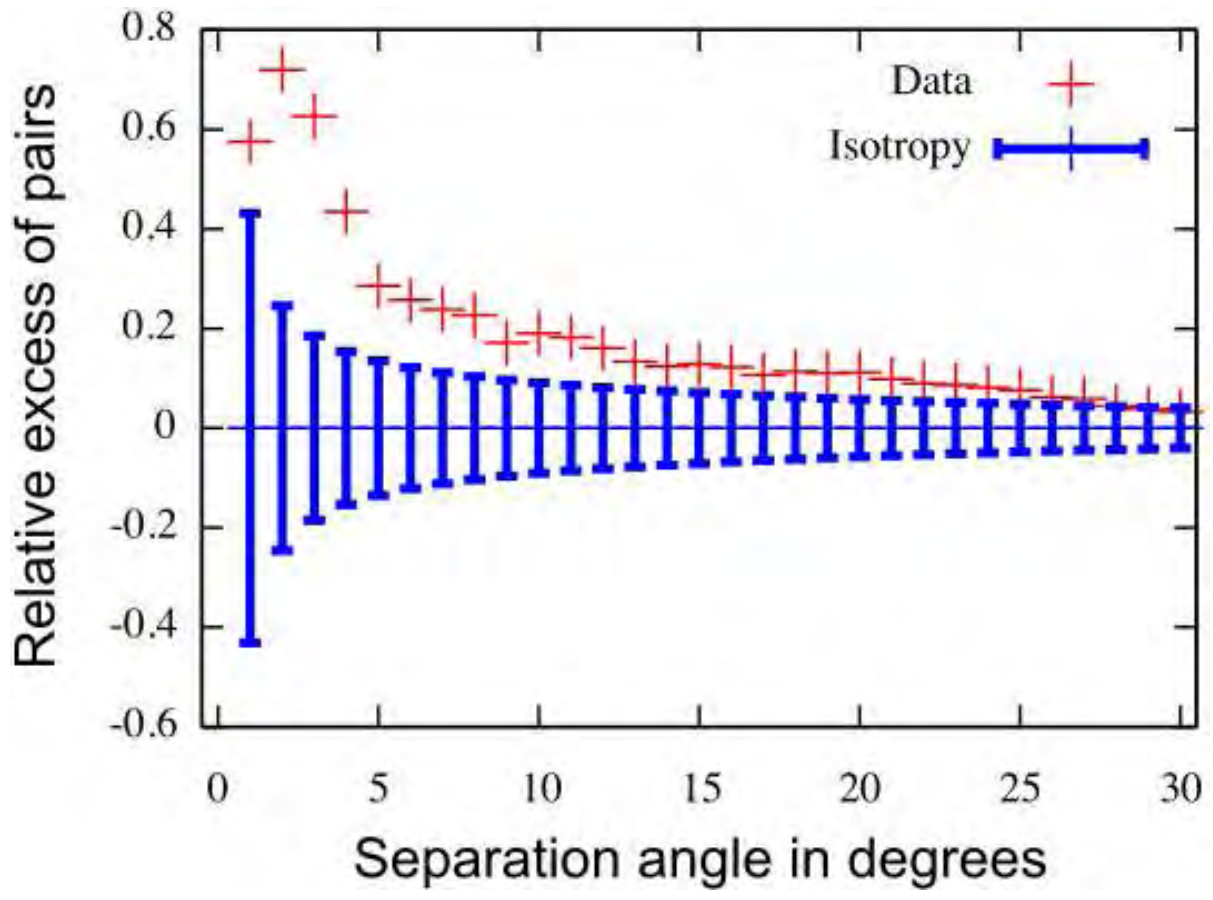

Fig. 1. Cross-correlation between Auger data and 2MRS VS galaxies: cumulative relative excess of pairs as a function of the separation angle. Error bars represent the dispersion in 68% of isotropic realizations.

The cross-correlation function with 2MRS VS is shown in figure 1. There is a clear excess of pairs for separation angles smaller than 30°. The most significant departure from isotropy is observed at 3°: the fraction of simulated samples that give a higher number of pairs than in the data is $f \simeq 1.5 \times 10^{-3}$. Under the basic assumption of an equal contribution of each source to the cosmic ray flux, this simple result is an indication that the arrival directions of the UHECRs are partially correlated to the local distribution of matter.

### B. Smoothed density maps

For each catalogue, we can build a density map with two free parameters: the smoothing angle and the fraction of isotropic background. Because of the lack of strong physical input for these parameters, we use the data to determine their best fit values for each catalogue. The smoothed maps are described by a function $F_c(\mathbf{n})$, which is normalized such that its value in a given direction $\mathbf{n}$ corresponds to the predicted probability of detecting a cosmic ray in that direction, according to the model.

We add an isotropic background in the density map as a free parameter to account for the missing flux, for it is very likely that the catalogue does not contain all the cosmic rays sources.

We write the function $F_c(\mathbf{n})$ as :

$$F_c(\mathbf{n}) = I^{-1}\,\varepsilon(\mathbf{n})\mu(\mathbf{n}) \left[\frac{f_{\rm iso}}{\Omega} + (1-f_{\rm iso})\frac{\phi_c(\mathbf{n})}{\langle\phi\rangle}\right]$$

where $\phi_c(\mathbf{n})$ is the flux coming from the catalogue objects and $f_{\rm iso}$ the fraction of isotropic background, the quantities $\Omega = \int {\rm d}\Omega'\mu(\mathbf{n})$ and $\langle\phi\rangle = \int {\rm d}\Omega\mu(\mathbf{n})\phi_c(\mathbf{n})$ accounting for the normalization. The relative exposure of the Pierre Auger Observatory is taken into account by a purely geometrical function $\varepsilon(\mathbf{n})$ computed analytically. The catalogue mask function $\mu(\mathbf{n})$ is equal to 0 in the regions of the sky that must be removed, and 1 elsewhere. Finally, the global normalization constant $I$ ensures that the integral of $F_c(\mathbf{n})$ is equal to unity.

The flux coming from the $N_{\rm cat}$ sources is given by:

$$\phi_c(\mathbf{n}) = \sum_{i=1}^{N_{\rm cat}} w(z_i)\; e^{-\frac{d(\mathbf{n_i},\mathbf{n})^2}{2\sigma^2}}$$

where $d(\mathbf{n_i},\mathbf{n})$ is the angle between the direction of the source $\mathbf{n_i}$ and the direction of interest $\mathbf{n}$. The free parameter $\sigma$ (smoothing angle) enables us to take into account the angular resolution of the Pierre Auger Observatory and the deflections experienced by cosmic rays, under the simplifying assumption that these deflections are purely random and gaussian. A weight $w(z_i)$ is attributed to the $i$th source located at redshift $z_i$. In this study, we assume a weight proportional to the flux $\phi_i$ of the source, measured in a given range of wavelengths (X-rays for SWIFT-BAT, radio for HIPASS and near IR for 2MRS), multiplied by an attenuation factor due to the GZK suppression, that is implemented following [13].

For each catalogue, we find the values of $\sigma$ and $f_{\rm iso}$ that maximize the log-likelihood of the data sample:

$$\mathcal{LL} = \sum_{k=1}^{N_{\rm data}} \ln F_c(\mathbf{n_k})$$

where $\mathbf{n_k}$ is the direction of the $k$th event. The results are shown in fig. 2 (a). We find $(\sigma, f_{\rm iso}) = (7.1°, 0.65)$ for SWIFT-BAT, $(1.4°, 0.7)$ for 2MRS, $(6°, 0.64)$ for HIPASS and $(5.3°, 0.67)$ for HIPASS HL. For the following analyses, we use these parameters, though they are not strongly constrained with the present statistics.

### C. Results of the likelihood test

Finding the values of $\sigma$ and $f_{\rm iso}$ that maximize the log-likelihood does not ensure that the model fits well the data. To test the compatibility between data and model, we generate $10^4$ simulated data samples, containing the same number of events as in the data. The points are either drawn from the model density map or isotropically, and we compare the distributions of the mean log-likelihood ($\mathcal{LL}/N_{\rm data}$) with the value obtained for the data. The results are illustrated in fig. 2 (b).

The data are in agreement with all models and significantly different from isotropic expectations. The fraction of simulated isotropic realizations that give a higher value than the data is around $10^{-5}$ for SWIFT, $10^{-3}$ for 2MRS and HIPASS HL and $10^{-2}$ for HIPASS.

The likelihood test is, by its intrinsic nature, only sensitive to the fact that data points lie or not in high density regions of the catalogue. We thus use a complementary method to test if all regions of the catalogue are fairly represented in the data set.

### D. The 2-fold correlation coefficients method

This test is based on the computation of two coefficients characterizing the compatibility between the smoothed density map of the model and a similar density map computed for the data. We apply a gaussian filtering to the $N_{\rm data}$ data points to obtain the following density map:

$$F_d(\mathbf{n}) = \frac{\mu(\mathbf{n})}{2\pi\sigma^2\, N_{\rm data}} \sum_{j=1}^{N_{\rm data}} \exp\left(-\frac{d(\mathbf{n_j},\mathbf{n})^2}{2\sigma^2}\right)$$

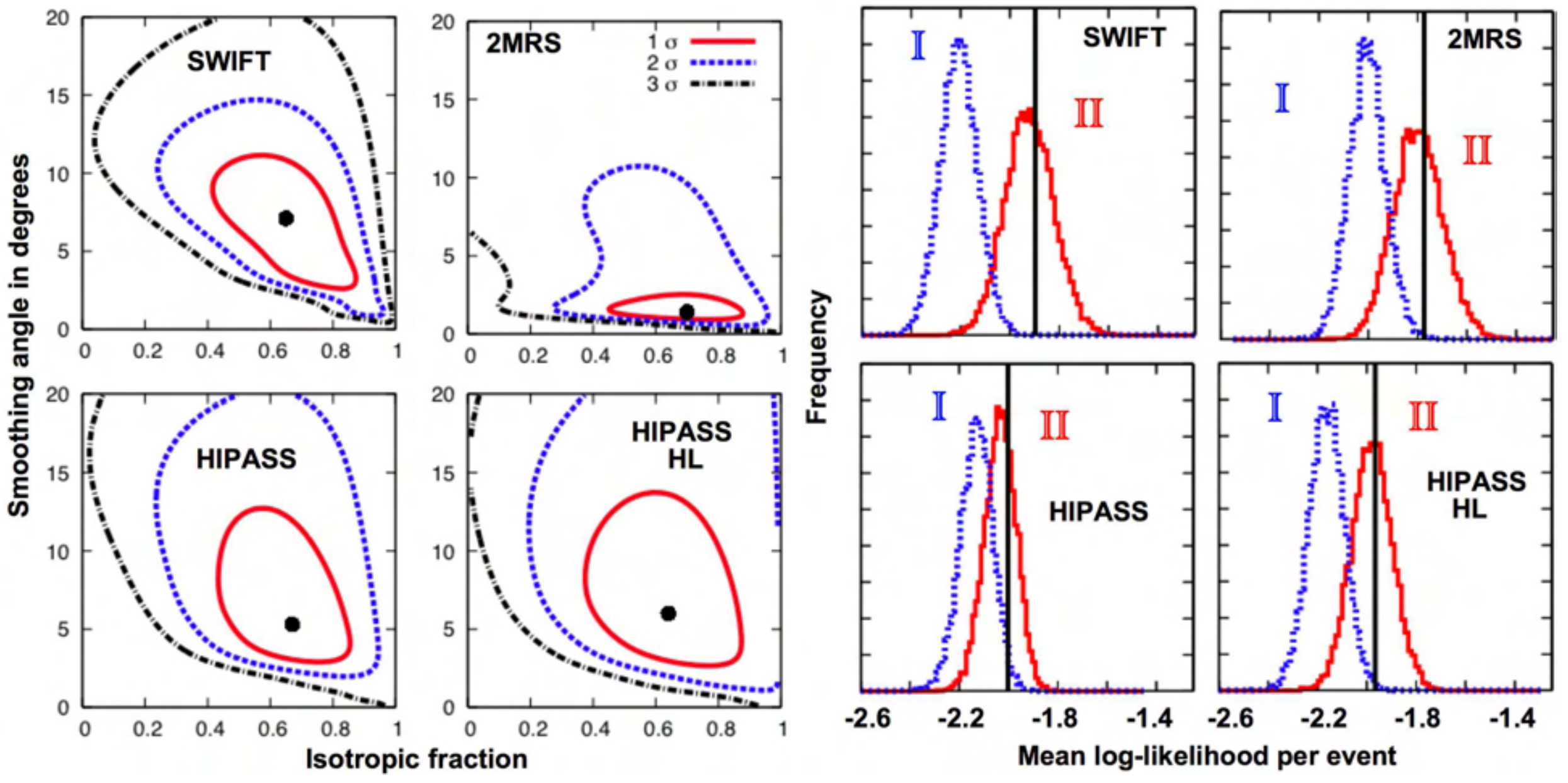

(a) (b)

Fig. 2. (a) Probability contours for the log-likelihood maximisation. The maximum is indicated by a black point. (b) Distributions of mean log-likelihood per event for the isotropy (labelled as I) and for the model (II). Data is indicated by a black vertical line.

The first coefficient, called ”correlation coefficient” is given by:

$$C(F_d,F_c) = \frac{\int F_c(\mathbf{n})F_d(\mathbf{n})\,\mathrm{d}\Omega}{\sqrt{\int \left(F_c(\mathbf{n})\right)^2 \mathrm{d}\Omega\ \int \left(F_d(\mathbf{n})\right)^2 \mathrm{d}\Omega}}$$

This coefficient ranges from 0 ($F_c$ and $F_d$ are anticorrelated) to 1 ($F_c$ and $F_d$ are identical). A high value of $C(F_d,F_c)$ indicates a good match between data and model distributions. The second coefficient, called ”concentration coefficient” is defined by:

$$I_{dd} = \int F_d^2(\mathbf{n})\,\mathrm{d}\Omega.$$

This second observable carries the information about the intrinsic clustering properties of the angular distribution of the data. The magnitude of $I_{dd}$ depends on the density map contrast: it is maximum if all the data points have the same position on the sky, and minimum if the points are uniformly distributed on the sphere. These coefficients are related to the standard two point cross and auto-correlation functions.

For each model, we generate $10^4$ simulated samples containing the same number of events as in the data. The points are either drawn from the model density map or isotropically. Fulfilling the test requires that both $C$ and $I_{\rm dd}$ distributions obtained with simulated sample are compatible with the values computed with the data.

The results of the test are shown in fig. 3. The data are compatible with all models, the map based on SWIFT-BAT gives, as in the likelihood test, the most discriminant test against isotropy. The fraction of isotropic simulations that have both a higher correlation and concentration coefficients than the data is $\sim 4{\times}10^{-3}$ for HIPASS and lower than $10^{-4}$ for SWIFT-BAT.

## V. Conclusion

The Pierre Auger Observatory has recorded 58 cosmic rays with energies $E > 55$ EeV between 1 January 2004 and 31 March 2009. Different complementary tests are applied to extract information about the compatibility between the arrival directions of Auger events and models based on catalogues of potential astrophysical sources or isotropic distributions.

When performing a cross-correlation analysis with the 2MRS VS catalogue, we find an excess of pairs over a range of angular scales, indicating that the UHECRs may be partially correlated to the distribution of local matter (under a basic ”equal flux” assumption). We then apply two other tests that require the computation of smoothed maps of expected cosmic ray flux. The maps have two free parameters, that are determined through the maximization of the likelihood of the data.

The log-likelihood and the 2-fold correlation coefficients tests show that our data are different from isotropic expectations and compatible with the models based on SWIFT-BAT, 2MRS and HIPASS catalogues with the parameters maximizing the likelihood. Within one standard deviation, these parameters are $\sigma \leq 10°$ and $f_{\rm iso} \in [0.4; 0.8]$. The map based on SWIFT-BAT gives the most discriminant test against isotropy.

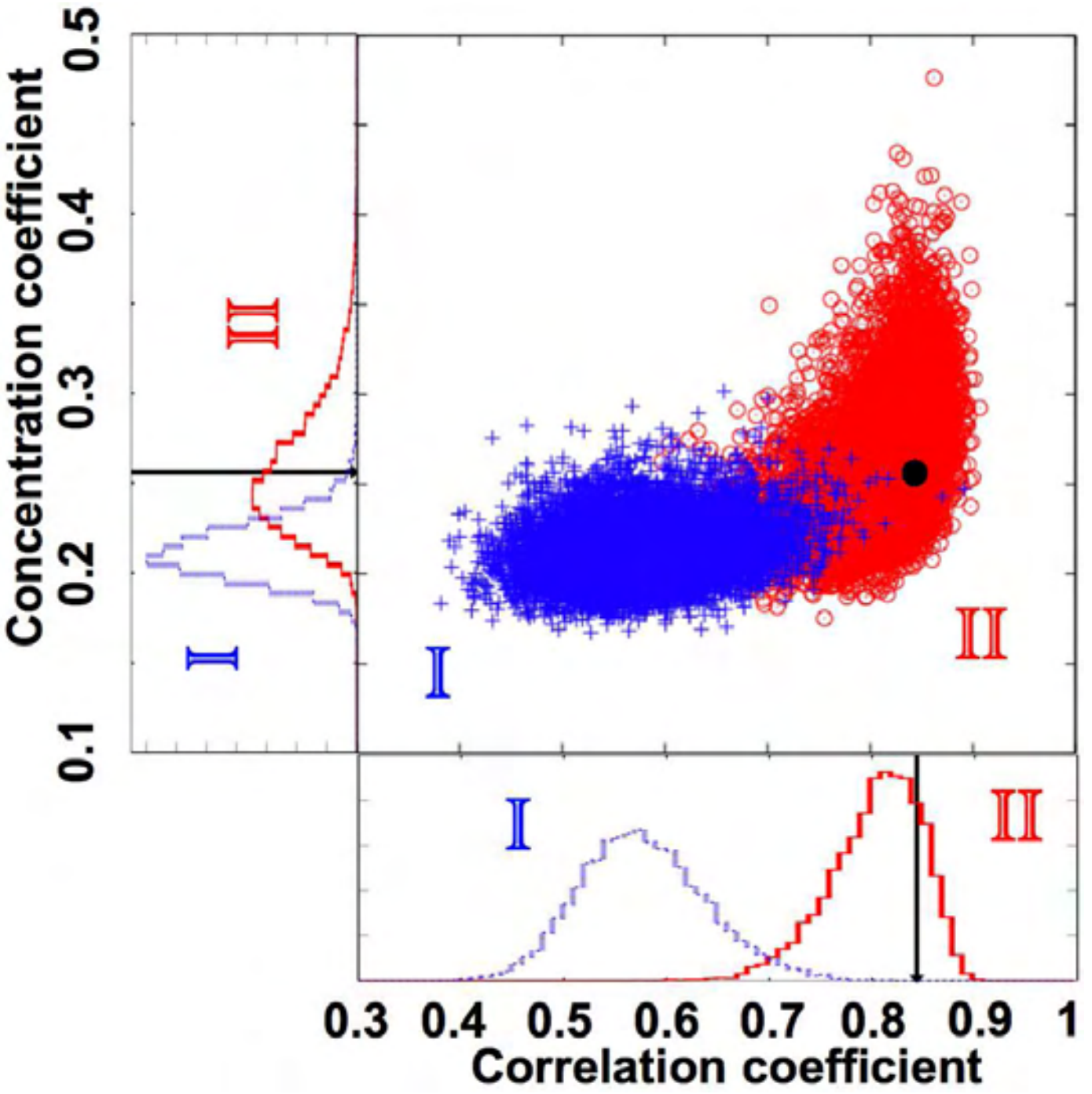

(a) SWIFT-BAT

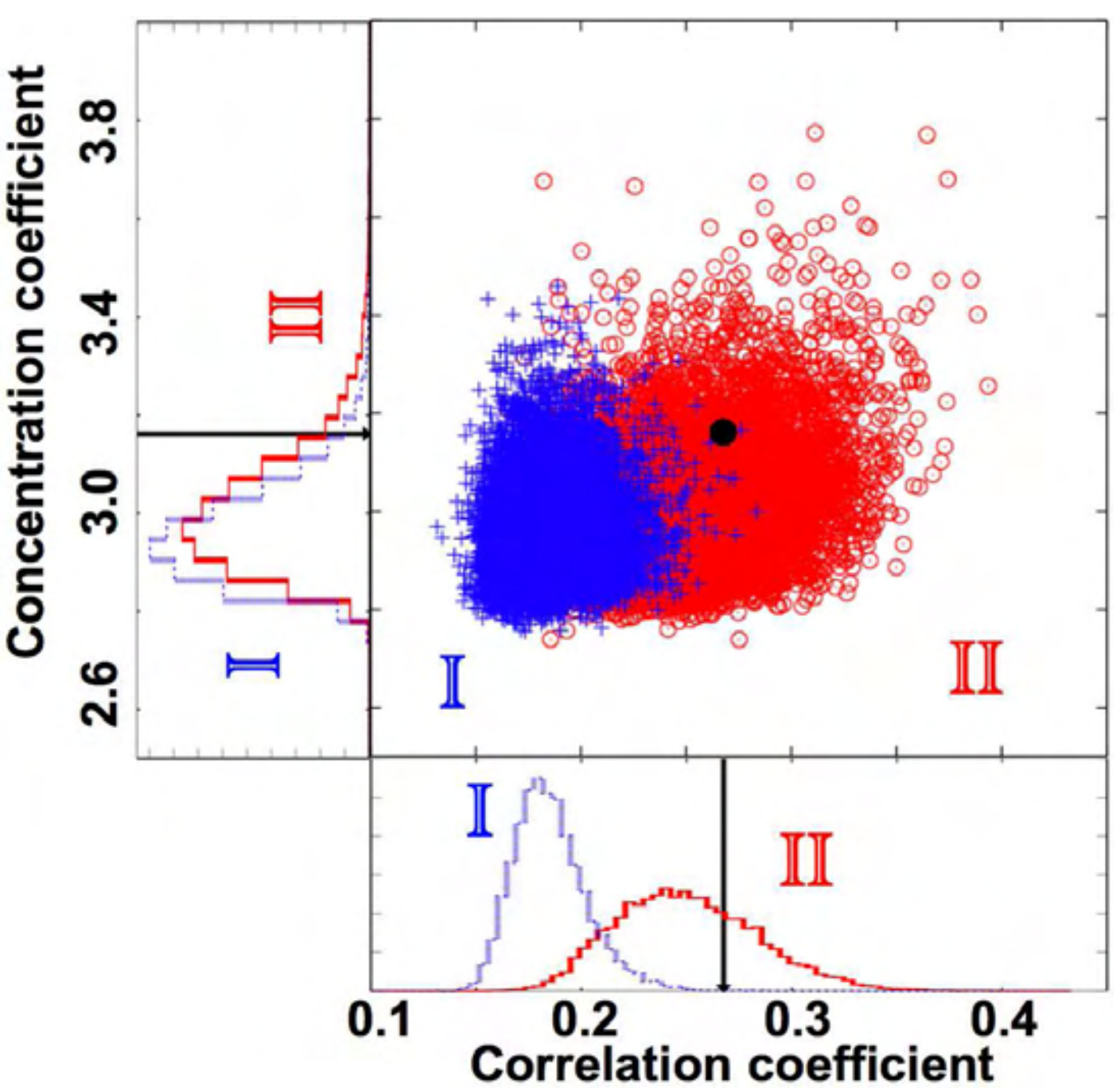

(b) 2MRS

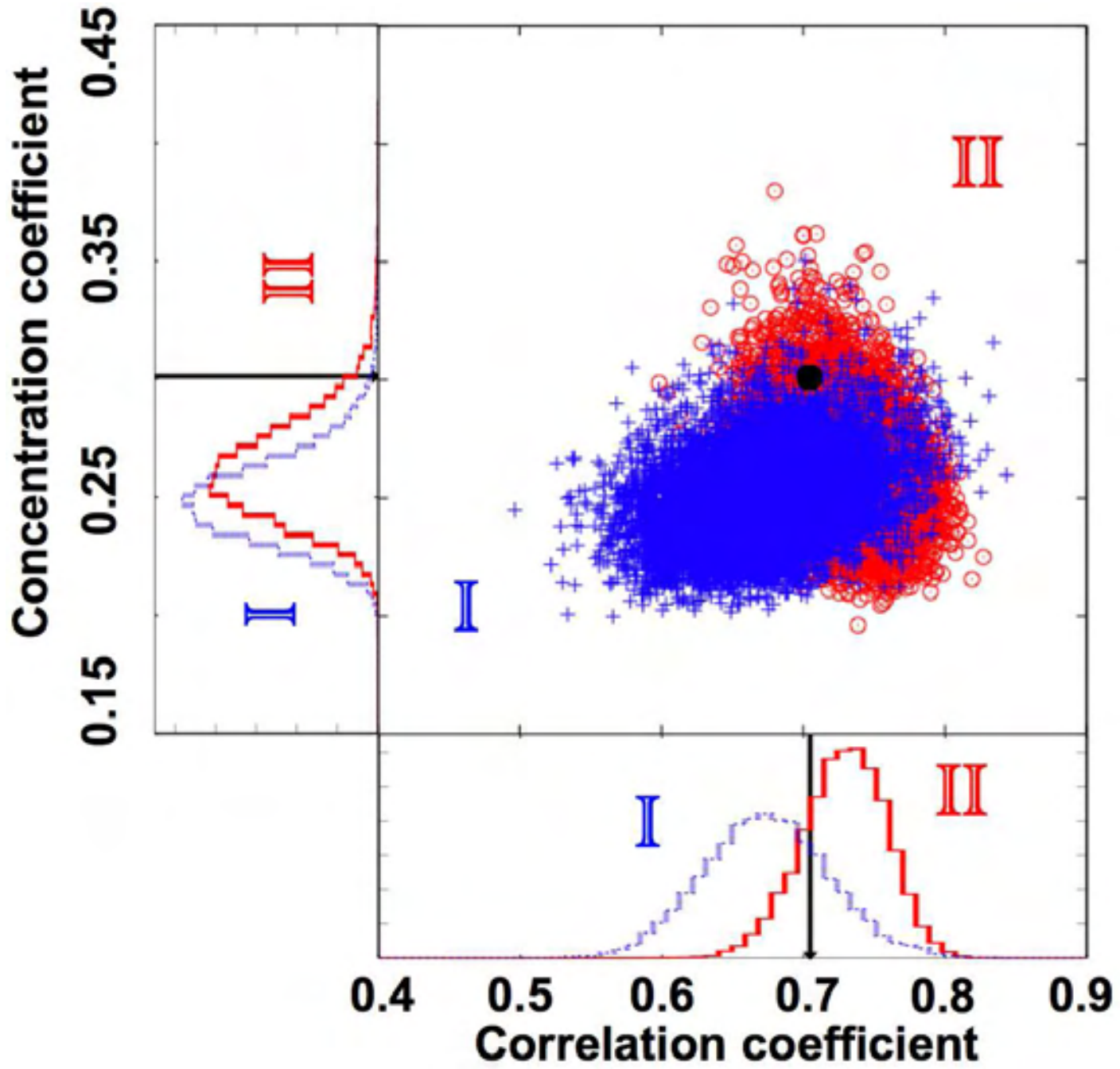

(c) HIPASS

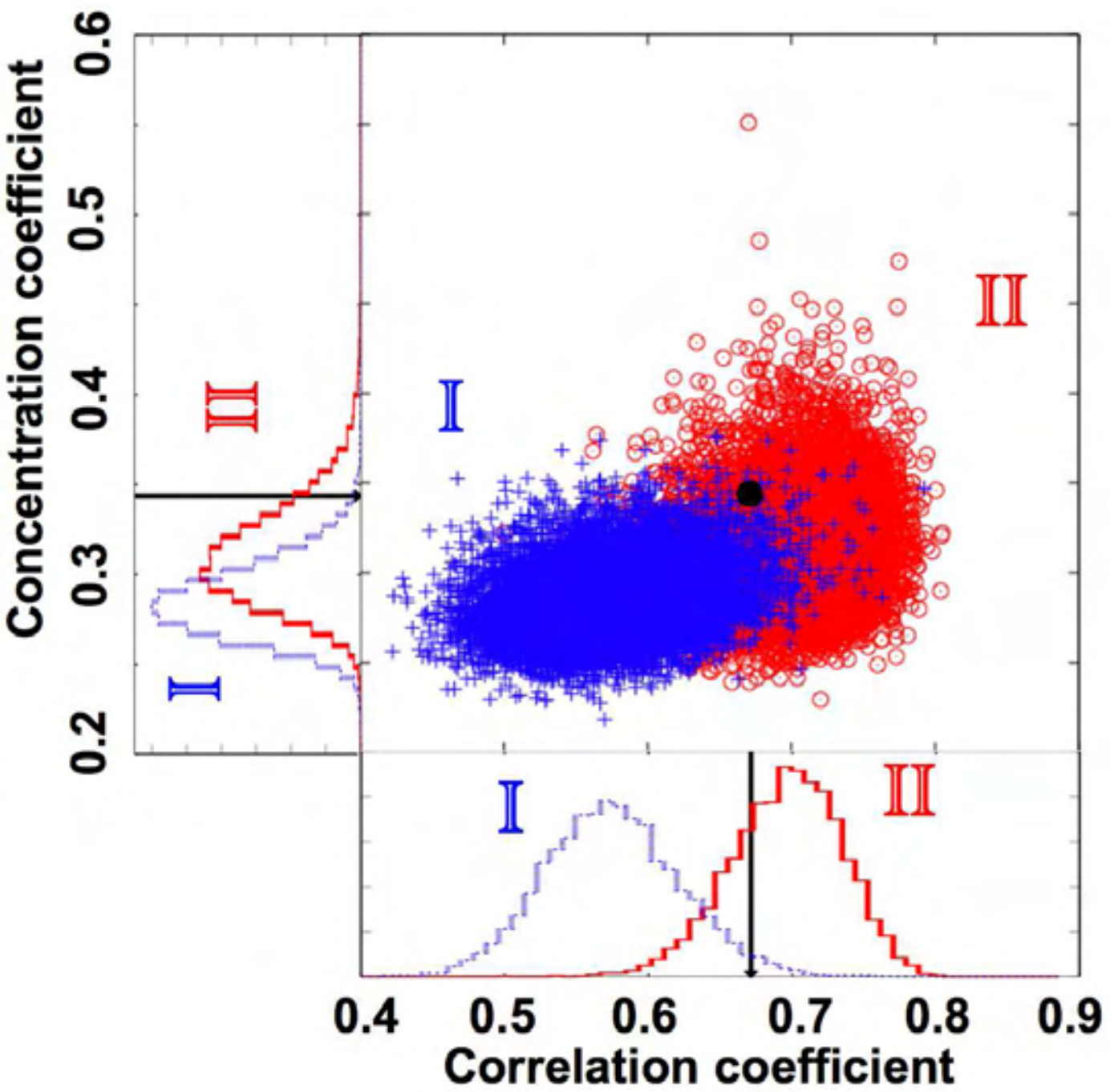

(d) HIPASS High Luminosity

Fig. 3. Two dimensional distributions of the correlation and concentration coefficient for isotropic simulations (labelled as I) and for the model (labelled as II). The value obtained with data is indicated by a black point. The individual distributions of the correlation and concentration coefficients are shown, the data being indicated by a vertical line.

# Search for intrinsic anisotropy in the UHECRs data from the Pierre Auger Observatory

**J. R. T. de Mello Neto***, **for the Pierre Auger Collaboration**†

**Instituto de Física, Universidade Federal do Rio de Janeiro, Ilha do Fundão, Rio de Janeiro, Brazil*
*† Pierre Auger Observatory, Av. San Martín Norte 304, (5613) Malargue, Argentina*

***Abstract.*** **We discuss techniques which have been developed for determining the intrinsic anisotropy of sparse ultra-high-energy cosmic ray datasets, including a two point, an improved two point and a three point method. Monte-Carlo studies of the sensitivities of these tests are presented. We perform a scan in energy above the 100 highest energy events (corresponding to $\simeq$43 EeV) detected at the Pierre Auger Observatory and find that the largest deviation from isotropic expectations occurs for events above 52 EeV.**

## I. INTRODUCTION

The origin of ultra-high energy cosmic rays (UHECRs) with energies greater then $10^{18}$ eV has been a longstanding mystery since their discovery about 50 years ago [1]. The Pierre Auger Collaboration has recently shown that the flux of cosmic rays is strongly suppressed above $4 \times 10^{19}$ eV [2], providing evidence for the 1966 prediction of Greisen [3] and of Zatsepin and Kuz'min [4] (GZK). The effect of energy losses combined with the anisotropic distribution of matter in the 100 Mpc volume around us suggests that cosmic rays at the highest energies are likely to be distributed anisotropically. This expectation of anisotropy above the GZK threshold was verified in 2007 [5], [6], when the Auger Collaboration reported an evidence for anisotropy at a C.L. of at least 99% using the correlation of the cosmic rays detected at the Pierre Auger Observatory with energies above $\sim 6 \times 10^{19}$ eV and the positions of the galaxies in the Veron-Cetty & Veron [7] (VCV) catalogue of active galactic nuclei (AGNs).

Here we report on tests designed to answer the question of whether the arrival directions of the highest-energy events observed by Auger are consistent with being drawn from an isotropic distribution, with no reference to an association with AGN or other extragalactic objects. The goal is to test for anisotropy using only the cosmic-ray data.

## II. STATISTICAL METHODS

At the highest energies, the steepening of the energy spectrum makes the current statistics so small that a measure of a statistically significant departure from isotropy is hard to establish, especially when using blind generic tests. This motivated us to test several methods by challenging their power for detecting anisotropy using simulated samples with few data points (typically less than 100) drawn from different kinds of anisotropies both in large and small scales. We report in this paper on auto-correlation analyses, using differential approaches based on a 2pt function, an extended 2pt function (refered to as 2pt+ in the following), and a 3pt function.

The standard 2pt function [8] was used as a reference. We histogrammed the number of event pairs within a given angular distance in bins of 5° and compared it to the isotropic expectation obtained from a large number (typically $10^6$) of Monte-Carlo samples. The departure from isotropy is then measured through a pseudo-log-likelihood $\Sigma_P$:

$$\Sigma_P^{data} = \sum_{i=1}^{N} \ln \mathcal{P}(n_{obs}^i | n_{exp}^i),$$

where $n_{obs}^i$ and $n_{exp}^i$ are the observed and expected number of event pairs in bin $i$ and $\mathcal{P}$ the Poisson distribution. The resulting $\Sigma_P^{data}$ is then compared to the distribution of $\Sigma_P$ obtained from isotropic Monte-Carlo samples. The probability $P$ for the data to come from the realisation of an isotropic distribution is calculated as the fraction of samples having $\Sigma_P$ lower than $\Sigma_P^{data}$.

A statistics was constructed to add the orientation information of the event pairs to the 2pt information[9]. The new estimator, 2pt+, is calculated on the data and on a large set of Monte-Carlo samples in the same way than in the 2pt case. Again, the departure from isotropy is measured by the fraction of samples giving a 2pt+ estimate smaller than the data.

Finally, we also constructed a 3pt method based on [10] where, for each triangle defined by a triplet of data points, a shape (round or elongated) and a strength (small or big) parameter can be calculated. The 2 dimensional distribution of these parameters from the data is then compared to the average expectation from a large set of Monte-Carlo samples of the same size by means of the same log-likelihood method with Poisson statistics than in the previous cases. More details can be found in [11].

## III. MONTE-CARLO STUDIES

A test is usually defined in terms of a *threshold* $\alpha$, which is the probability against the wrong rejection of the null hypothesis (in our case wrongly rejecting isotropy or claiming an anisotropy while there is not), and of a *power* $1 - \beta$, which is the probability to successfully claim anisotropy when it exits. A good test is a test that for a given number of events and a given threshold $\alpha$ has a high power $1 - \beta$. When the power of a test is less than 90%, the test may often miss a true signal. In this section, we present the power of the three tests at different thresholds $\alpha$ as a function of the number of events, based on mock samples inspired from the correlation of UHECRs with nearby extragalactic objects we reported in [5], [6].

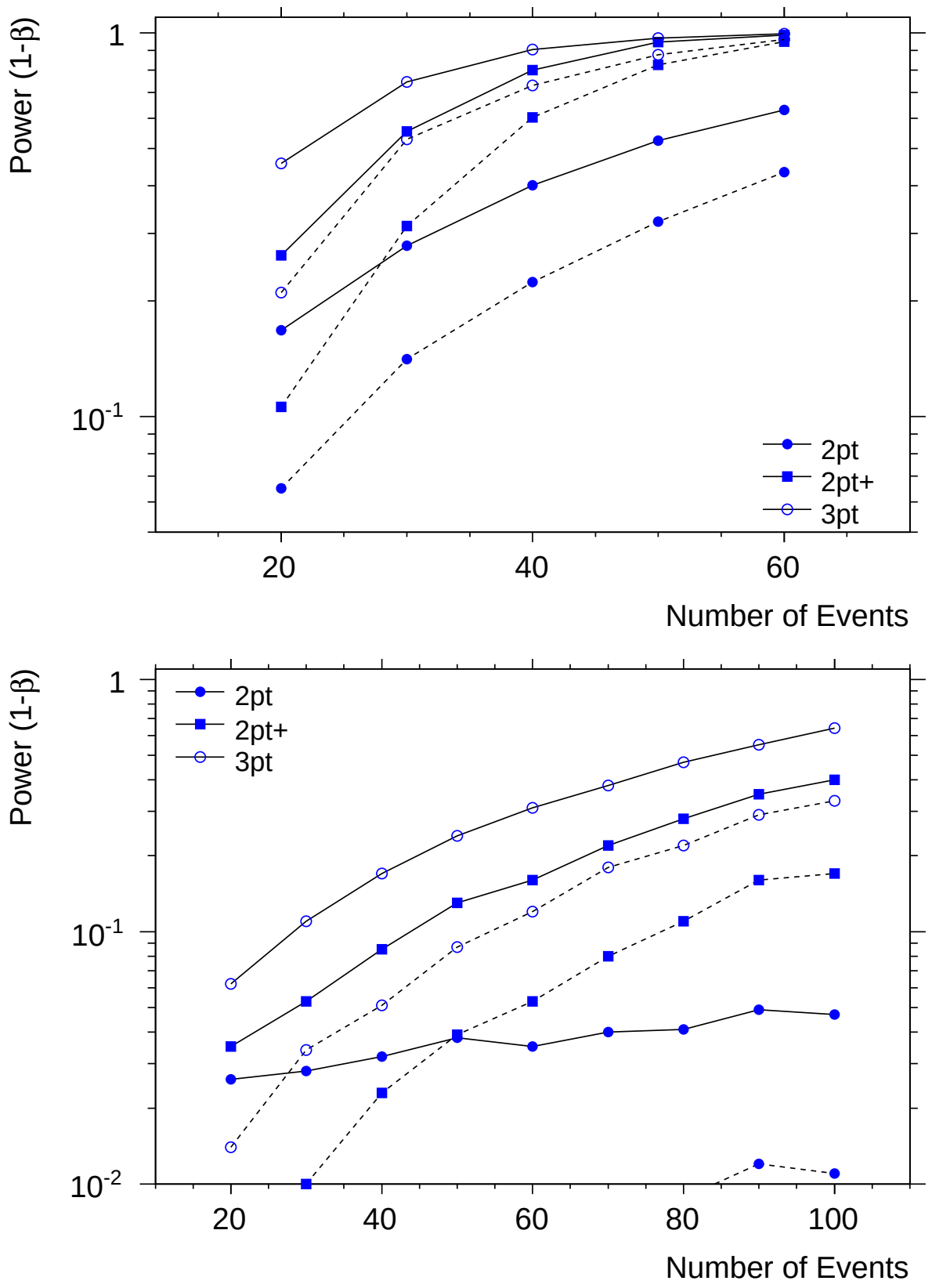

Fig. 1. Power of the 2pt (filled circles), 2pt+ (filled squares) and 3pt (empty circles) tests as a function of the number of events, for 2 threshold values: 1% (lines) and 0.1% (dotted lines). Events are drawn from nearby ($z \leq 0.018$) AGN from [7] with no isotropic background (upper panel) and 50% isotropic background (lower panel).

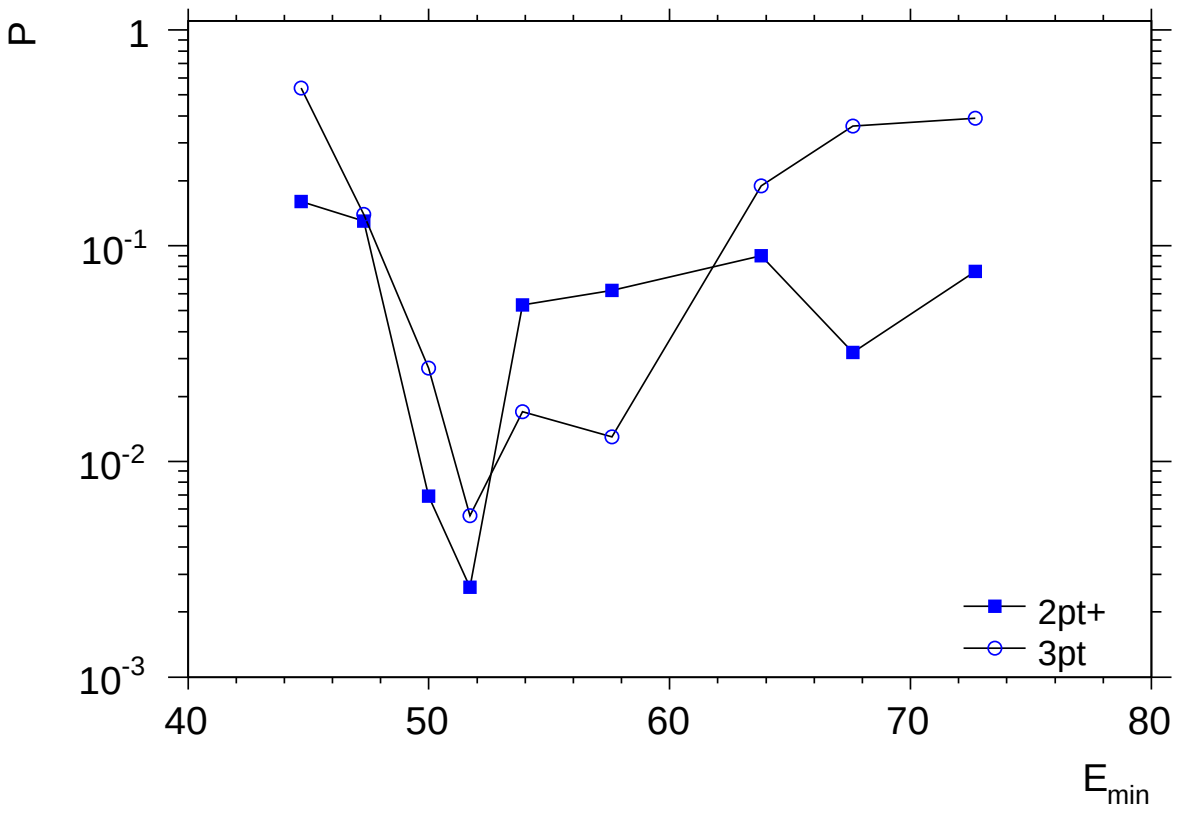

Fig. 2. Significance of the anisotropy in the highest energy events as a function of $E_{min}$. Filled squares (empty circles) are the probability values calculated using the 2pt+ method (3pt method). The largest departure from isotropy is found at energy of about 52 EeV.

We first built fair samples of the VCV catalogue of AGNs with redshift $z \leq 0.018$, accounting for the exposure function of the experiment. On the upper panel of Fig.1, we show the power of the 2pt test (filled circles), of the 2pt+ test (filled squares) and of the 3pt test (empty circles) as a function of the number of events. Two thresholds are illustrated: $\alpha = 1\%$ (lines) and $\alpha = 0.1\%$ (dotted lines). Whatever the number of events, the 2pt+ and 3pt tests are always more powerful than the standard 2pt one, and this is even more the case when the number of events decreases. Meanwhile, below 50 events, the power of each test is rapidly getting lower even in the case of a 1% threshold, reaching only less than 50% at best with 20 events.

We show the same analysis on the lower panel of Fig.1 but by adding a 50% mixture of isotropic events to the anisotropic signal. All tests are then less sensitive than in the previous case, the power of the best of them (3pt) being always below 90% even with a threshold of 1%, and reaching only few % when dealing with 20 events.

It thus turns out that at low statistics, the sensitivity of the tests gets rapidly diluted, their power never reaching 90% unless a strong signal of anisotropy is present in the data set.

## IV. Application to the data

The data set we use in this analysis consists of the 100 highest energy events (corresponding to energies greater than ≃43 EeV) with zenith angles smaller than 60° recorded by the surface detector of the Pierre Auger Observatory from January, $1^{st}$ 2004 to March, $31^{st}$ 2009. The energy resolution is 17%, with a systematic uncertainty fo 22% [2]. The angular resolution, defined as the angular radius around the true cosmic ray direction that would contain 68% of the reconstructed shower directions, is at these energies better than $0.9°$[13]. The fiducial cut implemented in the present analysis requires that at least 5 active nearest detectors surround the one with the highest signal when the event was recorded, and that the reconstructed shower core is inside an active equilateral triangle of detectors.

Applying both 2pt+ and 3pt estimators, we performed a scan in energy to search for intrinsic anisotropy. We show in Fig.2 the results of this scan, starting from the 20 highest energy events ($E_{min} \simeq 73$ EeV), and lowering the energy threshold by adding each time the 10 next events up to the 100 highest energy events ($E_{min} \simeq 43$ EeV). The filled squares are the results obtained using the 2pt+ method, while the empty circles are the results obtained using the 3pt method. The maximal departure from isotropy is observed to occur at $\simeq 52$ EeV (for the 70 highest energy events) using both methods: at this energy threshold, the probability $P$ for the data to be a realisation of an isotropic background is $P = 0.26\%$ using the 2pt+ estimator and $P = 0.56\%$ using the 3pt estimator. For higher energy thresholds, both methods give results above the % level. As expected from the second toy model described in the previous section, the relatively low power of the tests when lowering the number of events prevents us to conclude on the isotropic or anisotropic nature of the sky from these observations.

The numbers reported here do not take into account the penalties associated to the scan in energy. In any case, as all those analyses were performed *a posteriori*, this prevents us to rigorously report on probabilities that could be taken at face value.

In Fig.3, we illustrate the largest departure from isotropy we found in the data using the 3pt method, by showing the log-likelihood of individual bins in shape-strength parameter space of data above 52 EeV compared against isotropic expectations (upper panel). Because of bin-bin correlations, the method sums the log-likelihoods to obtain $\Sigma_P^{data}$ and compares them against isotropic skies to determine the probability that an isotropic distribution may produce this pattern at random. The distribution of $\Sigma_P$ for $2 \times 10^4$ isotropic skies is plotted with black hatching in the lower panel. As in the 2pt and the 2pt+ cases, the departure from isotropy is then obtained by counting the number of isotropic Monte-Carlo skies with a lower $\Sigma_P$ than the one observed in the data.

## V. Conclusion

We have reported three statistical methods to search for intrinsic anisotropy of the UHECRs data measured at the Pierre Auger Observatory. Despite of the sensitivity improvement that the 2pt+ and 3pt tests bring with respect to the standard 2pt test, they still show relatively low power at low statistics, as estimated on toy Monte-Carlo samples drawn with the help of catalogues of nearby astronomical objects. This makes difficult the detection of anisotropy independently of any catalogue of astronomical objects even at 99% confidence level with the current statistics we are dealing with at the highest energies. On the contrary, tests designed on correlation of UHECRs with the positions of nearby astronomical objects are more powerful to provide evidence for anisotropy [14], [15]. More statistics is clearly necessary to establish any anisotropy claim using the kind of blind generic tests we presented in this paper.

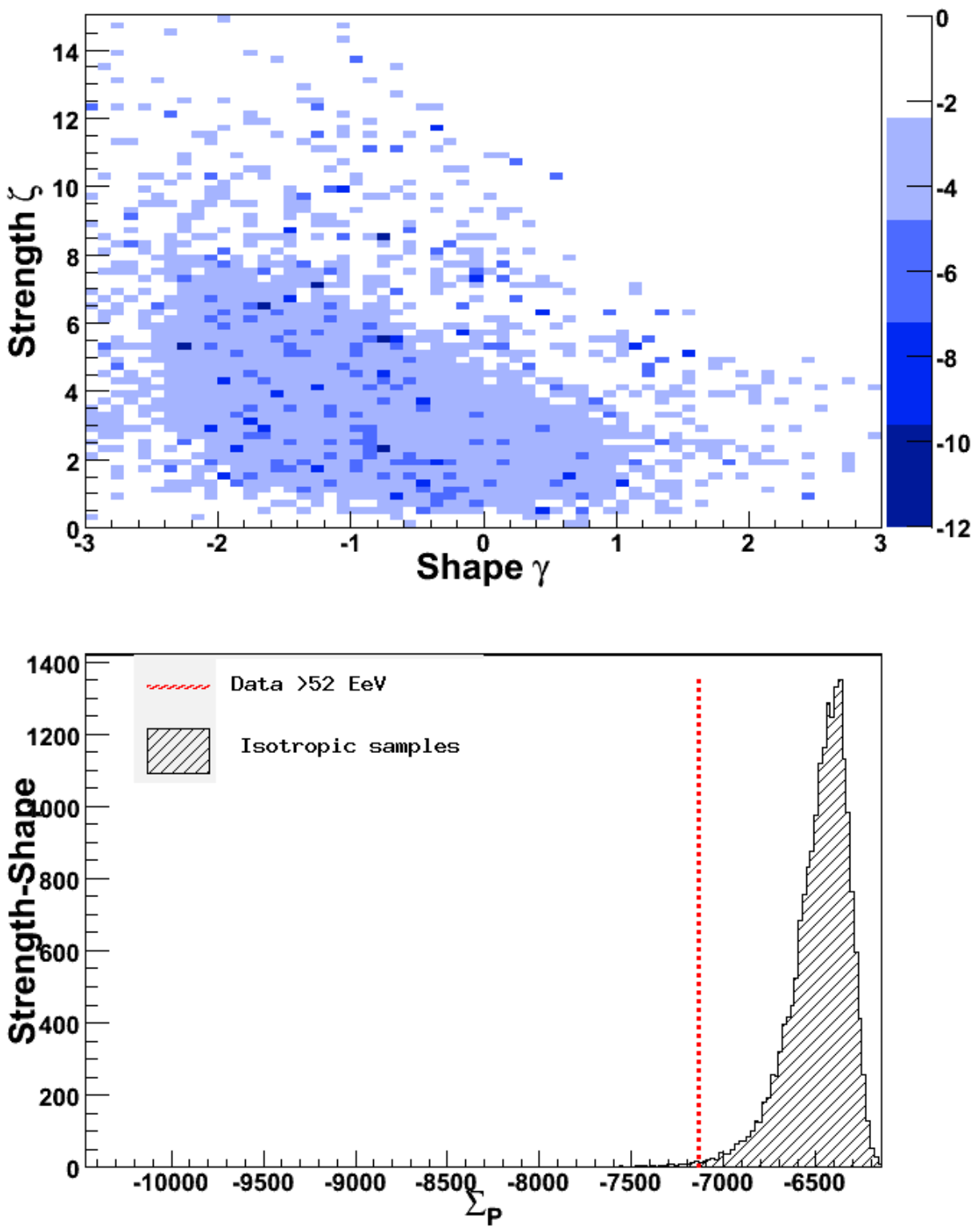

Fig. 3. Top: for each bin in shape and strength, we plot the natural-log of the Poisson probability to observe $n_{obs}$ triplets given $n_{exp}$ expected from an isotropic sky, in shades of blue. Bottom : The distribution of $\Sigma_P$ for $2 \times 10^4$ isotropic skies is plotted with black hatching. The significance is calculated by counting the number of isotropic Monte-Carlo skies to the left of the data (dashed red line).

# Ultra-high energy photon studies with the Pierre Auger Observatory

**Piotr Homola*** **for the Pierre Auger Collaboration**†

*H. Niewodniczański Institute of Nuclear Physics PAN, Radzikowskiego 152, 31-342 Kraków, Poland
†Av. San Martin Norte 304 (5613) Malargüe, Prov. de Mendoza, Argentina

***Abstract*. While the most likely candidates for cosmic rays above $10^{18}$ eV are protons and nuclei, many of the scenarios of cosmic ray origin predict in addition a photon component. Detection of this component is not only of importance for cosmic-ray physics but would also open a new research window with impact on astrophysics, cosmology, particle and fundamental physics. The Pierre Auger Observatory can be used for photon searches of unprecedented sensitivity. At this conference, the status of this search will be reported. In particular the first experimental limits at EeV energies will be presented.**

## I. Introduction

The composition of ultra-high energy cosmic rays (UHECR), i.e. those above $10^{18}$ eV, is still unknown. The Pierre Auger Observatory [1], the newly completed giant air shower detector, with its unprecedented event statistics, brings us closer than ever to resolving this issue. One of the theoretical candidates for UHECR are photons. The first photon searches based on Auger data resulted in upper limits on photon fractions and fluxes [2], [3]. So far, no primary CR photons were identified, the most significant upper limit on the photon fraction is 2% for photons of energies above 10 EeV, based on the data collected by the surface array of particle counters of the Pierre Auger Observatory. This limit severely constrains the family of 'top-down' models [4] which predict large photon contributions (up to 50%) to the observed CR flux. A smaller contribution with typical values around ∼0.1% is expected in 'bottom-up' models. Here, so-called 'GZK-photons' originate during the propagation of charged particles by photo-pion production with background radiation.

Until now, all UHE photon limits were placed at energies larger than 10 EeV. In this work the first limits for photons of energies down to 2 EeV are presented, based on the data collected by the Pierre Auger Observatory.

## II. Data set and selection cuts

The Pierre Auger Observatory collects data with two independent techniques: a surface array of water Cherenkov detectors (Surface Detector - SD) and a network of fluorescence telescopes (Fluorescence Detector - FD). The analysis presented in this work concerns the *hybrid* data (i.e. events recorded by both detectors) collected between December 2004 and December 2007.

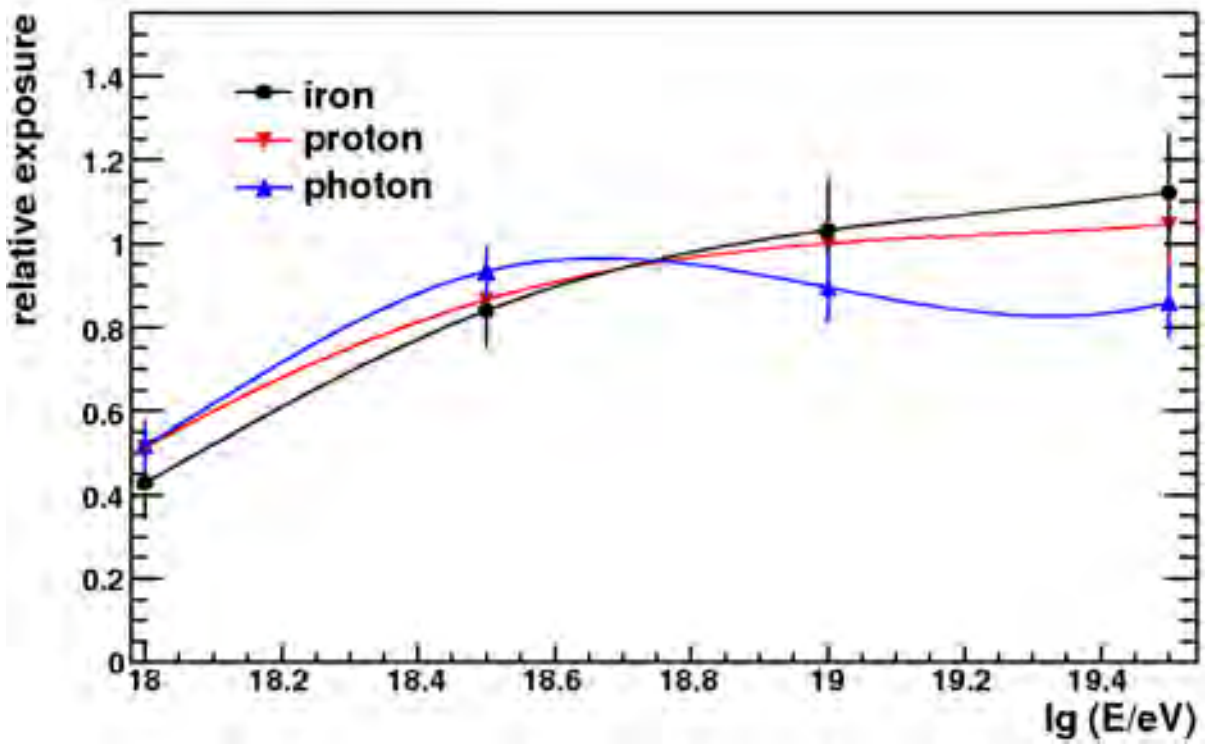

Fig. 1. Relative exposure to primary photons, protons and iron nuclei, normalized to protons at 10 EeV, after applying the quality and fiducial volume cuts with the requirement of the hybrid trigger (see text). In order to guide the eye polynomial fits are superimposed to the obtained values.

The hybrid data statistics are reduced comparing to the pure SD data because of the limited FD duty cycle (∼13% of the total time). On the other hand, the advantage of the hybrid technique is the direct observation of the longitudinal shower profile, reaching also to lower energies.

The requirements for the hybrid event selection include a good quality of shower longitudinal profiles (e.g. enough FD phototubes triggered, good quality of the profile fit, small contamination of direct Cherenkov light) and the shower maximum $X_{\max}$ within the FD field of view (see Ref. [5] and references therein). It has been proven before [2] that $X_{\max}$ is a powerful discriminating variable for photon searches (photon-induced showers in general reach their maxima deeper in the atmosphere than showers initiated by nuclei) and we make use of this fact here.

To avoid biases introduced by the above requirements a set of energy dependent fiducial volume cuts was introduced: nearly vertical showers and those landing too far from the detector were rejected from the analysis. Technical details and a complete list of the data selection cuts with explanations can be found in Ref. [5].

After applying the selection criteria the acceptances for photon and nuclear primaries are similar in the energy region of interest. This is shown in Fig. 1. The presented shower simulations were performed with CORSIKA [6] using QGSJET01 [7] and FLUKA [8] interaction models and processed through a complete detector simulation and reconstruction chain [9]. The application of all the cuts resulted in a data sample of $n_{\rm total}(E_{\rm thr})$ = 2063, 1021, 436 and 131 events above the predefined energy thresholds: $E_{\rm thr}$ = 2, 3, 5 and 10 EeV respectively. To account for the efficiency dependence on the primary energy, fiducial volume cut correction factors $\epsilon_{\rm fvc}(E_{\rm thr})$ = 0.72, 0.77, 0.77 and 0.77 were introduced for $E_{\rm thr}$ = 2, 3, 5 and 10 EeV respectively. These corrections are conservative and independent of the assumptions on the actual primary fluxes (see Ref. [5] for details).

The presence of clouds during shower detection could change the efficiencies shown in Fig. 1. In particular, the reconstructed values of $X_{\rm max}$ could be affected in case the measured longitudinal profile is partially obscured by clouds. In consequence, the primary particle could be misidentified. Thus, events are qualified as photon candidates only when IR cloud cameras could verify the absence of clouds. The fraction of events passing this cloud cut was determined by individual inspection of subsets of the data sample to be $\epsilon_{\rm clc} = 0.51$.

## III. The photon upper limits at EeV

To calculate the photon limit, the number of photon candidates $n_\gamma$ has to be specified for all the considered values of $E_{\rm thr}$. This is done by constructing the photon candidate cut as the median of the $X_{\rm max}$ distribution for photons. The relevant efficiency correction is then $\epsilon_{\rm pcc} = 0.5$. The values of the median were extracted with dedicated simulations performed for primary photons with geometry and energy corresponding to all the potential photon candidates. A parametrization for the typical median photon depth of shower maximum is shown as a solid line in Fig. 2, where the $X_{\rm max}$ values are plotted versus the reconstructed event energies above the lowest considered threshold (2 EeV) for all the events with $X_{\rm max} \geq 800$ g cm$^{-2}$ after executing all the cuts discussed before. Statistical uncertainties are typically a few percent in energy and $\sim$ 15-30 g cm$^{-2}$ in $X_{\rm max}$ while systematic uncertainties are $\sim$22% in energy and $\sim$11 g cm$^{-2}$ in $X_{\rm max}$. The photon candidates are located above the *pcc* line in Fig 2: $n_{\gamma-{\rm cand}} = 8, 1, 0, 0$ for the considered threshold energies $E_{\rm thr}$ = 2, 3, 5 and 10 EeV respectively. It has been checked that the observed number of photon candidates is within the expectations in case of nuclear primaries only. In Fig. 2 the 5% tail of the proton $X_{\rm max}$ distribution is shown. We therefore conclude that the observed photon candidate events may well be due to nuclear primaries only.

With the candidate number and the efficiency corrections defined above, the 95% c.l. upper limit for photon fraction can be calculated as

$$F_\gamma^{95}(E_{\rm thr}) = \frac{n^{95}_{\gamma-{\rm cand}}(E_{\rm thr})\frac{1}{\epsilon_{\rm fvc}}\frac{1}{\epsilon_{\rm pcc}}}{n_{\rm total}(E_{\rm thr})\epsilon_{\rm clc}} \quad (1)$$

where $n^{95}_{\gamma-{\rm cand}}(E_{\rm thr})$ is the 95% c.l. upper limit on the number of photon candidates. $n^{95}_{\gamma-{\rm cand}}(E_{\rm thr})$ was calculated using the Poisson distribution and conservatively assuming no background of nuclear primaries. The resultant 95% c.l. upper limits on the photon fractions are 3.8%, 2.4%, 3.5% and 11.7% for the primary energies above 2, 3, 5 and 10 EeV respectively.

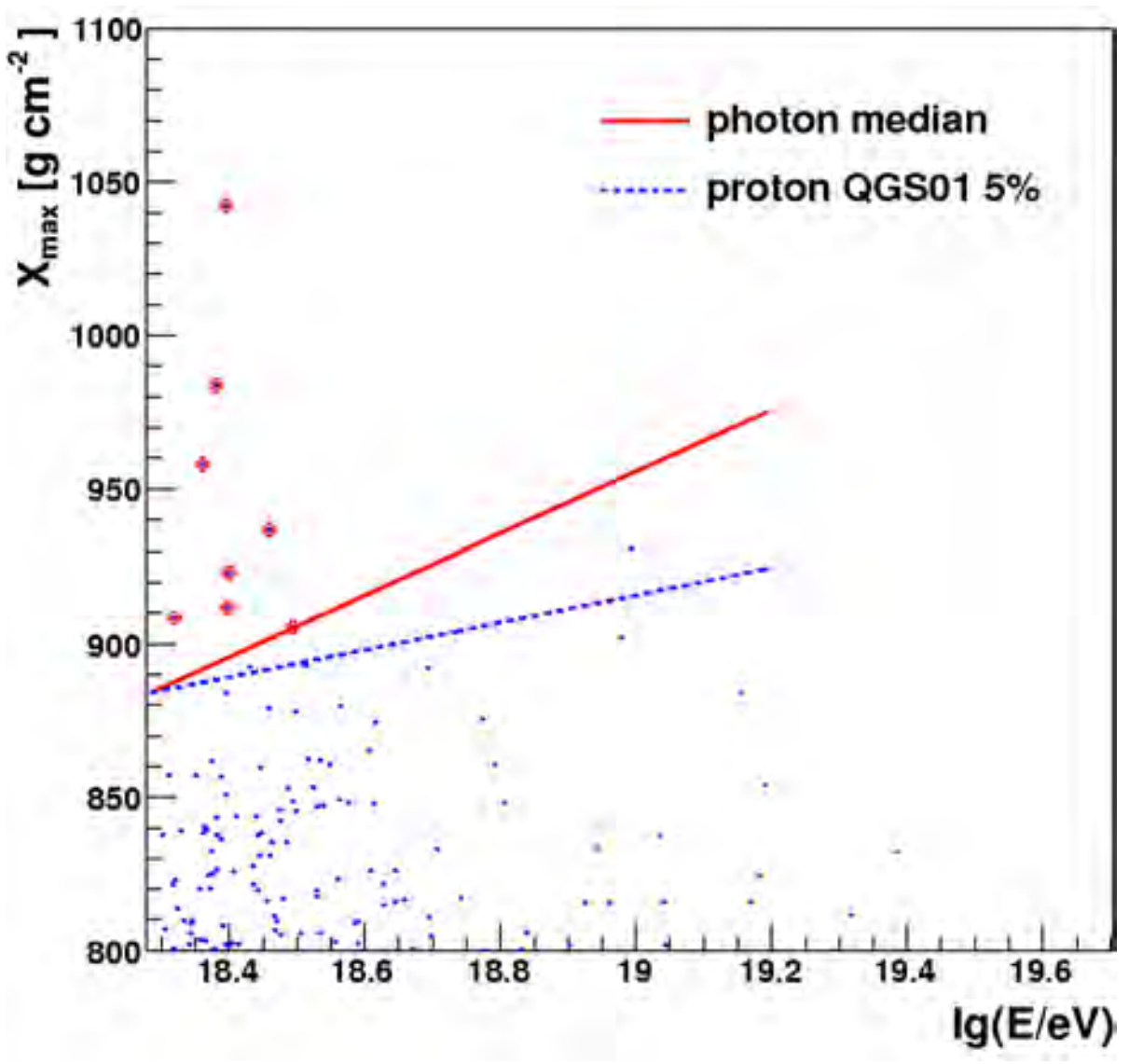

Fig. 2. Measured depth of shower maximum vs. energy for deep $X_{\rm max}$ events (blue dots) after quality, fiducial volume and cloud cuts. Red crosses show the 8 photon candidate events (see text). The solid red line indicates the typical median depth of shower maximum for primary photons.The dashed blue line indicates the 5% tail in the proton $X_{\rm max}$ distribution using QGSJET 01.

The robustness of these results was checked against different sources of uncertainties. The variation of the selection criteria within the experimental resolution essentially does not affect the results. The effective total uncertainty in $X_{\rm max}$ for this analysis amounts to $\sim$16 g cm$^{-2}$ (see Ref. [5] for details). Increasing (reducing) all the reconstructed $X_{\rm max}$ values by 16 g cm$^{-2}$ increases (reduces) the number of photon candidates only for the two lowest energy thresholds: 2 and 3 EeV. The corresponding variations of the photon upper limits are: $F_\gamma^{95}(E_{\rm thr} = 2$ EeV$) = 4.8\%$ (3.8% – no variation) and $F_\gamma^{95}(E_{\rm thr} = 3$ EeV$) = 3.1\%$ (1.5%).

## IV. Discussion

The current upper limits on photon fractions compared to theoretical predictions are plotted in Fig. 3. The Auger hybrid photon upper limits above 2, 3, and 5 EeV placed with this analysis are the first photon upper limits below 10 EeV. The limit above 10 EeV is an update of the previous Auger hybrid limit published in Ref. [2]. The predictions of 'top-down' models were tested here in a new energy range and the constraints from the Auger SD limits were confirmed by data taken with the fluorescence technique. It should be noted that the presented limits together with the one published in Ref. [2] are the only ones based on fluorescence data. It is also worth mentioning that the previous 10 EeV SD limit only marginally constrains the photon prediction at lower energies: even for $E_{\rm thr} = 5$ EeV as many as 75% events have the energies in previously untested 5-10 EeV range.

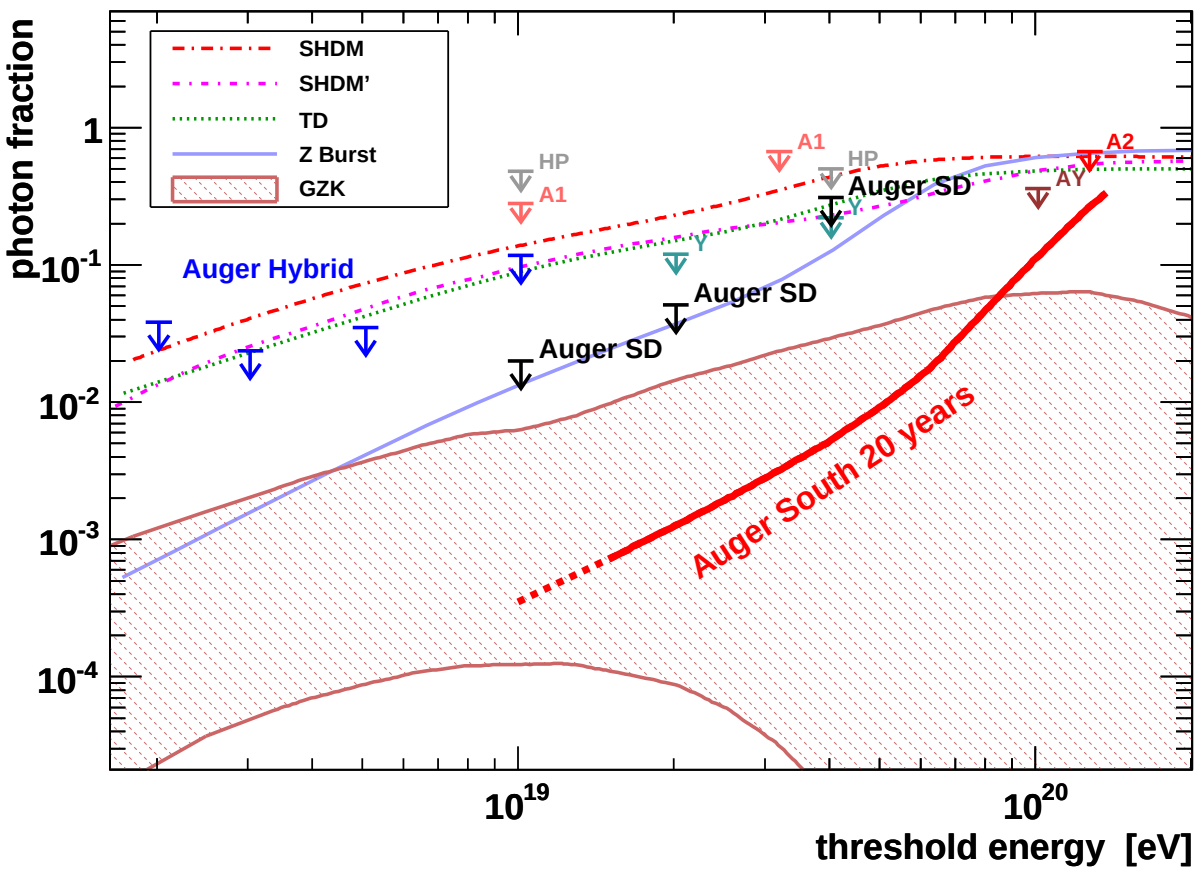

Fig. 3. Upper limits on the photon fraction in the integral cosmic-ray flux from different experiments. The limits from the Auger surface detector are labeled 'Auger SD' and the limits from this work – 'Auger Hybrid'. The thick red line indicates sensitivity of the southern site of the Auger Observatory to the photon fractions after 20 years of operation. The other lines indicate predictions from 'top-down' models and the shaded region shows the expected GZK photon fraction. The labels shown here are explained in [5].

The new limits reduce uncertainties related to the contamination of photons at EeV energies in other analyses of shower data. For instance, the possible contamination from photons was one of the dominant uncertainties for deriving the proton-air cross-section (see e.g. [10]). This uncertainty is now reduced to $\sim$50 mb for data at EeV energies, which corresponds to a relative uncertainty of $\sim$10%. Photon contamination is important also in the reconstruction of the energy spectrum or determination of the nuclear primary composition.

In future photon searches, the separation power between photons and nuclear primaries can be enhanced by adding the detailed information measured with the surface detectors in hybrid events.

## V. Perspectives

The current exposure of the Pierre Auger Observatory is already a factor $\sim$4 larger than the exposure used for the 2% photon limit at 10 EeV. Hence, the Observatory starts to be sensitive to photon fractions within the predicted range of GZK photons and specific GZK scenarios will be tested by UHE photon searches for the first time. Within 20 years of operation the southern part of the Observatory the detection of photon events at fractions below $\sim$0.1% (above 10 EeV) will be at hand (see Fig. 3). The sensitivity to UHE photons will be significantly strengthened with the advent of the northern site of the Observatory in Colorado (USA). This site is planned to cover a surface a factor 7 larger than the one in Argentina.

The northern site of the Observatory will bring another opportunity related to the UHE photon search. Thanks to the difference between the local geomagnetic fields at the two sites a possible detection of UHE photons at Auger South may be confirmed in an unambiguous way at Auger North by observing the well predictable change in the signal from geomagnetic cascading of UHE photon showers [11].

The photon upper limits placed by the Auger Collaboration also address fundamental physics questions. The GZK photons are expected to be absorbed on scales of a few Mpc by pair production with background photons if Lorentz symmetry holds. On the other hand, violation of Lorentz invariance could lead to the observation of an increased photon flux. The new constraints placed on the violation of Lorentz invariance based on our photon limits are substantially more stringent than previous ones [12]. A future *detection* of UHE photons will further impact fundamental physics and other branches of physics (see e.g. [13]).

# Limits on the diffuse flux of ultra high energy neutrinos set using the Pierre Auger Observatory

**Javier Tiffenberg***, **for the Pierre Auger Collaboration**†

*Facultad de Ciencias Exactas y Naturales, Univ. de Buenos Aires, Buenos Aires, ARGENTINA
†Av. San Martín Norte 304 (5613) Malargüe, Prov. de Mendoza, ARGENTINA

***Abstract*. The array of water-Cherenkov detectors of the Pierre Auger Observatory is sensitive to neutrinos of $>$ 1 EeV of all flavours. These interact through charged and neutral currents in the atmosphere (down-going) and, for tau neutrinos, through the "Earth skimming" mechanism (up-going). Both types of neutrinos can be identified by the presence of a broad time structure of signals in the water-Cherenkov detectors in the inclined showers that they induce when interacting close to ground. Using data collected from 1 January 2004 to 28 February 2009, we present for the first time an analysis based on down-going neutrinos and place a competitive limit on the all-flavour diffuse neutrino flux. We also update the previous limit for up-going tau neutrinos. Sources of possible backgrounds and systematic uncertainties are discussed.**

## I. Introduction

Essentially all models of Ultra High Energy Cosmic Ray (UHECR) production predict neutrinos as the result of the decay of charged pions, produced in interactions of the CRs within the sources themselves or in their propagation through background radiation fields [1], [2]. Neutrinos are also copiously produced in top-down models proposed as alternatives to explain the production of UHECRs [1].

With the surface detector (SD) of the Pierre Auger Observatory [3] we can detect and identify UHE neutrinos (UHE$\nu$s) in the EeV range and above.

Earth-skimming tau neutrinos [4], [5] are expected to be observed through the detection of showers induced by the decay products of an emerging $\tau$ lepton, after the propagation and interaction of a flux of $\nu_\tau$ inside the Earth. A limit on the diffuse flux of UHE $\nu_\tau$ was already placed using this technique with data collected from 1 Jan 04 to 31 Aug 07 [5].

The SD of the Pierre Auger Observatory has also been shown to be sensitive to "down-going" neutrinos of all flavours interacting in the atmosphere, and inducing a shower close to the ground [6]. In this contribution we present for the first time an analysis based on down-going neutrinos and place a competitive limit on the all-flavour diffuse neutrino flux using data from 1 Jan 04 up to 28 Feb 09. We also update the limit on the up-going tau neutrinos.

## II. Identifying neutrinos in data

Identifying neutrino-induced showers in the much larger background of the ones initiated by nucleonic cosmic rays is based on a simple idea: neutrinos can penetrate large amounts of matter and generate "young" inclined showers developing close to the SD exhibiting shower fronts extended in time (Fig. 1 right). In contrast, UHE particles such as protons or heavier nuclei interact within a few tens of $\mathrm{g\ cm^{-2}}$ after entering the atmosphere, producing "old" showers with shower fronts narrower in time (Fig. 1 left).

Although the SD is not directly sensitive to the nature of the arriving particles, the 25 ns time resolution of the FADC traces in which the signal is digitised in the SD stations, allows us to distinguish the narrow signals in time, expected from a shower initiated high in the atmosphere, from the broad signals expected from a young shower. Several observables can be used to characterise the time structure and shape of the FADC traces. They are described in [8] where their discrimination power is also studied.

Down-going neutrinos of any flavour interacting through charged (CC) or neutral (NC) current, may induce showers in the atmosphere that can be detected using the SD of the Pierre Auger Observatory (Fig. 2). Detailed simulations of UHE neutrinos forced to interact deep in the atmosphere were produced. Both CC and NC neutrino interactions were simulated using HERWIG [9] for the first interaction and AIRES [10] for the shower development. "Double bang" showers produced by tau neutrinos (CC interaction followed by the decay in flight of the tau lepton) are also generated using Tauola [8] to simulate the tau decay products.

The simulations indicate that only the signals in the first few triggered tanks are expected to be broader in time than those induced by a shower initiated high in the atmosphere [8]. This asymmetry is due to the larger grammage of atmosphere that the later portion of the shower front crosses before reaching ground [12], (Fig. 1 right).

A set of conditions has been designed to select inclined showers initiated by down-going neutrinos. As they are expected to be identified over a wide range of zenith angles, an identification criterion different from the one applied to search for up-going neutrinos [5], [8] has been developed. For this purpose data collected with the SD between 1 Jan 04 and 31 Oct 07 - corresponding to $\sim 1.2$ years of the full SD array - was used as "training" data. From the showers that trigger the SD array [3], those arriving during periods in which instabilities in data acquisition occur are excluded. After that the FADC traces are cleaned to remove segments that are due to accidental muons not belonging to the shower but arriving close in time with the shower front. Moreover, if 2 or more segments of comparable area appear in a trace the station is classified as ambiguous and it's not used. Then a selection of the stations actually belonging to the event is done based on space-time compatibility among them. Events with less than 4 tanks passing the level 2 trigger algorithm [3] are rejected. This sample is then searched for inclined events requiring that the triggered tanks have elongated patterns on the ground along the azimuthal arrival direction. A length $L$ and a width $W$ are assigned to the pattern [5], [8], and a cut on their ratio is applied ($L/W$ >3). Then we calculate the apparent speed of the signal in the event moving across the ground along $L$, using the arrival times of the signals at ground and the distances between tanks projected onto $L$ [13]. The average speed $\langle V \rangle$ is measured between pairs of triggered stations, and is required to be compatible with that expected in a simple planar model of the shower front in an inclined event with $\theta \geq 75°$, allowing for some spread due to fluctuations ($\langle V \rangle \leq 0.313 \text{ m ns}^{-1}$). Furthermore, since in inclined events the speed measured between pairs of tanks is concentrated around $\langle V \rangle$ [5] we require that the r.m.s. scatter of $V$ in an event to be smaller than $0.08 \cdot \langle V \rangle$. The zenith angle $\theta$ of the shower is also reconstructed, and those events with $\theta \geq 75°$ are selected. Exactly the same set of conditions is applied to the simulated neutrinos.

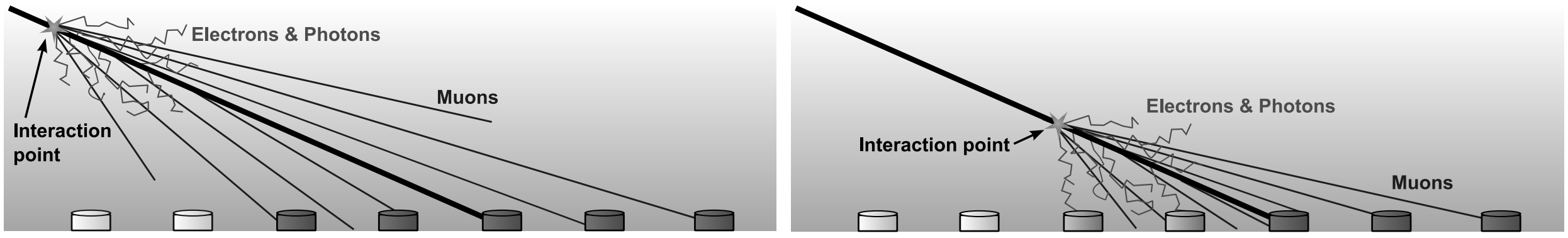

Fig. 1. Left panel: sketch of an inclined shower induced by a hadron interacting high in the atmosphere. The EM component is absorbed and only the muons reach the detector. Right panel: deep inclined shower. Its early region has a significant EM component at the detector level.

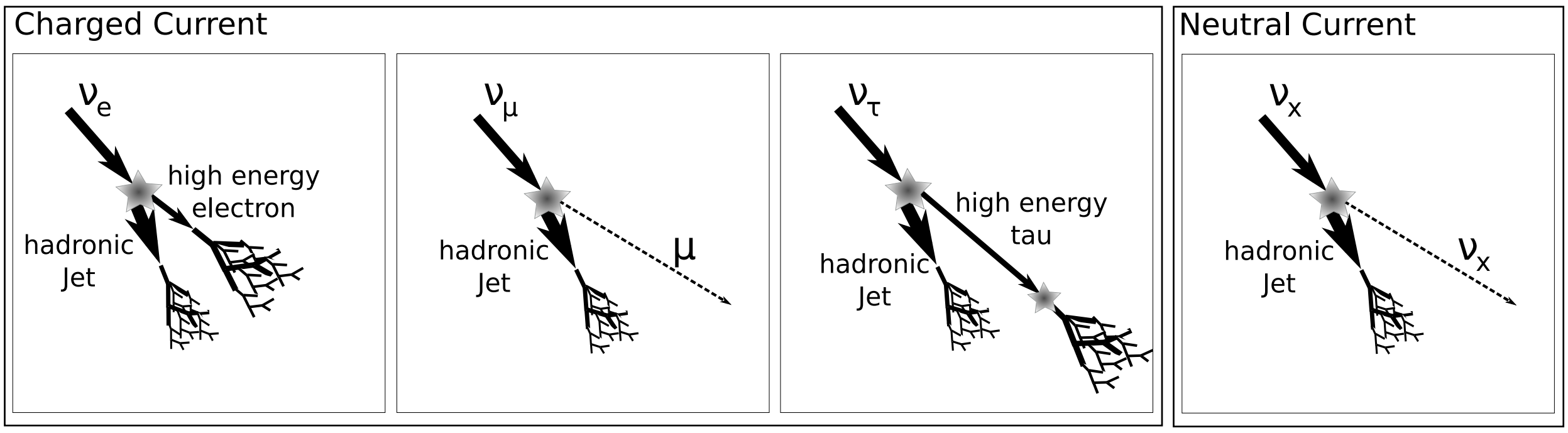

Fig. 2. Neutrinos can initiate atmospheric showers through charged (CC) or neutral (NC) current interactions. On $\nu_e$ CC interactions all the energy of the primary neutrino is transferred to the shower. This is not the case of the NC channel where the primary neutrino energy is only partially transferred to the shower while a significant fraction is carried away by the scattered neutrino. Similar behaviour is seen on the $\nu_\mu$ CC induced showers where the emerging high energy muon usually decays under the ground and doesn't produce a shower. Note that $\nu_\tau$ CC initiated showers may have a "double bang" structure due to the fact that the out-coming high energy $\tau$ may travel a long distance before decay producing a second displaced shower vertex.

The sample of inclined events is searched for "young" showers using observables characterising the time duration of the FADC traces in the early region of the event. To optimize their discrimination power we applied the Fisher discriminant method [7] to the training data – overwhelmingly, if not totally constituted of nucleonic showers – and to the Monte Carlo (MC) simulations – exclusively composed of neutrino-induced showers. Given two populations of events – nucleonic inclined showers and $\nu$-induced showers in our case – characterised by a set of observables, the Fisher method produces a linear combination of the various observables – $f$ the Fisher discriminant – so that the separation between the means of $f$ in the two samples is maximised, while the quadratic sum of the r.m.s. of $f$ in each of them is minimised. Since events with a large number of tanks $N$ (large multiplicity) are different from events with small multiplicity the sample of training data is divided into 3 sub-samples corresponding to events with number of tanks $4 \leq N \leq 6$, $7 \leq N \leq 11$ and $N \geq 12$, and a Fisher discriminant is obtained using each of the sub-samples as training data. We use the Area-over-Peak (AoP) [8] and its square of the first 4 tanks in each event, their product, and a global early-late asymmetry parameter of the event as the discriminant variables of the Fisher estimator. Distributions of these observables for real data and MC simulated neutrinos are shown in [8]. In Fig. 3 we show the distribution of the Fisher discriminant for data collected between 1 Jan 04 and 31 Oct 07 and for the neutrino simulations. A clear separation between the two samples is achieved. The expected number of background events can be computed by extrapolating the exponential tail of the distribution of the data. By this means three cut values $f_{\rm cut}$ – corresponding to each of the sub-samples– are chosen, so that we expect less than one background event every 20 years above its value. Events with $f > f_{\rm cut}$ are considered to be neutrino candidates. These cuts reject all real events in the training data samples while keeping a significant fraction of the neutrino simulations.

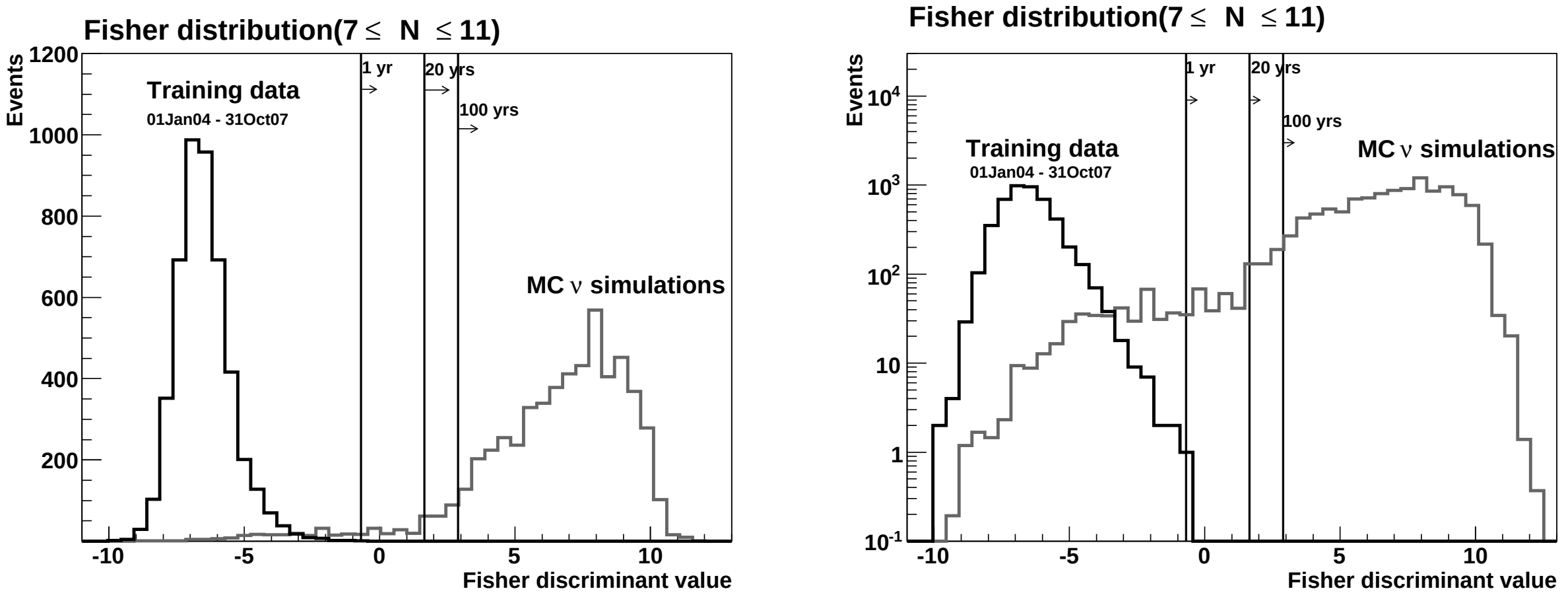

Fig. 3. Distribution of the Fisher discriminant (see text for details) in linear (left) and logaritmic (right) scale for real data in the training period (1 Jan 04 - 31 Oct 07) and Monte Carlo simulated down-going neutrinos for events with multiplicity $7 \leq N \leq 11$ .

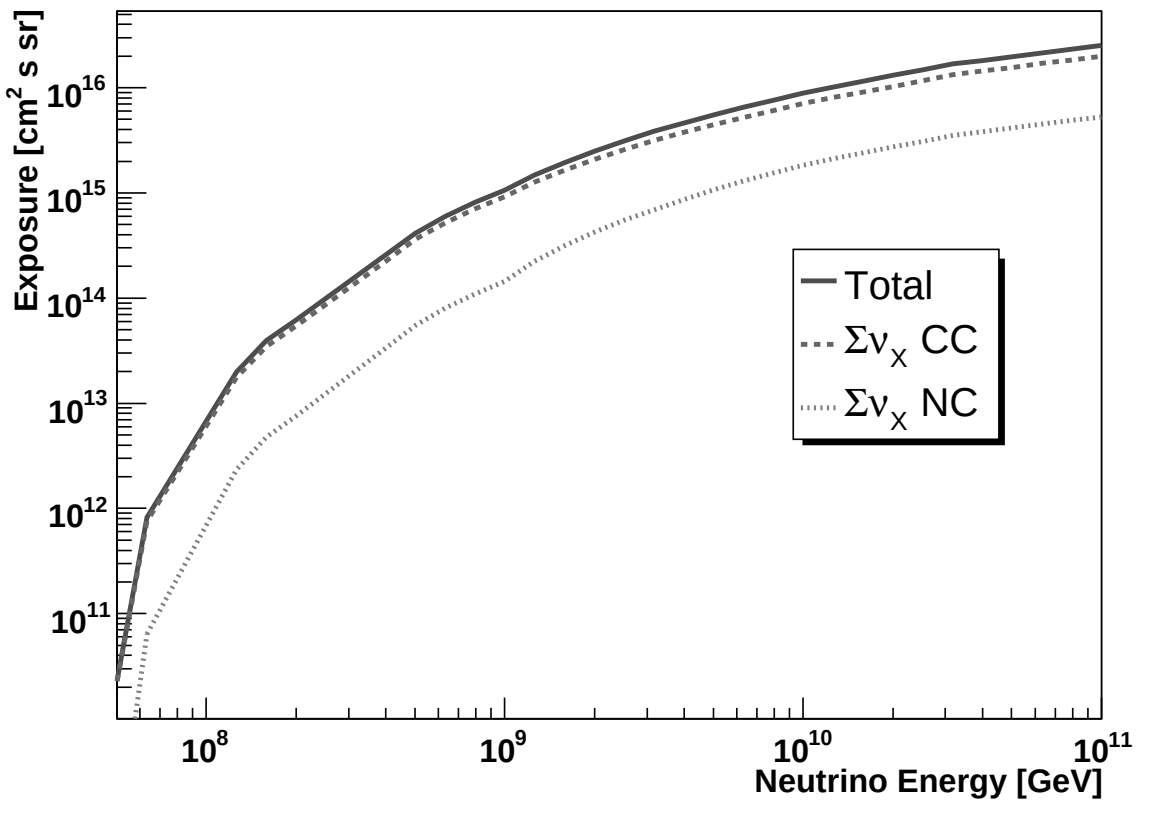

Fig. 4. Exposure of the SD array to down-going neutrinos in the search period (1 Nov 07 - 28 Feb 09).

## III. Exposure and limit on UHE neutrinos

Exactly the same selection procedure and cuts in $f$ are applied "blindly" to data collected between 1 Nov 07 and 28 Feb 09 – corresponding to $\sim 0.8$ yr of the full SD array[1]. These data were not used for training of the Fisher method. No neutrino candidates were found and an upper limit on the UHE diffuse flux of ultra-high energy neutrinos can be placed.

For this purpose the exposure of the SD array to UHE neutrinos is calculated. For down-going neutrinos this involves folding the SD array aperture with the interaction probability and the identification efficiency, and integrating in time taking into account changes in the array configuration due to the installation of new stations and instabilities in data taking. The identification efficiency $\epsilon$, for the set of cuts defined above, depends on the neutrino energy $E_\nu$, the depth along the atmosphere at which the neutrino interacts $D$, the zenith angle $\theta$, the position $\vec{r} = (x, y)$ of the shower in the surface $S$ covered by the array, and the time $t$ through the instantaneous configuration of the array. Moreover it depends on the neutrino flavour ($\nu_e,\ \nu_\mu$ or $\nu_\tau$), and the type of interaction – charged (CC) or neutral current (NC) – since the different combinations of flavour and interaction induce different type of showers. The efficiencies $\epsilon$ were obtained through MC simulations of the development of the shower in the atmosphere and the simulation of the surface detector array, see [8] for more details. The exposure can be written as:

$$\mathcal{E}(E_\nu) = \frac{1}{m} \sum_i \left[ \sigma^i(E_\nu) \int M^i_{\rm ap}(E_\nu, t)\, dt \right] \quad (1)$$

where the sum runs over the 3 neutrino flavours and the CC and NC interactions, and $m$ is the mass of a nucleon. In this equation $M^i_{\rm ap}$ is the mass aperture given by:

$$M_{\rm ap}(E_\nu) = 2\pi \iiiint \sin\theta \cos\theta \; \epsilon^i(\vec{r}, \theta, D, E_\nu, t)\, d\theta\, dD\, dx\, dy \quad (2)$$

[1]Although our current test sample is slightly smaller than the training one, with the SD fully commissioned, test data will grow fast and will rapidly surpass the acquired during the training period.

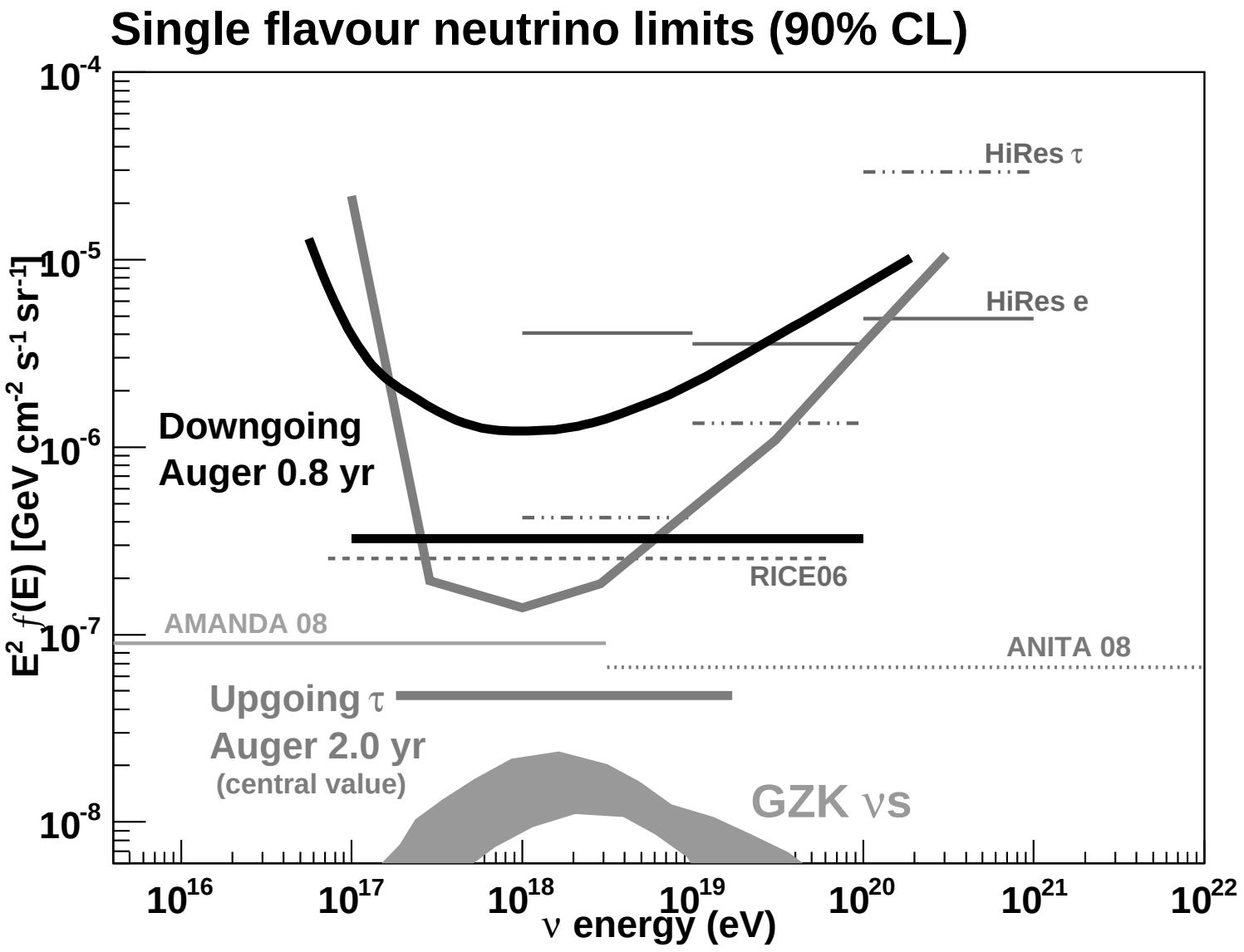

Fig. 5. Differential and integrated upper limits (90% C.L.) from the Pierre Auger Observatory for a diffuse flux of down-going $\nu$ in the period 1 Nov 07 - 28 Feb 09 and up-going $\nu_\tau$ (1 Jan 04 - 28 Feb 09). Limits from other experiments [14] are also plotted. A theoretical flux for GZK neutrinos Ref. [2] is shown.

The exposure was calculated using purely MC techniques and also integrating the neutrino identification efficiencies $\varepsilon$ over the whole parameter space [8]. All the neutrino flavours and interactions are accounted for in the simulations. In particular for $\nu_\tau$ we have taken into account the possibility that it produces a double shower in the atmosphere triggering the array – one in the $\nu_\tau$ CC interaction itself and another in the decay of the $\tau$ lepton. The exposure for the period 1 Nov 07 up to 28 Feb 09 is shown in Fig. 4 for CC and NC channels.

Several sources of systematic uncertainties have been taken into account and their effect on the exposure evaluated. We tentatively assign a $\sim 20\%$ systematic uncertainty due to the neutrino-induced shower simulations and the hadronic model (SIBYLL 2.1 vs QGSJETII.03). Another source of uncertainty comes from the neutrino cross section. Using [15] we estimate a systematic uncertainty of $\sim 10\%$. The topography around the Southern Site of the Pierre Auger Observatory enhances the flux of secondary tau leptons. In this work we neglected this effect. Our current simulations indicate that including it will improve the limit by roughly $\sim 15-20\%$.

Finally assuming a $f(E_\nu) = k \cdot E_\nu^{-2}$ differential neutrino flux we have obtained a 90% C.L. limit on the all-flavour neutrino flux using down-going showers:

$$k < 3.2 \times 10^{-7}\ \mathrm{GeV\ cm^{-2}\ s^{-1}\ sr^{-1}} \tag{3}$$

shown in Fig. 5. We also present the updated limit based on Earth-skimming up-going neutrinos:

$$k < 4.7^{-2.5}_{+2.2} \times 10^{-8}\ \mathrm{GeV\ cm^{-2}\ s^{-1}\ sr^{-1}} \tag{4}$$

where the upper/lower values correspond to best/worse scenario of systematics [13]. We have also include the limit in differential format to show the range in energies at which the sensitivity of the Pierre Auger Observatory to down-going and Earth-skimming $\nu$ peaks.

A preliminary limit on the flux of UHE neutrinos from the position of Centaurus A (Galactic coords. $\delta \sim -43.0°$, $l \sim -35.2°$) – assuming a point source at that position – was also obtained. For that purpose we have integrated the identification efficiency $\varepsilon$ over the fraction of the time ($\sim 15.6\%$) the source is seen in the SD array with $\theta$ between $75°$ and $90°$. The preliminary limit is $\sim 3 \times 10^{-6}$ neutrinos per $\mathrm{GeV\ cm^{-2}\ s^{-1}}$.

# Search for sidereal modulation of the arrival directions of events recorded at the Pierre Auger Observatory

**R.Bonino***, **for The Pierre Auger Collaboration**†

*Istituto di Fisica dello Spazio Interplanetario (INAF), Università di Torino and Sezione INFN Torino, Italy
†Observatorio Pierre Auger, Av. San Martín Norte 304, (5613) Malargüe, Mendoza, Argentina

***Abstract*. Using data collected by the Pierre Auger Observatory from 1 January 2004 until 31 March 2009, we search for large scale anisotropies in different energy windows above $2 \cdot 10^{17}$ eV. A Fourier analysis shows the presence of a $\sim 3\%$ modulation at the solar frequency, arising from modulations of the array exposure and weather effects on the showers. We study the sidereal anisotropies using a Rayleigh method which accounts for these effects, and the East-West differential method which is largely independent of them. No significant anisotropies are observed, resulting in bounds on the first harmonic amplitude at the 1% level at EeV energies.**

## I. Introduction

The large scale anisotropy, and in particular its dependence on primary energy, represents one of the main tools for discerning between a galactic or an extragalactic origin of UHECRs and for understanding their mechanisms of propagation. The transition from a galactic to an extragalactic origin should in fact induce a significant change in the CR large scale angular distribution, giving precious hints on their nature and on the magnetic fields that modify their trajectories.

Different theoretical models predict the transition to occur at different energies and consequently lead to dissimilar predictions for the shape and the amplitude of the corresponding anisotropy. A measure of the anisotropy or the eventual bounds on it are thus relevant to constrain different models for the CR origin.

## II. Data Analysis and results

The statistics accumulated so far by the Pierre Auger Observatory allows us to perform large scale analyses with a sensitivity that is already at the percent level. For this analysis we used data recorded from 1 January 2004 to 31 March 2009, removing the periods of unstable data acquisition ($\sim 3\%$ of the whole data set).

Searching for %-level large-scale patterns requires control of the sky exposure of the detector and of various acceptance effects, such as detector instabilities and weather modulations. The main effects are expected to appear at the solar frequency but may also be non-negligible at other frequencies. In particular, the combination of diurnal and yearly modulations of the acceptance may generate a spurious variation with similar amplitudes at both the sidereal and anti-sidereal frequencies [1]. The Fourier transform of the arrival times of the events is thus an ideal tool to analyse their frequency patterns and in particular the sidereal, solar and anti-sidereal modulations [2]. The resolution of this analysis is of the order of $1/T$, where $T$ is the exposure time. Therefore, if data are acquired over a few years, even with variable detector conditions, the resolution is sufficient to resolve the sidereal and the diurnal frequencies. For each frequency the associated Fourier amplitude is calculated using the distribution of the times $t_i$ of the events modified such that any sidereal modulation of $\tilde{t}_i$ is equal to the modulation of the event rate in Right Ascension:

$$\tilde{t}_i = t_i + \alpha_i - \alpha_i^0, \qquad (1)$$

where $\alpha_i$ is the RA of the event and $\alpha_i^0$ is the local sidereal time corresponding to UTC time $t_i$.

We show in Fig.1 the results of such analysis using the whole data set. The amplitude at the solar frequency largely stands out from the noise, whereas the amplitudes at all other frequencies stand at the level of the average noise (estimated to be 0.33% using data at all frequencies except for the solar band). In particular, the amplitudes at the sidereal and at the anti-sidereal frequencies are at a similar level. If a genuine large-scale pattern were present above the noise level, the amplitude at the sidereal frequency would clearly stand out the anti-sidereal one.

We repeated the same analysis for several energy ranges. The results are displayed in Tab.I, where the sidereal amplitudes ($r_{sid}$) are compared to the anti-sidereal ones ($r_{a-sid}$). It can be seen that there is no significant signal at the sidereal frequency within the available statistics.

TABLE I
SIDEREAL AMPLITUDES ($r_{sid}$) COMPARED TO THE ANTI-SIDEREAL ONES ($r_{a-sid}$) IN 6 ENERGY RANGES. ALSO INDICATED ARE THE TYPICAL STATISTICAL FLUCTUATIONS THROUGH THE AVERAGE NOISE AND ITS RMS.

| Energy Range [EeV] | $r_{sid}$ [%] | $r_{a-sid}$ [%] | Average Noise [%] | $\sigma_r$ [%] |
|---|---|---|---|---|
| 0.2 - 0.5 | 0.66 | 0.61 | 0.41 | 0.22 |
| 0.5 - 1 | 0.44 | 0.52 | 0.36 | 0.20 |
| 1 - 2 | 1.08 | 0.82 | 0.52 | 0.28 |
| 2 - 4 | 1.37 | 1.36 | 0.88 | 0.43 |
| 4 - 8 | 1.26 | 0.84 | 1.68 | 0.86 |
| >8 | 5.70 | 3.27 | 2.59 | 1.42 |

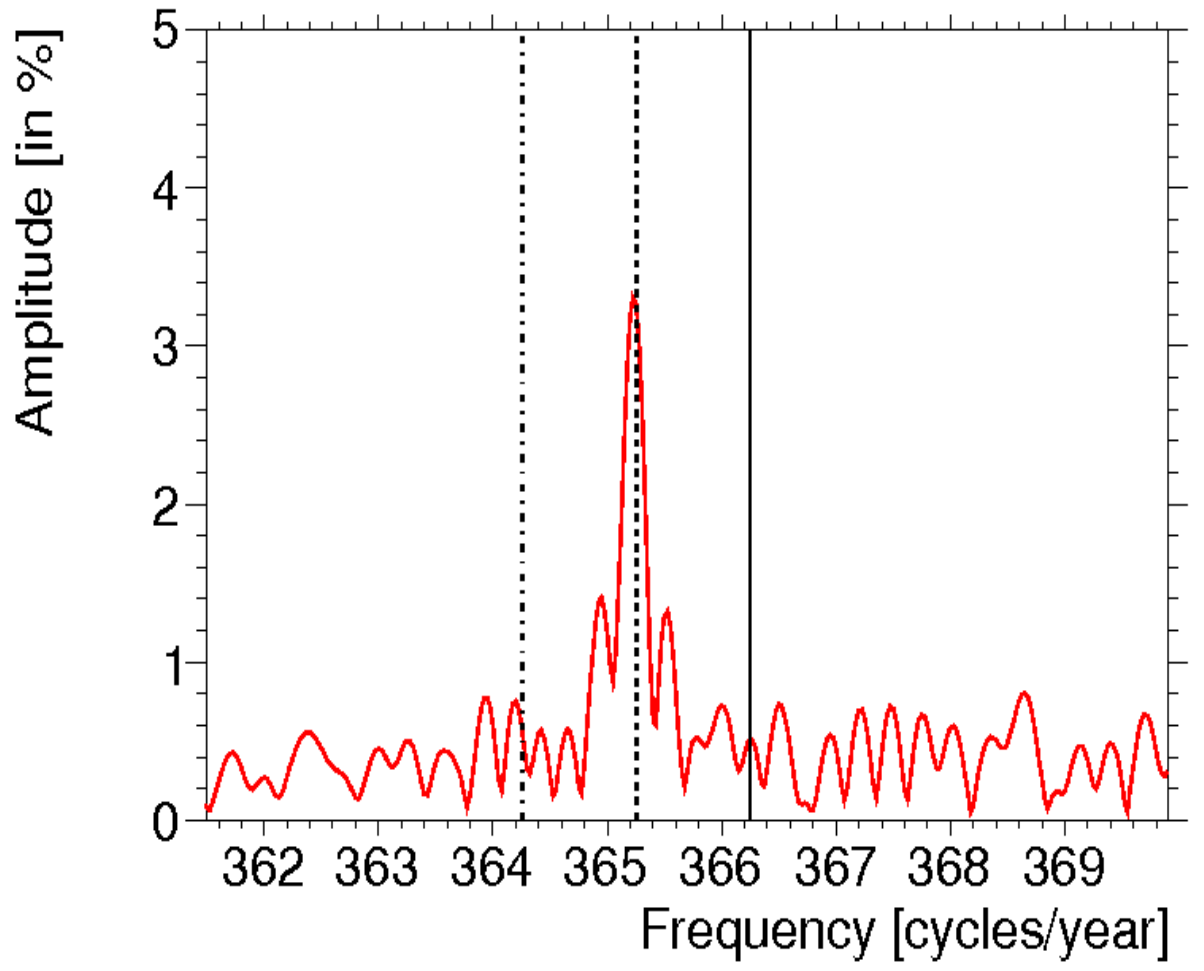

Fig. 1. Amplitude of the first harmonic as a function of the frequency, applying the Fourier time analysis on the whole data set. From left to right, the anti-sidereal and the solar frequencies are indicated by the *dotted vertical lines*, whereas the sidereal frequency is represented by the *continuous line*.

The variations due to the non-uniform detector on-times can be taken into account using a generalised Rayleigh analysis [3]. This method corrects for the effects of a non-uniform acceptance in right ascension by weighting each event with a factor $\omega_i$ inversely proportional to the relative exposure of the region of the sky observed at the arrival time of the event ($\alpha_{di}$ is the right ascension of the zenith of the detector at the time the event $i$ is detected) [4]. These factors are obtained from the detailed information about the individual detector stations on-times. Computing the coefficients:

$$A = \frac{2}{\Omega}\sum_i \omega_i(\alpha_{di})\cos\alpha_i \quad (2)$$

$$B = \frac{2}{\Omega}\sum_i \omega_i(\alpha_{di})\sin\alpha_i \quad (3)$$

where $\Omega = \sum_i \omega_i(\alpha_{di})$, the Rayleigh amplitude and phase are obtained through:

$$r = \sqrt{A^2 + B^2} \text{ and } \phi = \text{atan}\frac{B}{A} \quad (4)$$

In this case the deviations from a uniform exposure are small, so the probability that an amplitude larger or equal to $r$ arises from an isotropic distribution may be estimated with the standard expression $P = \exp(-k_0)$, where $k_0 = r^2 N/4$ (being $N$ the total number of events). In addition, we account for atmospheric effects, such as changes in the air density and pressure, in the energy estimation of each event [5]. This is the dominant weather effect above $\sim 1$ EeV, while below that energy the weather effects also start to affect the trigger efficiency in a significant way. Hence, with this method we present results only above 1 EeV.

After applying such corrections all the spurious modulations are removed. For instance, in the energy interval $1 - 2$ EeV a first harmonic in solar time of $3.33\%$ (corresponding to a chance probability $P \sim 10^{-20}$) is reduced to $0.88\%$ ($P \sim 2\%$) after all the corrections. The corresponding first harmonics in sidereal and anti-sidereal time are of the same order, being respectively $0.90\%$ and $0.71\%$, with a probability to result from a fluctuation of an isotropic distribution of $\sim 2\%$ and $\sim 8\%$.

An alternative method, which is largely independent of possible systematic effects, is the differential East-West method [6], which exploits the differences in the number of counts between the eastward and the westward arrival directions at a given time. Since the instantaneous eastward and westward acceptances are equal and the two sectors are equally affected by the instabilities of the apparatus, by making the difference in the East and West counts, this method allows us to remove direction-independent phenomena, such as atmospheric and acceptance effects, without applying any correction. The difference in the number of counts $E(t) - W(t)$ is related to the physical CR intensity $I(t)$ by $\mathrm{d}I/\mathrm{d}t = (E(t) - W(t))/\delta t$. The first harmonic analysis of $I(t)$, whose amplitude and phase are ($r_I$,$\phi_I$), can be derived from the first harmonic analysis of $E(t) - W(t)$, of amplitude and phase ($r_D$,$\phi_D$):

$$r_I = \frac{1}{\sin\delta t}\frac{n_{int}}{N} r_D \text{ and } \phi_I = \phi_D + \frac{\pi}{2} \quad (5)$$

where $N$ is the total number of events, $n_{int}$ is the number of intervals of sidereal time used to compute the first harmonic amplitude of $E(t) - W(t)$ and $\delta t$ is the average hour angle between the vertical and the events from sector $E$ (or $W$). The probability that an amplitude equal or larger than $r$ arises from an isotropic distribution is $P = \exp(-r^2 N \sin^2 \delta t/4)$.

Since this method is largely independent of spurious time variations, the analysis can be performed also on the whole data set (median energy $\sim 6 \cdot 10^{17}$ eV), even below the energy threshold for full efficiency. For the complete data set the amplitudes in solar and anti-sidereal time are respectively $0.29\%$ ($P \sim 55\%$) and $0.24\%$ ($P \sim 66\%$), showing that any spurious modulation has been removed (the amplitude in solar time with the standard Rayleigh analysis, without corrections, is $3.98\%$). The corresponding amplitude in sidereal time is $r = 0.48\%$, the probability for it to result from a fluctuation of an isotropic distribution is $\sim 20\%$ (see the first line of Tab.II).

In Fig.2 the results of the E-W and the Rayleigh analyses on all the events above increasing energy thresholds are shown. No significant modulation in sidereal time is detected throughout the scan. The two methods are complementary: while the Rayleigh analysis can only be reliably used above 1 EeV, the East-West analysis can be safely applied even below 1 EeV but it is affected by larger statistical uncertainties.

For completeness and because the points in Fig.2 are correlated, we repeated the two analyses in energy bins of 0.1 Log(E). The results are shown in Fig.3 and provide a further evidence about the lack of significant modulations in sidereal time.

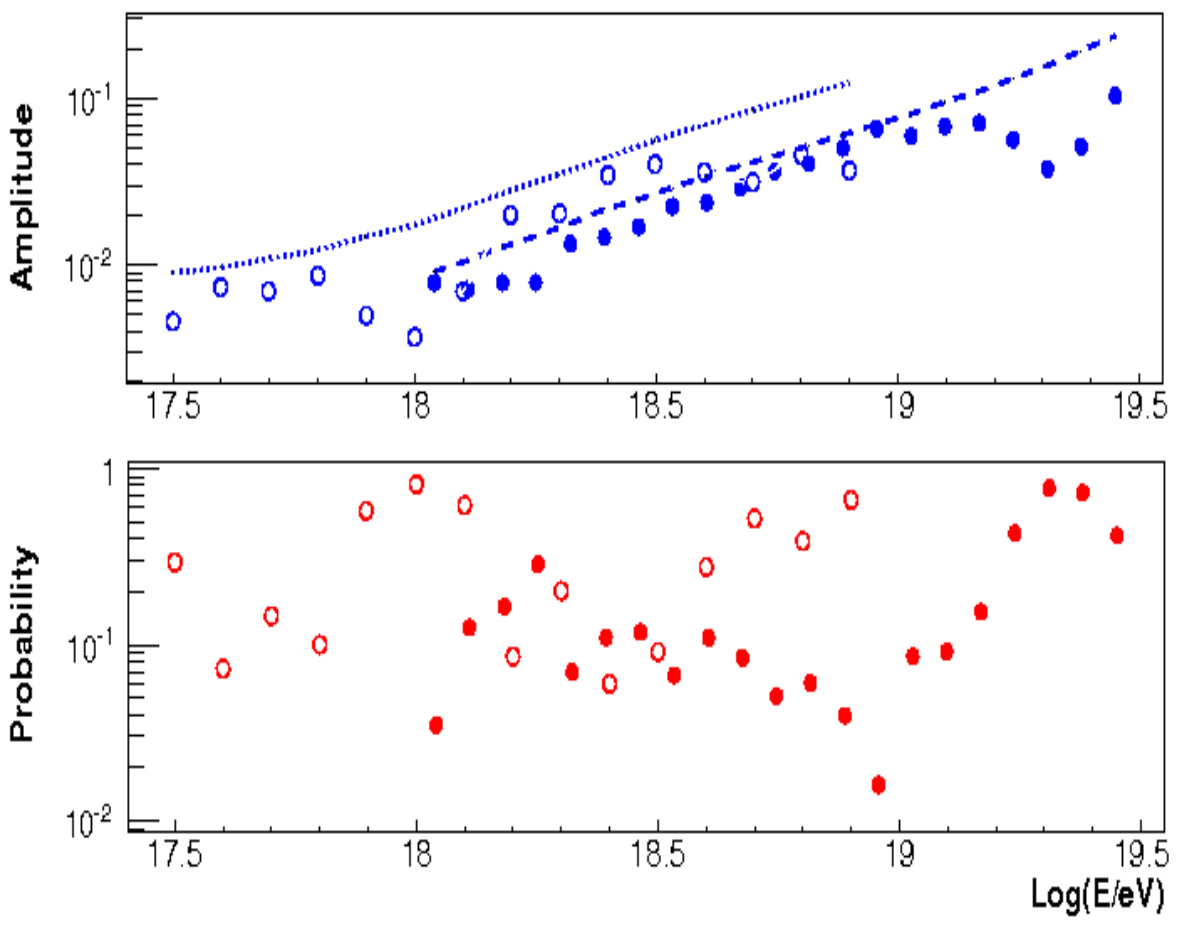

Fig. 2. Rayleigh amplitude (top) and probability for the amplitude to result from fluctuations of an isotropic background (bottom) as a function of increasing energy thresholds, obtained with both the generalised Rayleigh analysis, after correcting for non-constant acceptance and weather effects, (*filled circles*) and the East-West method (*empty circles*). The *dotted lines* indicate the 99% c.l. upper bound on the amplitudes that could result from fluctuations of an isotropic distribution.

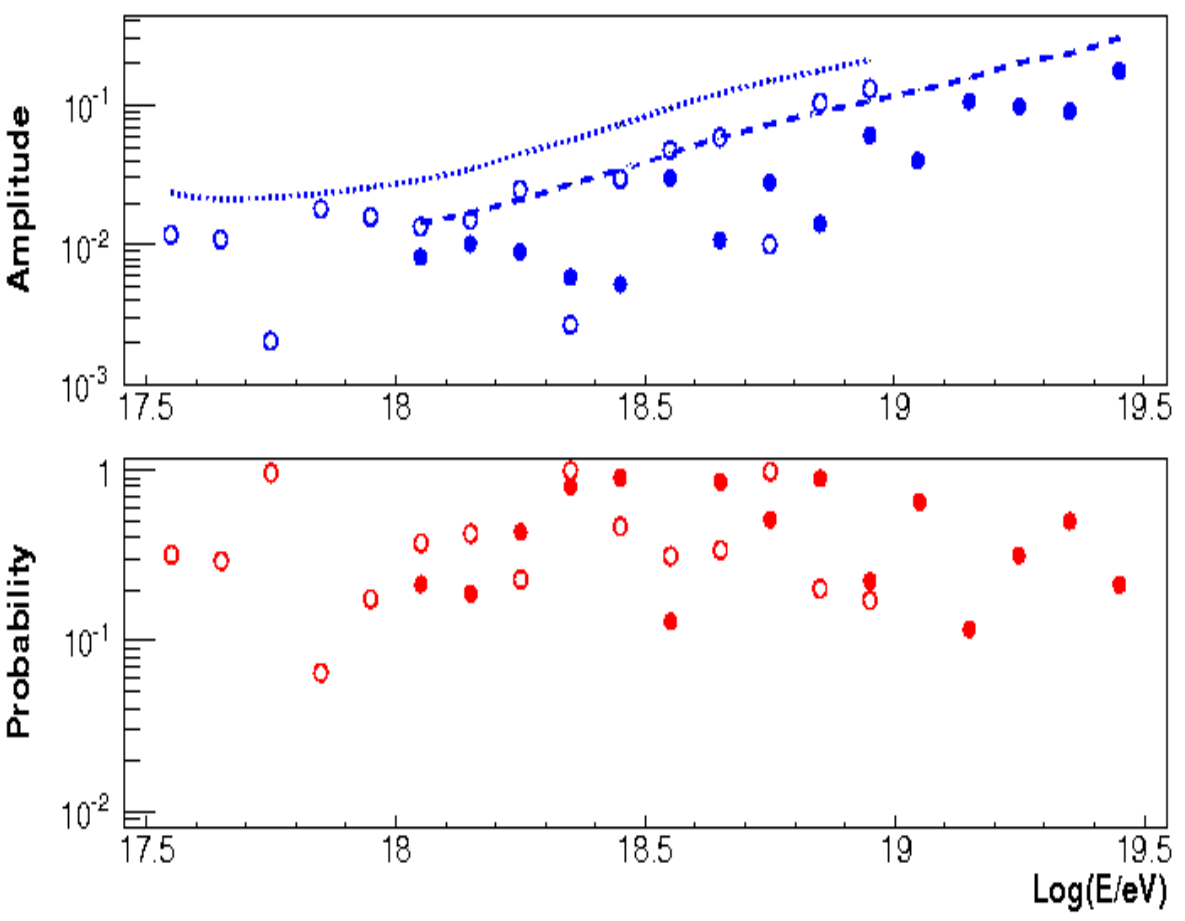

Fig. 3. The same as Fig.2 but here it is displayed for energy bins (instead of energy thresholds).

The statistics for some points of Fig.3 are obviously very low. Therefore we matched some of those energy intervals and repeated a first harmonic analysis using the two approaches. The results are collected in Tab.II. No significant departure from isotropy is observed with both methods. Having proved that both analyses account for the systematic effects, upper limits at 99% c.l. can thus be derived using only the statistical uncertainties. Such upper bounds, reported in the last column of Tab.II, have been calculated according to the distribution drawn from a population characterised by an anisotropy of unknown amplitude, as derived by J. Linsley in his $3^{rd}$ alternative [3].

## III. Discussion

Studying large scale anisotropies as a function of energy may give a handle to study the galactic/extragalactic transition. We show in Fig.4 the upper limits obtained in this study, together with some predictions for the anisotropies arising from both galactic and extra-galactic models.

If the transition occurs at the ankle energy [7], cosmic rays at $10^{18}$ eV are predominantly galactic and their escape from the galaxy by diffusion and drift motions could induce a modulation at the percent level at EeV energies. The exact value strongly depends on specific models: Ptuskin et al. [8], considering different orientations of the local magnetic field and different positions of the observer, predict anisotropy amplitudes ranging from $10^{-6}$ up to $10^{-2}$. We show in Fig. 4 the models discussed by Candia et al. [9], in which the predictions for the anisotropies up to EeV energies arising from the diffusion in the Galaxy are obtained. As these predictions depend on the assumed galactic magnetic field model as well as on the source distribution, two illustrative examples are shown. The bounds obtained here already exclude the predictions from the particular model with an antisymmetric halo magnetic field ($A$) and are starting to become sensitive to the predictions of the model with a symmetric field ($S$).

On the other hand, a second possible scenario considers the transition taking place at lower energies, i.e. around the so-called “second knee”, at $\sim 5 \cdot 10^{17}$ eV [10]. In this case, at $10^{18}$ eV cosmic rays are dominantly of extra-galactic origin and their large scale distribution could be influenced by the relative motion of the observer with respect to the frame of the sources. For instance, if the frame in which the CR distribution is isotropic coincides with the CMB rest frame, the resulting anisotropy due to the Compton-Getting effect (*C-G Xgal* in Fig. 4) would be about 0.6% with a phase $\alpha \simeq 168°$ [11]. This amplitude is very close to the upper limits set in this analysis (the statistics required to become sensitive to such amplitude at 99% c.l. is $\sim 3$ times the present statistics).

In the same figure we also display previous results from KASCADE, KASCADE-Grande and AGASA. A proper comparison of the results from different observatories should take into account the particular sky coverage of each experiment. All the anisotropy amplitudes have thus been divided by the mean value of the cosine of the declination of the observed sky, giving a direct measurement of the component of the dipole in the equatorial plane. The results presented here do not confirm the $\sim 4\%$ anisotropy reported by AGASA in the $1-2$ EeV energy bin [12] (however a proper comparison should take into account the peculiarities of the two experiments).

TABLE II

RESULTS OF THE TWO ANALYSES IN DIFFERENT ENERGY RANGES (THE EVENTS IN THE DIFFERENT ENERGY INTERVALS ARE SLIGHTLY DIFFERENT BETWEEN THE TWO METHODS BECAUSE THE RAYLEIGH ANALYSIS, UNLIKE THE EAST-WEST METHOD, CORRECTS THE ENERGY OF THE EVENTS FOR THE WEATHER EFFECTS). THE STATISTICAL UNCERTAINTIES ARE CHARACTERISED BY THE QUANTITIES $s_R = \sqrt{2/N}$ AND $s_{EW} = \sqrt{2/N}/\sin\delta t$. RAYLEIGH PROBABILITIES AND 99%C.L. UPPER LIMITS ARE ALSO GIVEN. SINCE ALL THE MEASURED AMPLITUDES ARE COMPATIBLE WITH BACKGROUND, THE PHASES ARE NOT SIGNIFICANT AND ARE NOT REPORTED HERE.

| | **Rayleigh analysis** | | | **E-W method** | | | **upper limits** |
|---|---|---|---|---|---|---|---|
| Energy range [EeV] | r [%] | $s_R$ [%] | P [%] | r [%] | $s_{EW}$ [%] | P [%] | $r_{99\%}$ [%] |
| all energies | | | | 0.48 | 0.27 | 19.5 | 1.05 |
| 0.2 - 0.5 | | | | 0.25 | 0.43 | 84.2 | 1.19 |
| 0.5 - 1 | | | | 1.08 | 0.44 | 4.8 | 2.03 |
| 1 - 2 | 0.90 | 0.32 | 1.8 | 0.77 | 0.65 | 49.9 | 1.59 |
| 2 - 4 | 0.79 | 0.64 | 45.8 | 1.65 | 1.33 | 46.3 | 2.12 |
| 4 - 8 | 0.71 | 1.33 | 86.6 | 5.05 | 2.73 | 18.0 | 3.66 |
| >8 | 5.36 | 2.05 | 3.3 | 2.76 | 4.08 | 79.5 | 9.79 |

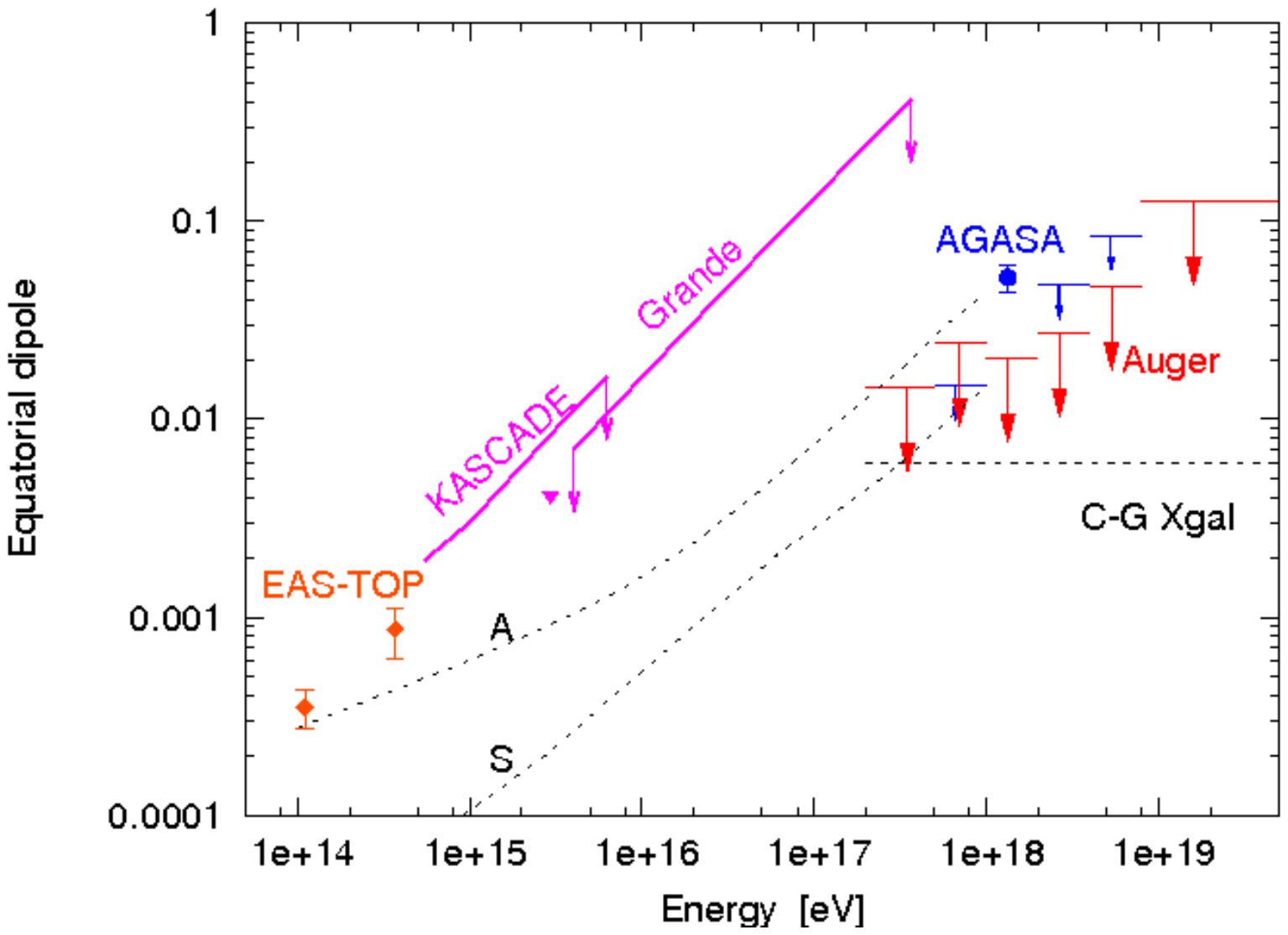

Fig. 4. Upper limits on the anisotropy amplitude as a function of energy from this analysis. Results from EAS-TOP, AGASA and KASCADE/Grande experiments are displayed too. Also shown are the predictions from two different galactic magnetic field models with different symmetries ($A$ and $S$) and the expectations from the Compton-Getting effect for an extra-galactic component isotropic in the CMB rest frame (*C-G Xgal*).

## IV. CONCLUSIONS

We have searched for large scale patterns in the arrival directions of events recorded at the Pierre Auger Observatory using two complementary analyses.

We have set 99% c.l. upper limits at the percent level at EeV energies, constraining some theoretical models. In particular, we can already exclude all those models that predict anisotropy amplitudes greater than $\sim 2\%$ below 4 EeV. Further statistics will obviously be useful, and the sensitivity will be improved in the coming years using data from the Pierre Auger Observatory.

Finally we do not confirm the 4% modulation detected by AGASA at 4 s.d. between 1 and 2 EeV.

# Cosmic Ray Solar Modulation Studies at the Pierre Auger Observatory

**Hernán Asorey*** **for the Pierre Auger Collaboration**†

*Centro Atómico Bariloche, Instituto Balseiro (CNEA-UNCuyo-CONICET)
and Universidad Nacional de Río Negro, San Carlos de Bariloche, Río Negro, Argentina
†Pierre Auger Observatory, (5613) Malargüe, Prov. de Mendoza, Argentina

***Abstract*. Since data-taking began in January 2004, the Pierre Auger Observatory has been recording low energy threshold rates for the self-calibration of its surface detectors. After atmospheric corrections are applied, solar modulation and transient events are observed. In this study, we present an analysis of the available data, with special emphasis on the observation of Forbush Decreases. A strong correlation with neutron monitor rates is found. The high total count rates allow us to determine temporal variations of solar origin with high accuracy.**

## I. Introduction

The Pierre Auger Observatory [1] has been designed to study the physics of cosmic rays of the highest energies. The Observatory combines two detection techniques in a hybrid design: the observation of the fluorescence light produced by the secondary particles as they propagate through the atmosphere; and the measurement of particles reaching ground level. The Pierre Auger Observatory has been taking data in a stable way since January 2004. The Surface Detector (SD) [2] is an array of more than 1600 water-Cherenkov detectors in a triangular grid with a spacing of 1500 m, covering 3000 km$^2$.

Each water-Cherenkov detector consists of a $10\,\mathrm{m}^2$ area polyethylene tank containing 12 tonnes of high-purity water in a highly-reflective liner bag. Cherenkov radiation generated by the passage of charged particles through the detector is collected by three 9” photomultipliers (PMTs). The water-Cherenkov detector is also sensitive to high energy photons, as they convert to $e^+e^-$ pairs in the water volume. The signals in the PMTs are processed by a fast analog-to-digital converter with a sampling rate of 40 MHz. A GPS system is used for timing and synchronization. Each detector is powered by a solar panel and batteries, working as an autonomous station linked to the central data acquisition system in Malargüe through a dedicated WAN-like radio network. Typically, more than 98% of the stations are operational at any time.

Previous work reported on the sensitivity of the Pierre Auger Observatory to Gamma Ray Bursts (GRBs) [3] using the *single particle technique* [4]. This method, in use in other cosmic ray experiments, consists in recording low threshold rates with all the surface detector stations, and looking for significant excesses in these rates.

In March 2005, a first set of scalers was implemented in each station of the Surface Detector of the Pierre Auger Observatory, mainly intended for the search of GRBs and for long term stability and monitoring studies. They consist in counters which register signals above a very low threshold, corresponding to an energy of $\sim 15\,\mathrm{MeV}$ deposited by individual particles in the detector. The typical station rate is $3.8\,\mathrm{kHz}$. In order to remove signals produced by muons and to improve the signal to noise ratio for GRB searches, an upper threshold of $\sim 100\,\mathrm{MeV}$ was introduced in September 2005, reducing the station rate to $2\,\mathrm{kHz}$ [3]. These rates are read every second and sent to the Central Data Acquisition System for their storage and further analysis.

## II. Scaler data treatment

In addition to the fact that the rate of low energy particles is not intrinsically constant, some instrumental instabilities and the atmospheric weather conditions are known to modify further this rate. They have to be taken into account before searching for transient events that last longer than a few minutes, or for effects such as the solar modulation of the galactic cosmic ray flux.

A data cleaning procedure optimized for the search of GRBs has been previously reported [3]. However, due to the very different time scales involved here, different cleaning procedures are needed.

First, stations with rates lower than $500\,\mathrm{Hz}$ are removed, as such a low rate may be an indication of temporary malfunction. For each individual second we also discard data from those stations with extreme rate counting (upper and lower 2.5%). The second step is to remove those periods where less than 97% of the array is in operation, resulting in a loss of less than 10% of data. This step is needed because individual stations have different average counting rates, due to many factors, from detector calibration to pressure effects coming from the different altitudes at which detectors are deployed.

Data for the two different periods (before and after September 2005, when the upper threshold was implemented) was analyzed independently. For both periods the average scaler rate was computed for each station over the lifetime of the detector. Detectors showing an RMS of more than twice the square root of the rate were excluded, keeping more than 90% of the stations after this cut. Brief spurious events (such as high frequency noise produced by lightnings) were removed by computing the average scaler rate for each detector over a 5 minutes period in which data is available for at least 4 minutes, removing the upper and lower 25% extreme values. Figure 1 shows the scaler data obtained for both the first and second scaler mode periods.

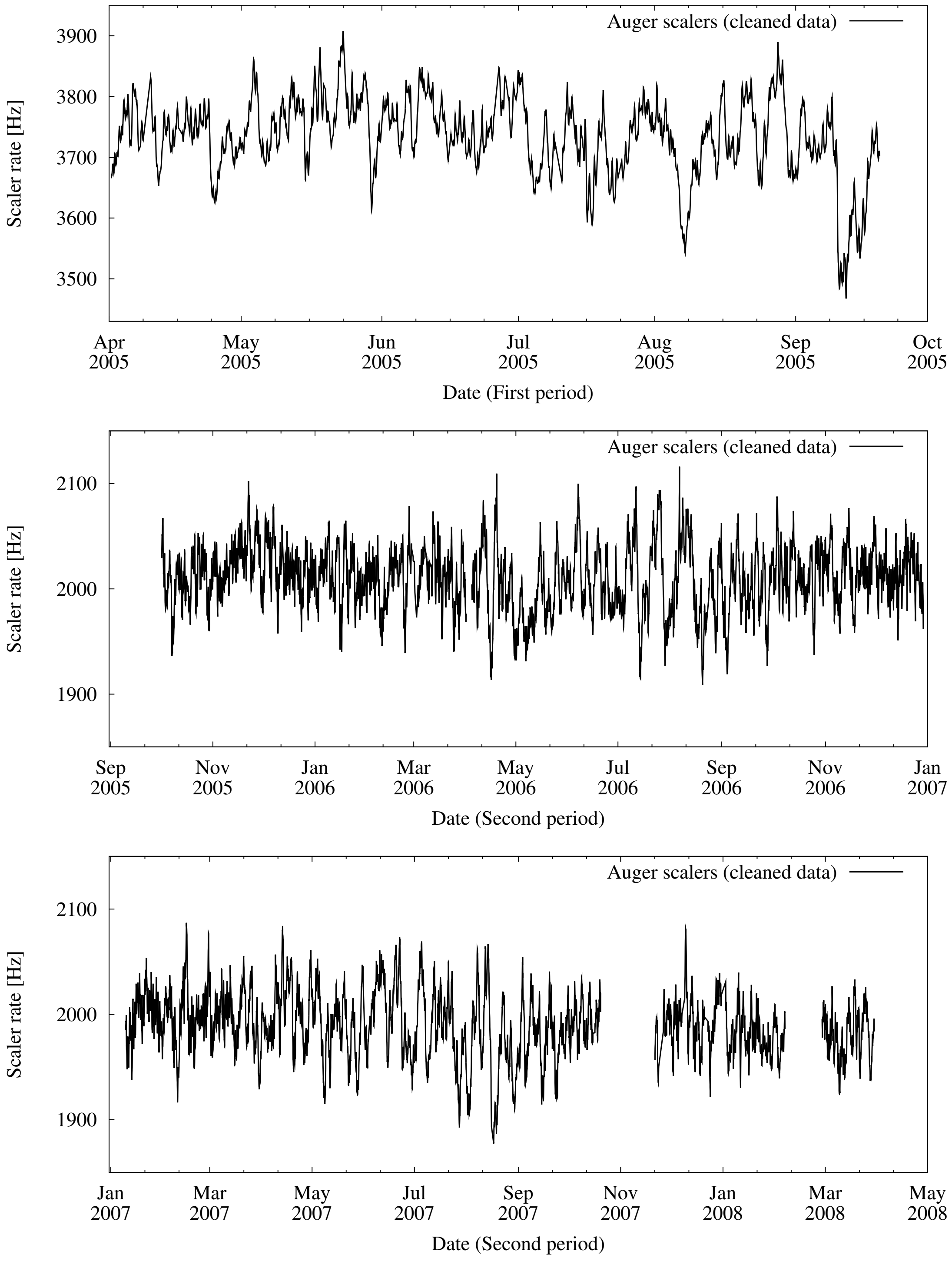

Fig. 1. Pierre Auger Observatory scaler rates after the data cleaning procedure described in the text, for the first period (before September 2005, top), and for the second period (after September 2005, middle and bottom). Each data point is the average scaler rate over one hour.

Atmospheric pressure variations are known to modify the flux of secondary particles at ground level, due to the different mass of atmosphere above the detector: an increase in the atmospheric pressure is correlated with a reduction in the background rate. Figure 2 shows the average scaler rate after the cleaning procedure, correlated with the flux of secondary particles at ground level, for the first ten days of May 2007, compared with the atmospheric pressure, as measured by several weather stations monitoring the array.

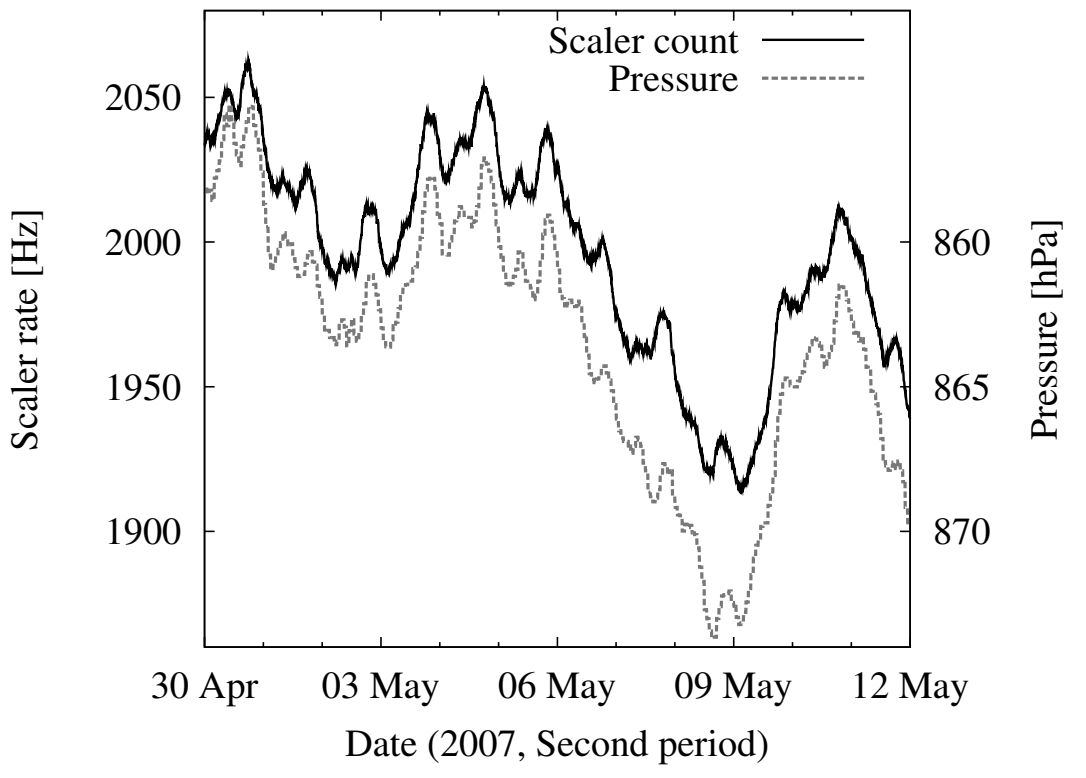

Fig. 2. Average scaler rate (solid line) and atmospheric pressure (dashed line, please note the reversed scale for pressure) for the first ten days of May of 2007.

The aforementioned correlation is observed and corresponds to about -0.27% (-0.36%) per hPa before (after) the upper threshold implementation. This also implies that an additional correction of 0.03% (0.04%) per metre of difference in altitude between stations has to be included.

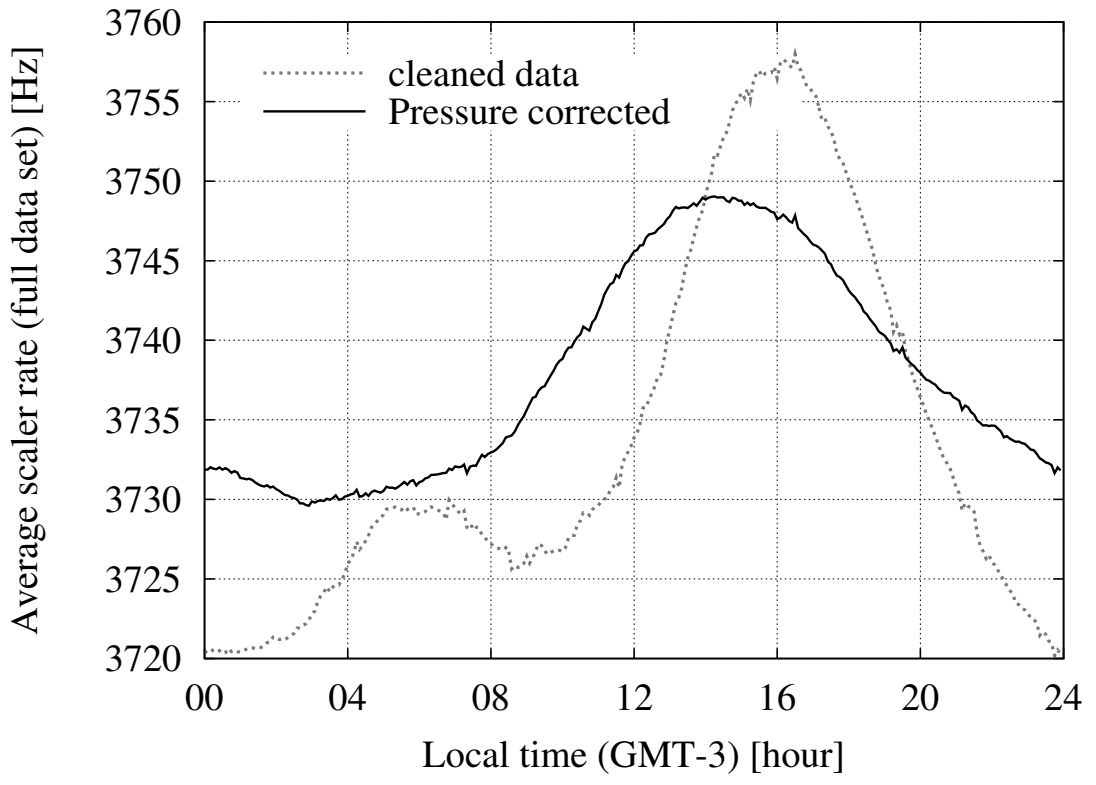

Fig. 3. Averaged scaler rates for the first period (before September 2005) as a function of local hour of day (ARS, GMT-3) for the cleaned (dotted line) and for the atmospheric pressure corrected data (solid line).

## III. Solar Modulation

Figure 3 shows the average daily dependence. After atmospheric pressure correction, a 0.25% modulation remains, peaked at 17h45 UT (14h45 local time).

To show that the cleaned, pressure corrected, data set is of relevance for solar studies [5], we compare the Pierre Auger Observatory SD scaler rates with data from McMurdo neutron monitor of the Bartol Research Institute [6]. Figure 5 shows the excellent agreement found: Forbush decreases are clearly visible in the scaler data for both periods. The upper threshold introduced on September 2005, intended to optimize signal over noise ratio for GRB detection, is probably not the ideal one for these studies, where the muon flux at ground level might be better correlated to the primary cosmic ray flux than the electron one.

As an example, Figure 4 shows the evolution of the scaler rates during the 11 Sep 2005 Forbush decrease [7]. A 4% variation is observed, in agreement with the 14% measured at McMurdo once taken into account the rigidity cut-off of $9.5$ GeV at Malargüe, Argentina (35.3$^o$S, 69.3$^o$W).

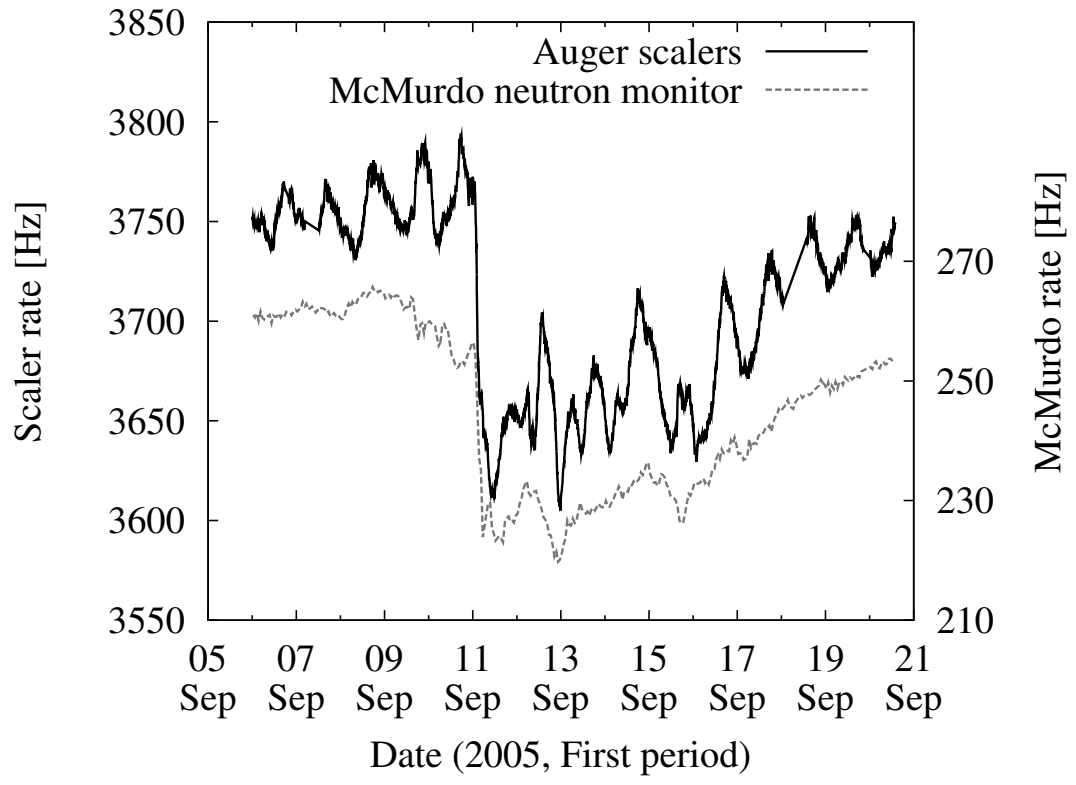

Fig. 4. Averaged, cleaned and pressure corrected scaler rate for the 11 Sep 2005 Forbush decrease, compared with the McMurdo neutron detector rate (dotted line). A 4% variation is observed.

## IV. Conclusions

Low energy radiation rates are registered with high statistics at each surface detector station of the Pierre Auger Observatory since March 2005. A data cleaning method, based on a previous one intended for the search of GRBs, has been implemented and optimized to study solar modulation effects. After correcting for pressure, an excellent agreement with data from McMurdo neutron monitor is found, evidencing the high sensitivity that water-Cherenkov detectors, operating in scaler mode, have for the detection and measurement of Forbush decreases and other transient events related with the solar modulation of Galactic cosmic rays. Additional analyses of scaler rates at individual detectors and scaler thresholds optimisation are underway, together with low energy cosmic ray simulations in order to determine the energies of the primaries responsible for the observed modulation.

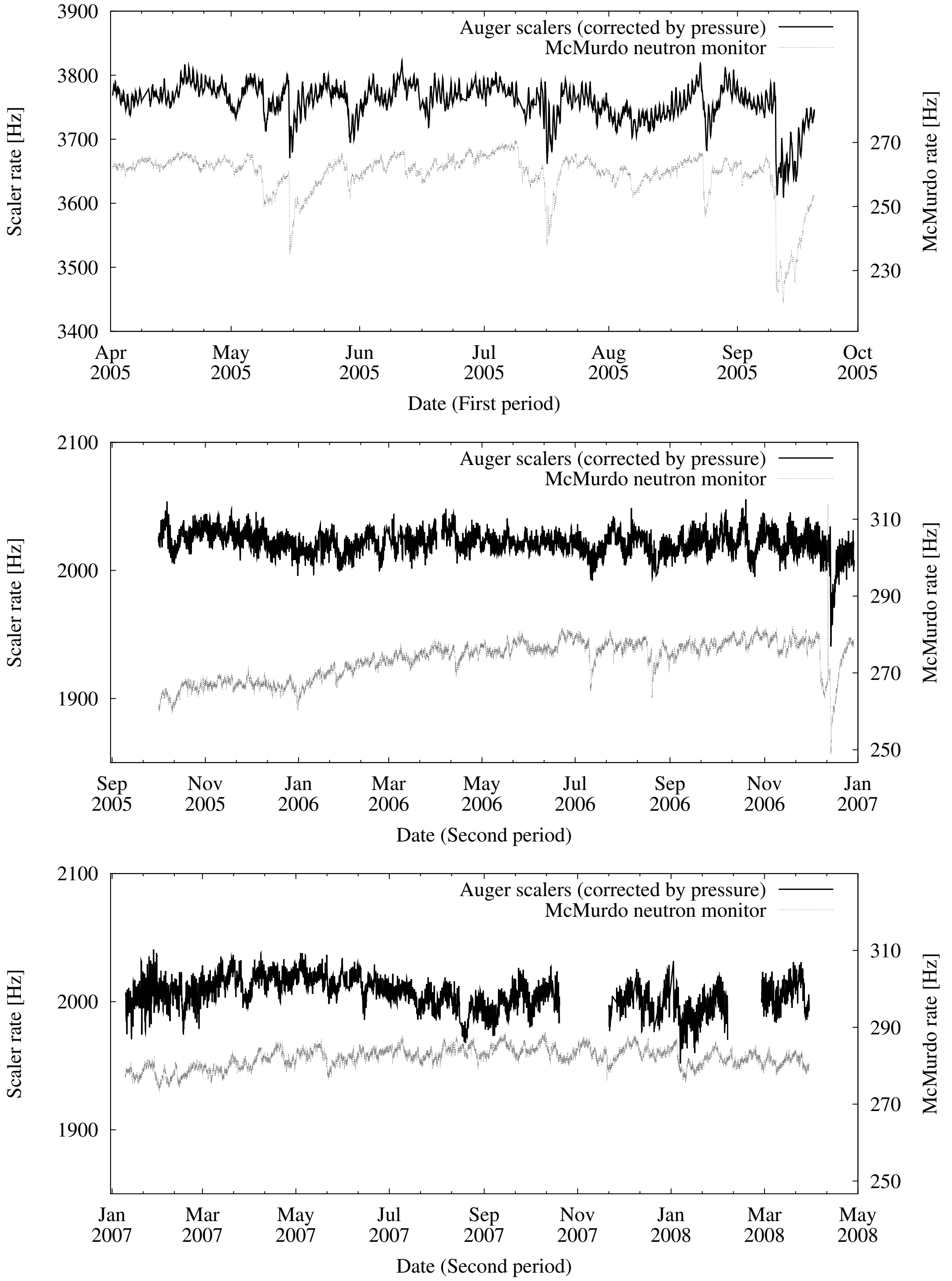

Fig. 5. Pierre Auger Observatory scaler rates (solid lines) after cleaned procedure and atmospheric pressure correction, compared to McMurdo neutron detector data (dotted line), for both periods, before (top) and after September 2005 (middle and bottom).

# Investigation of the Displacement Angle of the Highest Energy Cosmic Rays Caused by the Galactic Magnetic Field

**B. M. Baughman*, for the Pierre Auger Collaboration†**

*Ohio State University, 191 W. Woodruff Ave., Columbus, OH 43210-1061, United States

†Observatorio Pierre Auger, Av. San Martín Norte 304, (5613) Malargüe, Mendoza, Argentina

***Abstract.*** **Ultra-high energy cosmic rays (UHECR) are deflected by magnetic fields during their propagation. Different theoretical parameterizations of the Galactic magnetic field are examined using a numerical tool which simulates their propagation through models of these fields. We constrain the possible parameter space of the models by comparing data on UHECR obtained with the Pierre Auger Observatory with the results of the simulations.**

## I. Introduction

Current knowledge on the strength and shape of the Galactic magnetic field is limited [1], [2], [3], [4], [5], [6], [7]. While there are hints [3], [4], [5], [6], [7] at the form and magnitude of the regular component of the Galactic magnetic field there is currently no consensus on its general form or magnitude. Presented here is a method for determining regions of magnetic field model parameter space which are compatible with a given set of assumptions. Combining this method with complementary methods [8] employing multi-wavelength observations will improve the global delimiting power. This method is applied to determine regions of parameter space for two Galactic magnetic field models and under two dramatically different sets of assumptions: 1) the Pierre Auger Observatory [9] (Auger) anisotropy result with a pure proton composition and 2) Centaurus A as a dominant source with a pure iron composition.

## II. Data Set

This analysis uses data recorded by the southern site of the Auger Observatory between 1 January 2004 and 31 March 2009 with energies greater than 55 EeV and zenith angles smaller than 60°.

## III. Galactic Magnetic Field Models

Logarithmic symmetric spiral field models [1], [10] are used to model the large scale regular Galactic magnetic field in this work. Axisymmetric [Bi-symmetric] (denoted herein as ASS_* [BSS_*]) fields exhibit [anti-]symmetry under rotations of $\pi$ around the Galactic pole. Fields [anti-]symmetric under reflection across the Galactic plane are denoted with *SS_S [*SS_A]. A precise description of the models used, including the nominal parameter values, is given by Harari et al. [1].

These models are likely not a complete description of the Galactic magnetic field [2], [3], [5], [7], [11]. There is evidence of turbulent and halo fields in the Galaxy [2], [3], [4], [7], [6], [12] and others [13], [14], [11]. Furthermore, extra-galactic magnetic fields [15], [16], [17] will also have an effect on the results of this work. However, logarithmic symmetric spirals represent a priori reasonable models for the functional form of the regular component of the Galactic magnetic field [4], [14][1]. Here we present results using two logarithmic symmetric spiral models: BSS_S and ASS_A.

## IV. Method

Regions of compatible parameter space are determined by testing the Auger Observatory data under a set of assumptions. The general method presented here requires a hypothesis which defines the following: 1) A charge for each observed event, 2) a catalog which traces the true source distribution and parameters for correlation ($\Psi_{max}$), 3) a model for the magnetic field. A region of interest in parameter space is then gridded and all events are backtracked through the model at each grid point. The region around the direction of the exiting velocity vector for each event is then searched for the nearest catalog object. Finally, the number of events ($N_{corr}$) correlating ($\Psi \leq \Psi_{max}$) with the hypothesized source catalog are interpreted relative to the number of correlations determined without backtracking through the magnetic field model.

The statistical significance of $N_{corr}$ is determined by assuming that the number of correlations with the non-backtracked arrival directions ($N^0_{corr}$) is an estimate of the true number of correlations for the parameters given. We take the number of correlations with the non-backtracked arrival directions as an estimate of the mean of a Poisson distribution. We expect this to be a conservative estimator of the true probability mass function[2]. A region of compatibility is defined as lying centered on the mean and containing 68% of random $N_{corr}$ values sampled from this distribution. Scan points with $N_{corr}$ values lying outside this region are considered incompatible with the full set of assumptions (field, source, and composition). Figure 1 shows two Poisson distributions with shaded regions found using the above method.

[1]The results presented herein may depend on the specifics of turbulent, halo, or extra-galactic magnetic fields which are poorly constrained, as such the effects of such fields have been ignored in favor of presenting a clear description of the method used.

[2]We have verified this by smearing the data with a 2D gaussian and examining simple Monte Carlo realizations.

TABLE I: Example Assumptions

| | VCV | CenA |
|---|---|---|
| Composition | Pure proton | Pure iron |
| Source Distribution | VCV Catalog | Centaurus A |
| Correlation Scale ($\Psi_{max}$) | 3.1° | 15° |
| Minimum Energy ($E_{max}$) | 55 EeV | 55 EeV |
| Poisson Mean ($N^0_{corr}$) | 27 | 9 |

## V. EXAMPLES

Here we present two examples for clarity. Each example explicitly assumes a composition, a source distribution, and angular correlation window which can be found in Table I.

The first example is inspired by the Auger Observatory anisotropy result [9] and will be referred to as the VCV example throughout. By correlating the arrival directions of UHECR with Active Galactic Nuclei (AGN) from the Veron Catalog of Quasars & AGN, 12th Edition [18] (VCV catalog) and comparing the results with what one would expect from a truly isotropic source distribution, the Auger Observatory found that their observations were consistent with a sampling of a distribution similar to the VCV catalog. The Auger Observatory scanned over three parameters: maximum redshift ($z_{max}$), maximum angular correlation scale ($\Psi_{max}$), and minimum energy ($E_{min}$) finding the strongest correlation signal for $z_{max} = 0.018$, $\Psi_{max} = 3.1°$ and $E_{max} = 55$ EeV (equivalent to 57 EeV as reported in [9] due to a small change in our energy calibration). This analysis uses VCV catalog objects with a redshift $\leq 0.018$.

The second example assumes that Centaurus A (Cen A) is the dominant source in a region centered around it. This example is referred to as the Cen A example throughout. We have arbitrarily chosen $\Psi_{max} = 15°$.

Both examples use the minimum energy ($E > 55$ EeV) found in the Auger Observatory anisotropy result [19], [20]. Compatible regions for both BSS_S and ASS_A Galactic magnetic field models are presented. Neither the overall strength nor the scale height of the Galactic magnetic field is well constrained [1], [2], [4], [10]. Thus for each of the two models two parameters are varied: the strength of the field in the solar vicinity ($B_\odot$) and the dominant scale height ($z_1$). $B_\odot$ is varied from $-2.5$ $\mu$G to $2.5$ $\mu$G in steps of $0.01$ $\mu$G and from $0.05$ kpc to $3.05$ kpc in $z_1$ in steps of $0.1$ kpc.

## VI. RESULTS

### A. VCV

Figures 2a and 2b show the regions of $B_\odot$- $z_1$ parameter space compatible with the Auger Observatory anisotropy result. The regions of compatibility are defined using the method described in Sec. IV above. This example uses a $N^0_{corr} = 27$ to define the mean of the Poisson distribution from which we determine our compatibility regions. The shaded regions correspond to the similarly shaded regions in Fig. 1a. There are some general features to both field models which should be noted. First, $z_1$ is not well constrained as both models have regions compatible with the Auger Observatory anisotropy result extending across all reasonable scale lengths. Second, that connected statistically compatible regions lie within the range over which $B_\odot$ is varied.

There is marked difference in the compatible parameter space between ASS_A (Fig.2a,) and BSS_S (Fig.2b) models. ASS_A has large connected compatible regions only for a small range in $B_\odot$ centered around zero. Furthermore, no regions appear to improve the number of correlations, thus it is likely that the compatible regions are simply where the correlation with the arrival directions have yet to be destroyed. ASS_A is compatible only with $|N_\odot| < 1$ $\mu$G which is lower than the expected value [1], [3], [4], [5]. Overall, the ASS_A model is incompatible with the VCV correlation hypothesis in the regions deemed acceptable due to other constraints.

The connected compatible (gray and dark gray) region for BSS_S model is much larger and extends over a significant range in $B_\odot$. Furthermore, extended regions exist where $N_{corr}$ is above 27, the nominal value, indicating candidate models where the correlations are preserved and new correlations are made.

### B. Cen A

Figures 3a and 3b show the regions of $B_\odot$- $z_1$ parameter space compatible with the Cen A hypothesis. This example uses a $N^0_{corr} = 9$ to define the mean of the Poisson distribution from which we determine our compatibility regions. The shaded regions correspond to the similarly shaded regions in Fig. 1b.

This parameter space shares features with the previous example, mainly that $z_1$ is not well constrained and compatible regions of parameter space tend to prefer positive field strengths. The most striking feature of this parameter space is the very limited range in field strength compatible with the underlying hypothesis; both models require a field strength in the solar vicinity of less than $0.24$ $\mu$G.

## VII. CONCLUSION

Presented is a method which can be used to determine compatible regions of parameter space for magnetic field models using UHECRs. Regions of parameter space are shown for two common Galactic magnetic field models compatible with either the Auger Observatory anisotropy results and a pure proton composition or with Cen A as a dominant source on a $15°$ scale with a pure iron composition. These regions are determined through a statistical comparison between the results of correlations between Auger Observatory data and either the VCV catalog or Cen A and similar correlations between particles backtracked through BSS_S and ASS_A Galactic magnetic field models.

By assuming a pure proton composition and VCV catalog correlation parameters, it is found that the ASS_A model has a smaller compatible parameter space while BSS_S model is largely compatible with the Auger Observatory anisotropy result. The BSS_S model appears to improve the correlation with VCV catalog objects for a wide range of parameters. While the limited number of events available for this analysis limits our ability to make concrete predictions, there are hints that regions of parameter space for BSS_S model in particular should be examined for consistency with other measurements. When Cen A is assumed to be a dominant pure iron source, the compatible regions of parameter space for both BSS_S and ASS_A Galactic magnetic field models are much smaller than expected from other measurements. Both examples do not constrain the scale height of the models.

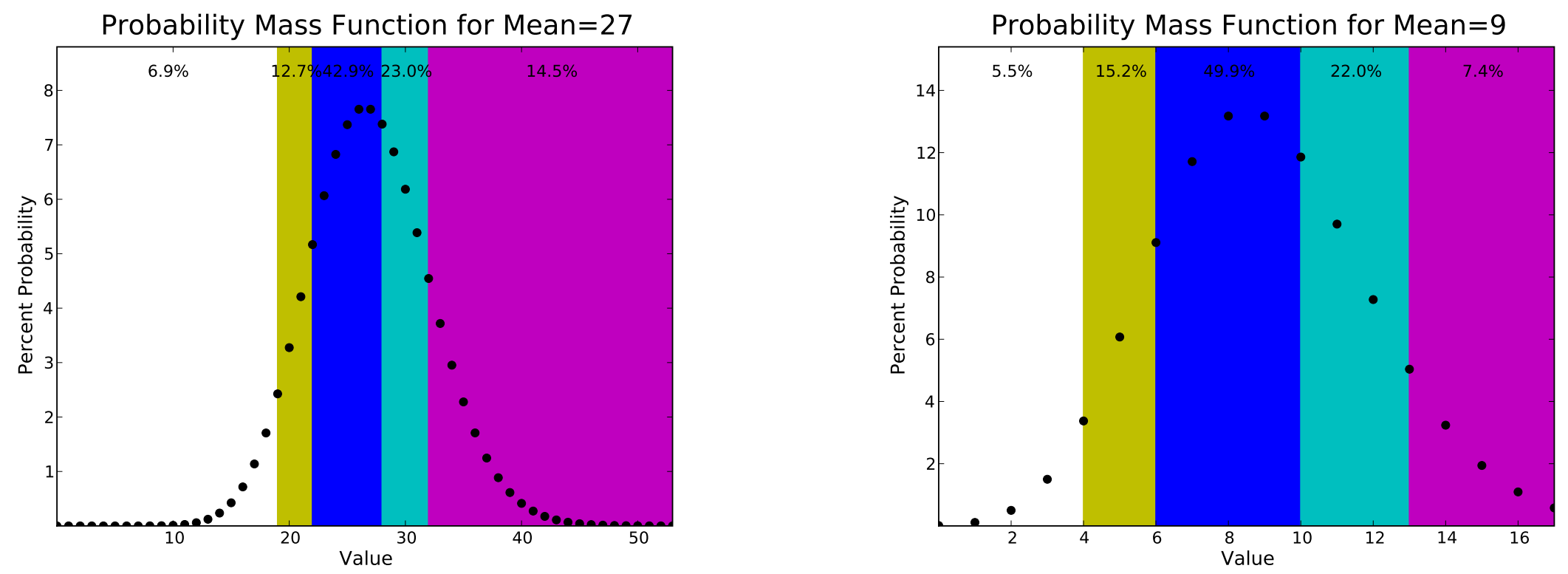

(a) Poisson distribution corresponding to the VCV example. (b) Poisson distribution corresponding to the CenA example.

Fig. 1: Circles represent the percent probability of obtaining each value. Regions shaded to indicate the portion of probability space occupied by corresponding regions in Figs. 2 and 3

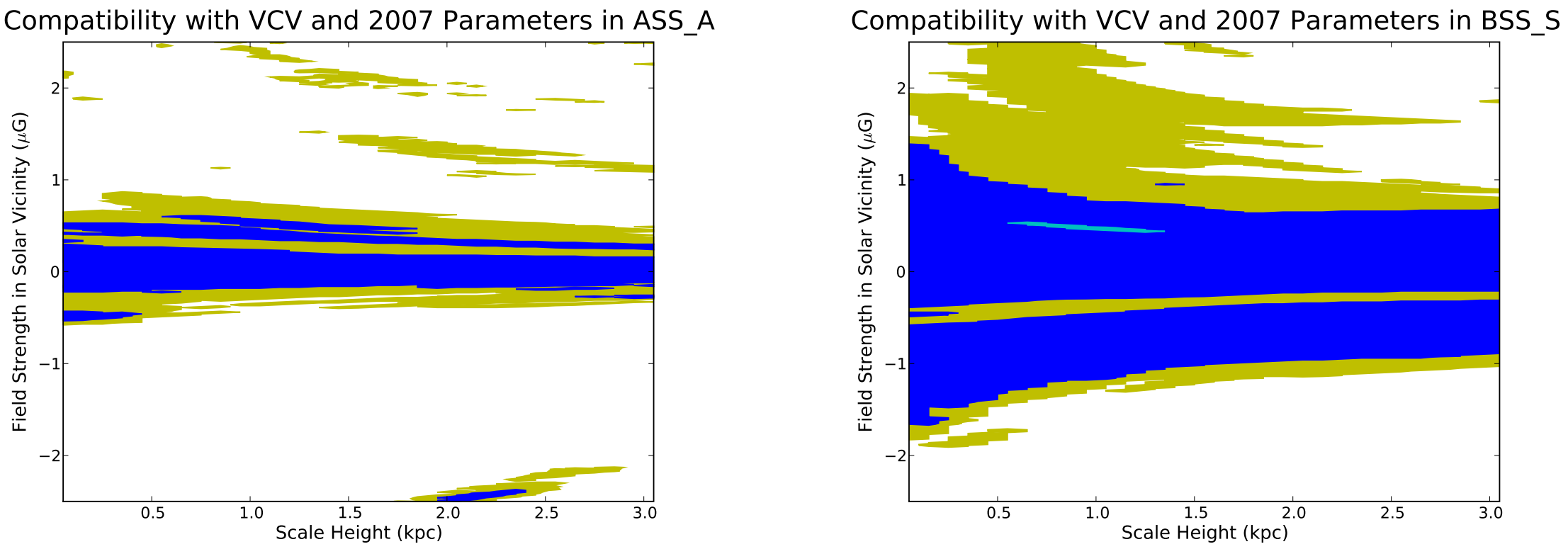

(a) ASS_A Galactic magnetic field model. (b) BSS_S Galactic magnetic field model.

Fig. 2: Regions shaded in: white has [0-19), light gray has [19,22), gray has [22,28), and dark gray has [28,32) correlations. The ASS_A model reaches a maximum of 28 correlations while the BSS_S model has a maximum of 29.

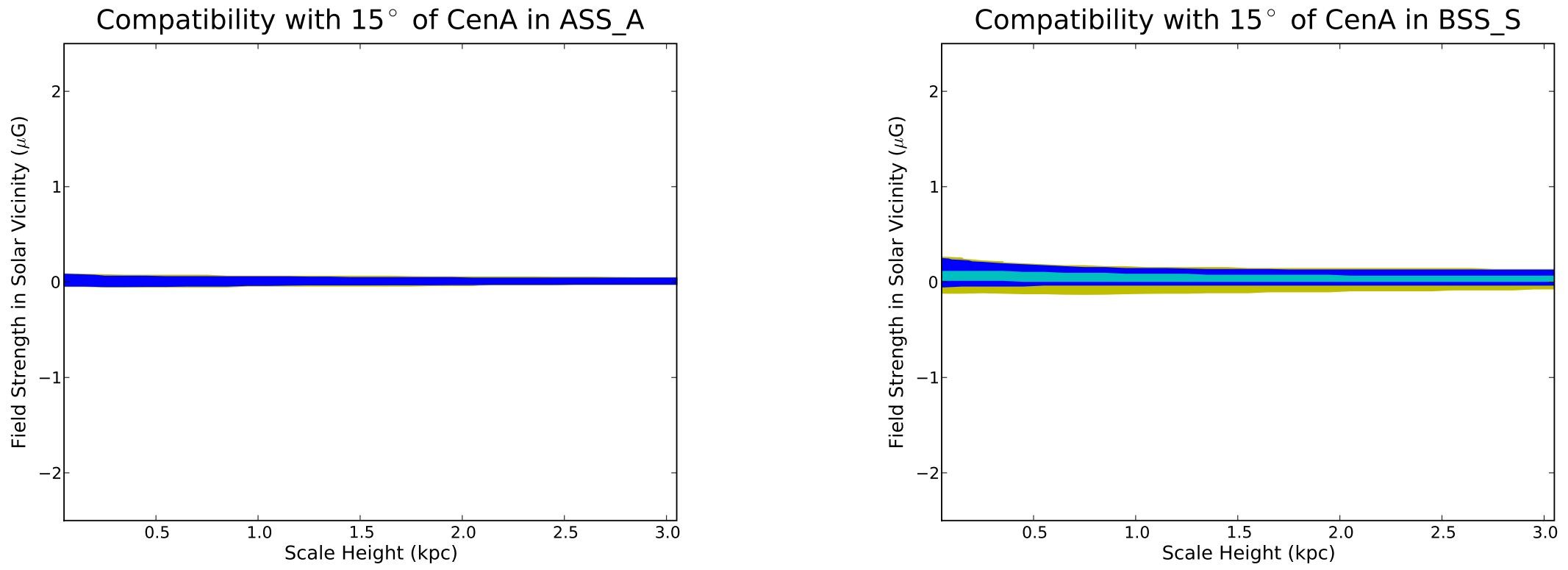

(a) ASS_A Galactic magnetic field model. (b) BSS_S Galactic magnetic field model.

Fig. 3: Regions shaded in: white has [0-4), light gray has [4,6), gray has [6,10), and dark gray has [10,13) correlations. The ASS_A model reaches a maximum of 10 correlations while the BSS_S model has a maximum of 13.

# Search for coincidences with astrophysical transients in Pierre Auger Observatory data

**David Thomas*** **for the Pierre Auger Collaboration**†

*Colorado State University, Fort Collins CO, USA
†Pierre Auger Observatory, Av. San Martin Norte 304 (5613) Malargüe, Prov. de Mendoza, Argentina

***Abstract*. We analyse extensive air shower data collected by the Pierre Auger Observatory to search for coincidences between the arrival directions of ultra-high energy cosmic rays and the positions of gamma-ray bursts. We also analyse the trigger rate data from individual surface detector stations to search for an increase of the average trigger rate over the entire surface detector array in correlation with gamma-ray bursts.**

## I. Introduction

Since their discovery at the end of the 1960s [1], gamma-ray bursts (GRBs) have been of high interest to astrophysics. A GRB is characterised by a sudden emission of gamma rays during a very short period of time, typically lasting between 0.1 and 100 seconds. The equivalent isotropic energy emission during a burst is typically between $10^{51}$ and $10^{55}$ ergs. Good source candidates for these bursts are the merger of neutron stars, for short bursts of less than 2 seconds, and core-collapse supernovae type Ib/c, for long bursts. See [2] for a recent review.

A large data set of GRBs was provided by the BATSE instrument on board the Compton Gamma Rays Observatory (1991-2000). More GRBs were then detected by BEPPO-SAX (1997-2002) and HETE (2001-2006). Currently, GRBs are registered by Swift, INTEGRAL, IPN, the Fermi Gamma Ray Telescope. In the last several years, afterglows have been observed allowing us to precisely measure their direction, and giving us a much better understanding of the GRB phenomena. Most observations have however been done below a few GeV of energy, and the highest energy observations have been around 10 GeV [3].

Using the recently completed surface detector array (SD) of the Pierre Auger Observatory [4], [5], an array of more than 1600 water Cherenkov detectors, we look for evidence of GRBs in our data with two separate techniques. For the first study, we use the regular Auger cosmic ray (CR) events to look for an excess of ultra-high energy CRs (UHECRs) in the direction of GRBs. We presented a similar analysis at the last ICRC [6].

In the second study, the "single particle technique" [7] is used with low-level trigger rate data from individual surface detector stations to look for a possible extension of the photon spectrum of GRBs in the 1 GeV – 1 TeV range, which is not well-studied by satellites. When photons in this energy range reach the atmosphere, they produce cosmic ray cascades that can be detected, although the energies are too low to produce a shower detectable at ground level, even at high altitudes. However, a lot of these photons are expected to arrive during a GRB in a short period of time, which would be detectable as an overall increase of the trigger rate in all the detectors [8] in the array.

This technique was first applied in EAS-TOP [7], using scintillators, and subsequently in INCA [9] and ARGO [10], using scintillators and RPCs, respectively. The main advantage to using water Cherenkov detectors [11], [12] is their sensitivity to photons, which represent up to 90% of the secondary particles at ground level for high energy photon initiated showers such as these. This study is also an update to one we presented at the last ICRC [13].

## II. Search for coincidence of UHECRs and GRBs

For this analysis, we use CR data collected with the Pierre Auger Observatory surface detector. We consider events from 1 January 2004 to 31 March 2009, passing quality criteria discussed elsewhere [14] with zenith angle $\theta < 60°$. We perform no energy cut on the data.

We use a catalogue of 511 GRBs [15] observed with an accuracy better than 1° compiled using data primarily from the Swift mission complemented by measurements from additional GRB observing satellites, including HETE, INTEGRAL, IPN, and Fermi. Out of the total GRB sample, 115 bursts are within the field of view at the time of their bursts, i.e. $\theta_{GRB} < 60°$.

We look for excess CRs in the direction of GRBs by determining the differences in the observation times of the GRBs and the arrival times of CR events that fall within a window of radius $\psi$ around each GRB position. We compare this with the expected rate of events, which is calculated in the same manner, using the declination band of width $2\psi$ around the GRB (excluding the circular window around the GRB). The expected rate of events is normalized by multiplying by the on-source solid angle of the window divided by the total solid angle (declination band minus the window.)

A value of $\psi = 2°$ corresponds roughly to the angular resolution when all SD events are used. To be consistent with the previous analysis, we also used a window of $\psi = 5°$. The results are presented in Figure 1. There

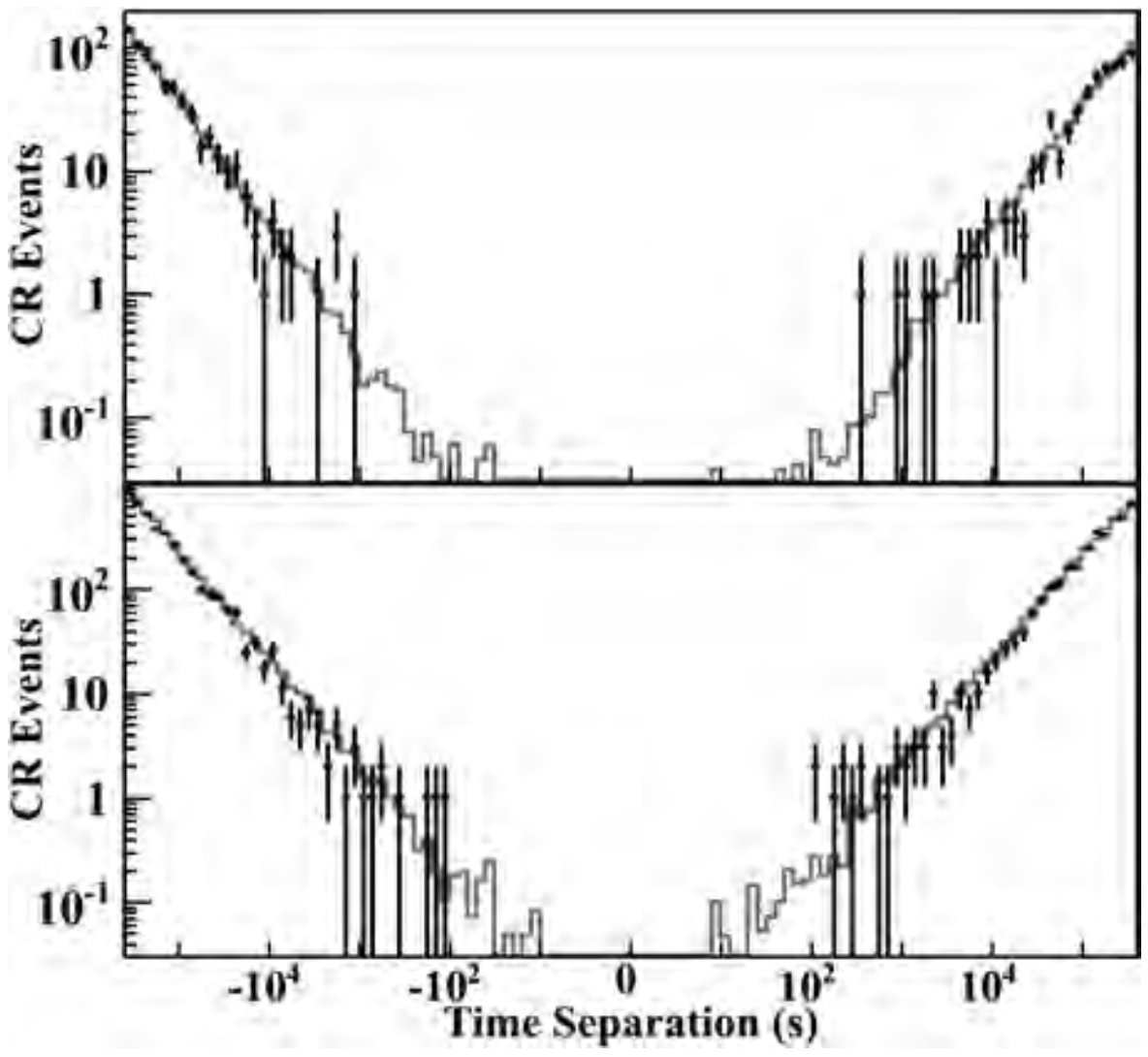

Fig. 1. Number of CR events as a function of the difference between the GRB time and the CR arrival time, for windows of $\psi = 2°$ (top) and $\psi = 5°$ (bottom). Data falling within the windows of radius $\psi$ are indicated by the points. The expected rate of events is indicated by the solid line. For clarity, statistical errors are only shown for the circular windows.

is no evidence for an excess of CRs coming from the direction of GRBs.

## III. Single particle technique

### A. Scaler data

In addition to the regular data acquisition system used to detect cosmic rays, the surface detectors are equipped with scalers, simple counters that can be set like any other trigger. They record the counting rates of signals more than 3 ADC counts above baseline and less than 20 ADC counts, or approximately between 15 and 100 MeV deposited in the detector. This trigger level has been set to optimise the signal to noise ratio given the expected signal extracted from simulations [16], and the background signal derived from real data histograms. With these cuts, the average scaler rate over the array is of about 2 kHz per detector. Note that the scaler data is completely separate from the regular cosmic ray data, as it only reports trigger rates in individual surface detector stations, and cannot be used to reconstruct cosmic ray events.

To use the scaler data, it must be cleaned. Individual detectors often experience increases in their counting rates due to noisy or unstable baseline, unstable PMTs, or bad calibration. Detectors with less than 500 Hz of scaler counts are discarded. This removes a few badly calibrated detectors. Additionally, for each individual second, only 95% of detectors are kept, removing the 5% with extreme rate counting (2.5% on each side). This removes outliers which could impact the average rate of a specific second, without affecting the GRB detection capability, as GRB would appear as an increase of counting rates in all the detectors.

One then needs to have the array operating properly. Suddenly losing a significant fraction of the array will cause jumps in the scaler rate, as this rate is not uniform over the whole Observatory. Consequently, we only use data periods for which more than 97% of the stations are operating. This keeps 83% of the data. We also require at least 5 continuous minutes with data, to be able to compute reasonable averages and see eventual bursts. This removes 12% of the remaining data set. Some artificial bursts are found in the cleaned scaler data set due to lightning strikes. Therefore, we do not use the data taken during lightning storms, removing 4% of the data. This scaler data cleaning process is different from the one used in the analysis of the modulation of solar CRs [17] due to the difference in time scales.

### B. $\sigma - \delta$ method

To search for bursts, the average rate for each second as well as a longer term average rate are computed. As a burst would produce a similar increase in all stations, a good estimator of the average rate for each second, $r$, is the median of the rates of all the stations. It is much less sensitive to misbehaving detectors than the arithmetic average. Then, to estimate a long term average $R$, a $\sigma - \delta$ method is used with $\sigma = 0$ and $\delta = 0.1$ Hz, meaning that every second the average rate $R$ is moved by 0.1 Hz towards the current rate $r$. After 30 seconds of data, this average converges to the expected average value, and one can compute the variation $\Delta$ of the rate $r$ of a specific second using:

$$\Delta = \frac{r - R}{\sqrt{r/N}}$$

where $N$ is the number of active detectors at that second.

The $\Delta$ parameter can be used directly to search for bursts, and its histogram can be seen in Figure 2. The underlying Gaussian has a width of 1.4. It would have a width of 1 if the arriving flux of particle was poissonian, the fluctuations of each detector were independent, the baselines of the detectors were not fluctuating, and the $\sigma - \delta$ method gave the true average at each moment. One sigma of deviation corresponds roughly to 1.5 particles per detector, i.e. a flux at ground level of 0.15 m$^{-2}$ s$^{-1}$.

### C. Search for self-triggered bursts

Once all the cuts defined above have been applied, a total of 70% of the data period (21 September 2005 - 31 March 2009) is available for a search for bursts. The resulting $\Delta$ histogram is shown in Figure 2.

Only three candidate bursts are observed significantly outside of the Gaussian distribution. To be related to a GRB, the increase of the rate should be uniformly distributed over all the detectors. One can therefore check that each individual detector has on average an increase at the moment of the burst with respect of the previous seconds. The observed outliers do not present such a feature, as only a fraction of the array sees a significant excess. We show in Figure 3 the histogram of the difference of the scaler rate from the average over the previous seconds for one of these candidate bursts. Most of the rate increase is attributable to a small portion of the array. The excess is therefore artificial and must be due to lightning. All other candidate bursts display this same attribute.

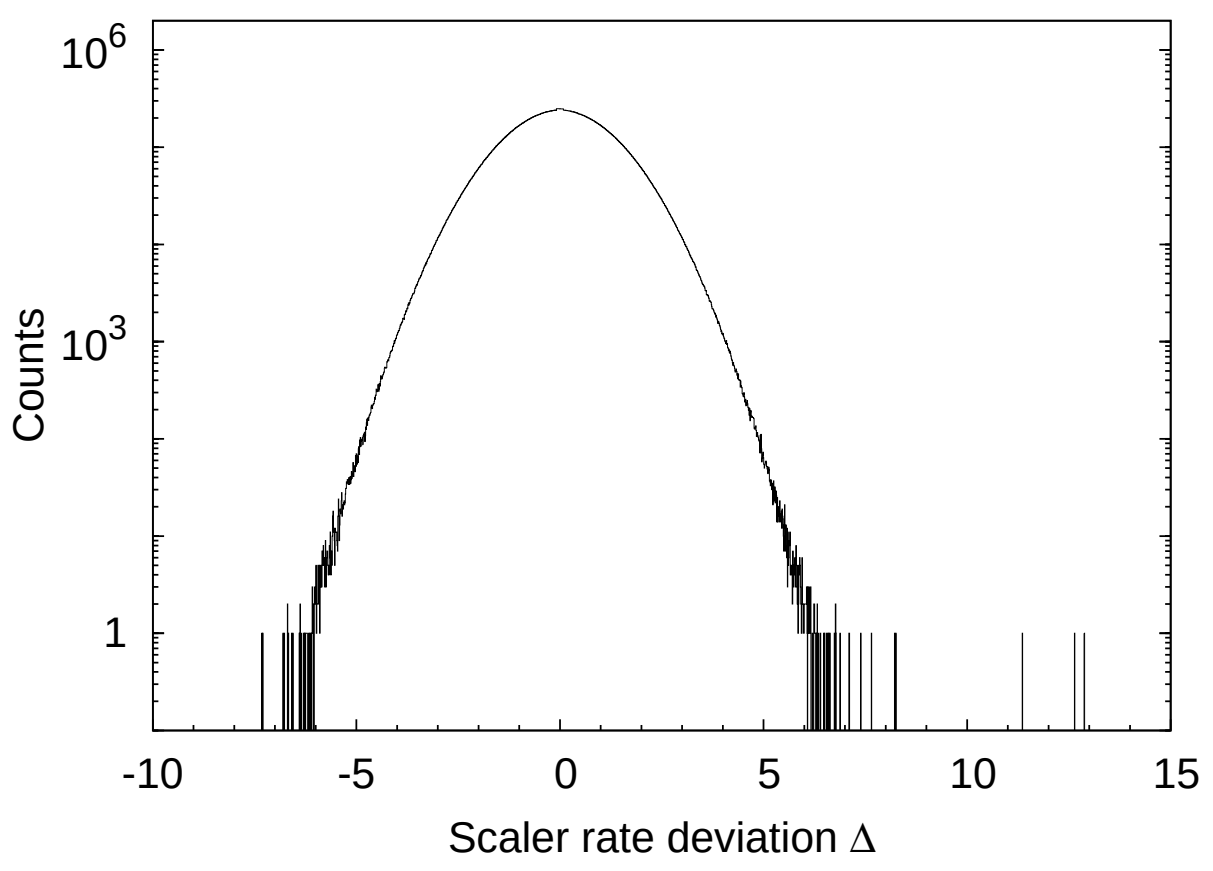

Fig. 2. Histogram of the deviations $\Delta$ of the scaler counting rates. The distribution is Gaussian with a width of 1.4. Three candidate bursts are present above $10\sigma$ which we looked at more closely to determine if they are due to GRBs.

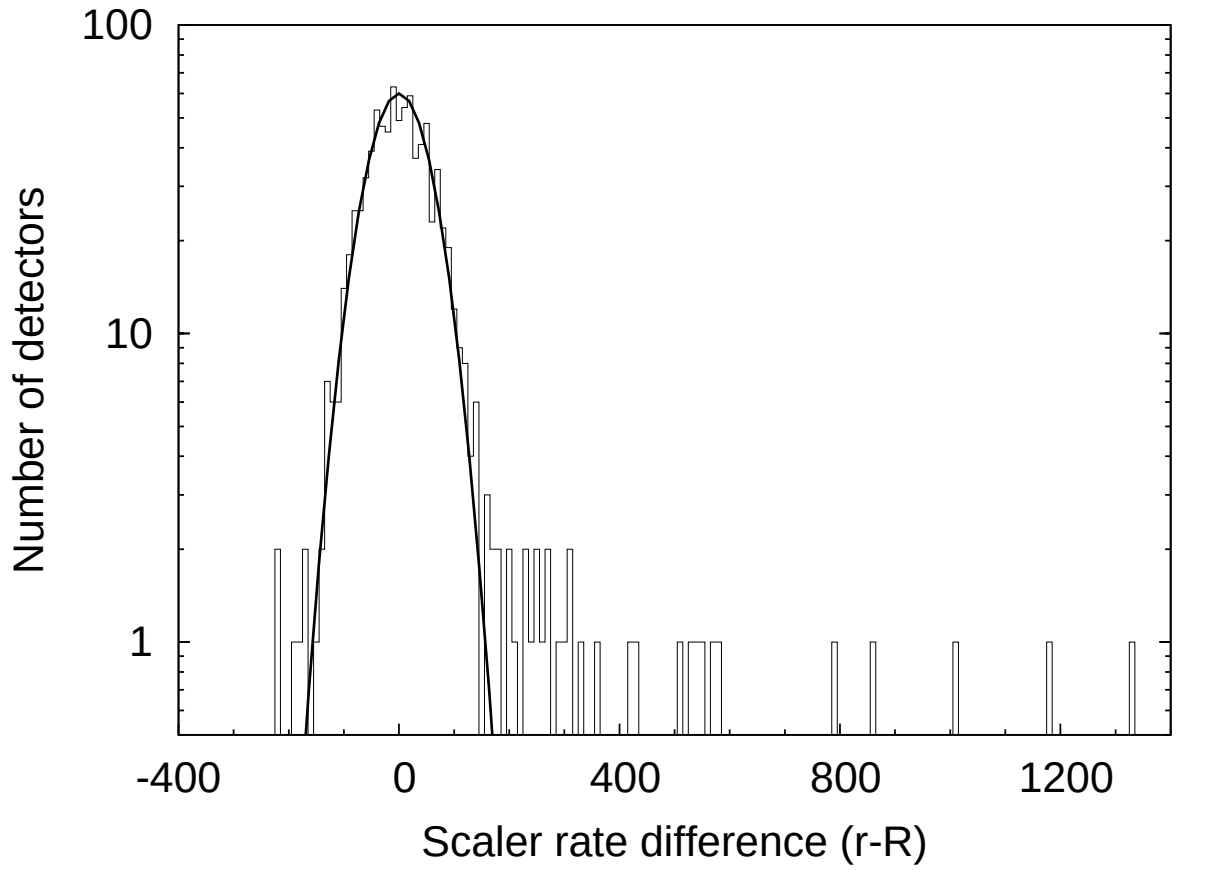

Fig. 3. Histogram of the difference of scaler rates from the average over the previous seconds for one of the three candidate bursts. The scaler rate increase is due to an increase in the rate for a small fraction of the array. Thus, this candidate burst is due to lightning. The other candidate bursts display this same characteristic.

### D. Search for satellite-triggered bursts

In the period studied, 129 bursts detected by satellites occurred in the field of view of Auger ($\theta_{GRB} < 90°$). For all these bursts, the scaler data were checked within 100 seconds of the burst for a one second excess. No excesses were found and the resulting 5 $\sigma$ fluence limits were computed assuming a GRB spectra $dN/dE \propto E^{-2}$ in the 1 GeV – 1 TeV energy range (as in [9]). The limits are reported in Figure 4.

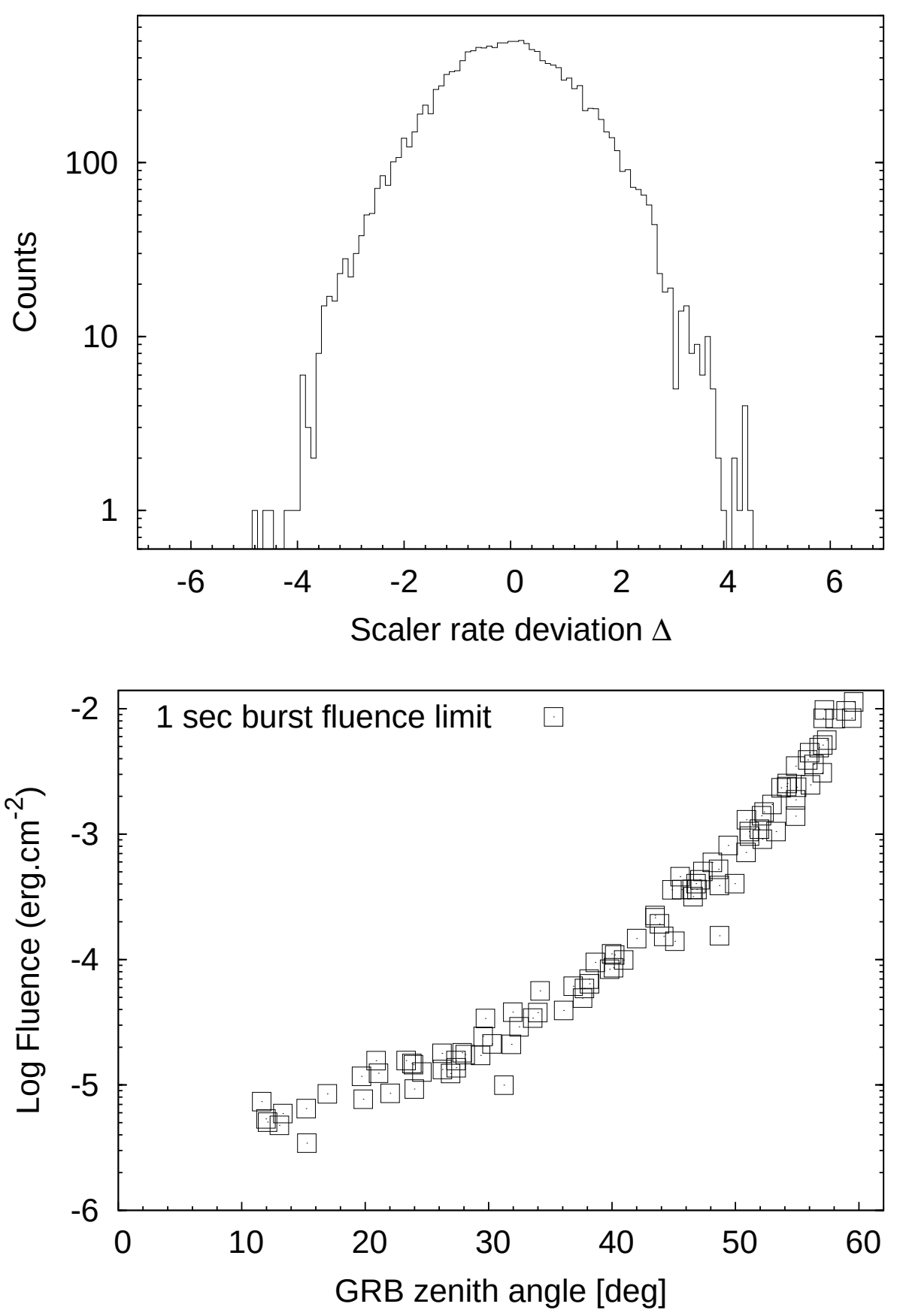

Fig. 4. Top: histogram of the deviations $\Delta$ of the scaler counting rates within 100 seconds of the bursts reported by satellites. No significant excess is observed. Bottom: 5-$\sigma$ fluence limits in the 1 GeV – 1 TeV energy range from Auger for these bursts, for a single second burst, assuming a spectral index of -2.

## IV. Conclusion

We have used the cosmic ray data from the surface detector of the Pierre Auger Observatory to search for cosmic rays that correlate with the time and position of GRBs. No such correlations were found.

As a separate analysis, we used the scaler data to look for increases in the average trigger rate of the surface detectors, which would indicate the occurrence of a GRB. No burst with characteristics similar to those expected for GRBs was observed.

Fluence limits of up to $3.4 \times 10^{-6}\,\mathrm{erg\,cm^{-2}}$ (depending on the burst zenith and duration), were deduced for the 1 GeV – 1 TeV energy range. Note that models do not generally favor fluences above $10^{-6}\,\mathrm{erg\,cm^{-2}}$ in the energy range considered [18], [19]. To reach such a sensitivity, a detector is needed that covers a significant surface at higher altitude. Such a detector could be an extension of the LAGO project [20] that has been taking data since 2007.

# An alternative method for determining the energy of hybrid events at the Pierre Auger Observatory

**Patrick Younk*** **for the Pierre Auger Collaboration**†

*Colorado State University, Fort Collins, Colorado 80523, United States
†Observatorio Pierre Auger, Av. San Martin Norte 304, (5613) Malargue, Mendoza, Argentina

***Abstract*****. An important feature of the Pierre Auger Observatory is the detection of hybrid events; i.e., extensive air showers simultaneously detected with at least one water-Cherenkov detector (surface detector) and one fluorescence telescope. Here we describe an alternative method of estimating the energy of these events. The shower axis is determined using data from both detector systems. The shower energy is determined from the integrated surface detector signals and the distance of each detector from the shower axis. The energy estimate is facilitated by the characterisation of an average lateral distribution function as a function of the shower energy and zenith angle. The method requires only the signal from one surface detector. Thus, it is useful for estimating the energy of hybrid events for which the geometry cannot be estimated with the surface detectors alone and the longitudinal profile measured by the fluorescence telescopes is not well determined. In the energy range $0.4 < E < 1$ EeV, the method doubles the number of hybrid events that can be given an energy estimate. The statistical uncertainty of this energy estimate is dependent on the shower energy and geometry. For events with energy greater than 0.4 EeV, the median statistical uncertainty is 26% and the 90% quantile is 44%.**

## I. Introduction

The Pierre Auger Observatory detects the highest energy cosmic rays with over 1600 water-Cherenkov detectors arranged as an array on a triangular grid with 1500 m spacing. The 3000 km$^2$ array is collectively called the surface detector array (SD). The SD is overlooked by the fluorescence detector (FD), which consists of 24 fluorescence telescopes grouped in units of 6 at four locations on the periphery of the SD. The Auger Observatory was designed so that events recorded by the FD are generally recorded also by the SD. Cosmic ray showers detected with both detectors are referred to as hybrid events.

There are two standard methods for reconstructing the geometry (shower axis) and energy of hybrid events. The first method, SD reconstruction, uses only data from SD stations. The shower geometry and station signal at 1000 m from the shower axis $S(1000)$ are estimated. The shower energy is derived from S(1000) and zenith angle through a set of calibration equations (e.g., [1]). The requirements for SD reconstruction are that the intersection of the shower axis with the ground (core position) be contained within an equilateral triangle of operating stations and that at least three stations record shower particles [2].

The second method, hybrid reconstruction, is best described as a two-step sequence. First, data from the FD telescopes and one SD station are used to estimate the shower geometry [3]. This estimate is referred to as the hybrid geometry. Second, the hybrid geometry and the signal levels recorded by the FD telescopes are used to determine the shower longitudinal profile and the shower energy [4]. This energy estimate is referred to as the FD energy. The requirements for the hybrid geometry estimate are that at least one telescope and one SD station record shower particles. The requirements for the FD energy estimate are much more restrictive. One of the most critical requirements is that the FD records a significant fraction of the longitudinal profile.

Many hybrid events do not meet the requirements for SD reconstruction or the FD energy estimate. Events in this category are mostly low energy events where the three SD stations closest to the axis did not all record shower particles and the FD recorded only a small fraction of the longitudinal profile. For events in this category, it is still possible to obtain an energy estimate with an alternative method.

This alternative method proceeds in three steps. First, the shower geometry is estimated with the hybrid geometry method. Second, $S(1000)$ is estimated based on the hybrid geometry and the integrated signals from the SD station(s). Third, the shower energy is derived from $S(1000)$ and zenith angle following the same procedure used in standard SD reconstruction. The requirements for this alternative method are the same as for the hybrid geometry method. We call this energy estimate the alternative-SD (alt-SD) energy estimate. In this paper, we describe the details of the alt-SD energy estimate and motivate its utility.

## II. The Lateral Distribution Function

The $S(1000)$ parameter has been shown to be correlated with shower energy [5]. To estimate $S(1000)$ from the signal levels in one or more SD stations, we must know the average shape of the lateral distribution function (LDF), i.e., the station signals as a function of distance from the shower axis. We have previously shown [6] that the LDF for Auger events is well described by the modified Nishimura Kamata Greisen (NKG) function

$$S = S(1000)(r/1000)^{\beta}(r + 700/1700)^{\beta},$$

where $S$ is the SD station signal calibrated in vertical equivalent muons (VEM), $r$ is the distance the station is from the shower axis, and $\beta$ is the slope parameter. The slope parameter $\beta$ describes the LDF shape, i.e., the rate at which station signals decrease with distance. The slope parameter is a function of $S(1000)$ (or shower energy) and shower zenith angle $\theta$. We have parameterised $\beta = \beta(S(1000), \theta)$ for $S(1000) > 3$ VEM and $\theta < 60°$ using SD data. For a 0.4 EeV shower with $\theta = 38°$, $S(1000) \approx 3$ VEM.

For an unbiased parameterisation, it is important that the SD have 100% trigger efficiency for the showers used in the parameterisation. The main SD array has a detector spacing of 1500 m and has near 100% trigger efficiency above 3 EeV [7]. However, a small area of the SD array, part of the AMIGA enhancement, has a detector spacing of 750 m and has near 100% trigger efficiency above 0.4 EeV [8]. The 750 m array was used to obtain an unbiased sampling of $\beta$ for showers with $0.4 < E < 3$ EeV.

## III. Details of the Alt-SD Method

Given $\beta(S(1000), \theta)$ and the hybrid geometry, it is possible to estimate $S(1000)$ with a minimum of one SD station. We calculate the total integrated signal $S$ of each station, and estimate the statistical uncertainty on $S$ as $\Delta S = 1.06\sqrt{S}$ [9]. Using the hybrid geometry, we calculate the distance $r$ each station is from the shower axis. The station radius uncertainty $\Delta r$ is obtained from the axis uncertainty (i.e., the uncertainty on core position and arrival direction) via a bootstrap method [10].

To obtain an estimate of $S(1000)$, we minimize the following function

$$\chi^2 = \sum_{i=1}^{N} \frac{\left(S_i - S(1000)\left(\frac{r_i}{1000}\right)^{\beta}\left(\frac{r_i+700}{1700}\right)^{\beta}\right)^2}{(\Delta S_i)^2 + (\Delta r_i dS/dr)^2}.$$

where $N$ is the number of stations. The statistical uncertainty on $S(1000)$, i.e., $\Delta S(1000)$, is obtained by varying $S(1000)$ until $\chi^2$ increases by 1.

Since the number of stations and average signal per station increases with energy, $\Delta S(1000)/S(1000)$ tends to decrease as shower energy increases. However, $\Delta S(1000)/S(1000)$ also depends on the shower geometry. For example, the value(s) of $r$ is important. There are actually two opposing trends. First, $S$ increases as $r$ decreases, which tends to decrease $\Delta S(1000)/S(1000)$. Second, $|dS/dr|$ increases as $r$ decreases, which tends to increase $\Delta S(1000)/S(1000)$. This implies that for a given shower energy and $\Delta r$, there is an optimum value for $r$. This value increases with shower energy. In the energy range $0.4 < E < 3$ EeV, the optimum range of $r$ is approximately 500 m to 1000 m.

The process of obtaining an energy estimate from $S(1000)$ proceeds exactly as in the standard SD reconstruction algorithm. First, the dependence of $S(1000)$ on zenith angle is removed by calculating $S_{38}$ for each event, i.e., the value of $S(1000)$ if the zenith angle of the event was 38°. The attenuation function used to calculate $S_{38}$ is derived directly from the data using the constant intensity cut (CIC) technique.

Second, $S_{38}$ is converted to energy through a calibration equation. This calibration equation is derived from the subset of events with an estimate of FD energy and an alt-SD estimate of $S_{38}$. In this way, the FD energy estimate sets the energy scale for the alt-SD method. For the details of the CIC and energy calibration procedure, see [11].

## IV. Energy Resolution

In Fig. 1, we show the results of the energy calibration step. We applied the alt-SD method to Auger hybrid events recorded from 2004 through 2008 using the following SD station selection criteria: signal not saturated, $S > 10$ VEM, and $r > 200$ m. We selected events with an $S_{38}$ uncertainty $< 20\%$ and which met the strict FD energy criteria reported in [1]. Then, we fit the $S_{38}$ and FD energy data with a broken power-law function:

$$E = \begin{cases} b(S_{38})^a & : S_{38} \le S_B \\ b(S_B)^{a-c}(S_{38})^c & : S_{38} > S_B \end{cases},$$

where $a = 1.245 \pm 0.005$, $b = 0.100 \pm 0.001$, and $c = 1.030 \pm 0.007$. The break point was fixed at $S_B = 20$ VEM. This function is shown in Fig. 1. The reduced $\chi^2$ of the fit was 1.49. During the fitting process, we rejected events below an anti-bias cut line shown as a dotted line in Fig. 1. The line intersects the fitted function at log(3 VEM).

A broken power law describes the data better than a single power law. This shows that the calibration equation flattens slightly (i.e., the $S_{38}$ exponent becomes larger) as shower energy decreases. We are currently investigating this phenomenon.

Fig. 2 shows the fractional difference between the alt-SD energy and the FD energy estimates for events that passed the above selection criteria and with $E > 3$ EeV. The RMS is 21%. This is similar to the difference between the FD and standard SD energy estimates in the same energy range. The main contributions to the width of the distribution are the statistical uncertainty on the FD and alt-SD energy estimates and inherent shower-to-shower fluctuations (including the lack of knowledge of the true LDF shape). At lower energy, the width of this distribution increases slightly. For example, in the energy range $0.4 < E < 3$ EeV the RMS is 26%.

For the energy calibration procedure, it was necessary to use events which passed a set of strict selection rules. We have also examined the energy uncertainty of events that passed a set of less restrictive selection rules. To do this, we selected events with the following criteria: FD track length $> 15°$, at least one station within 850m of the shower axis, at least one unsaturated station with $r > 200$ m and $S > 10$ VEM, $\theta < 60°$, and $E > 0.4$ EeV. We derived the uncertainty on the energy estimate from the uncertainty on $S(1000)$. The median uncertainty is 26%, and the 90% quantile is 44%.

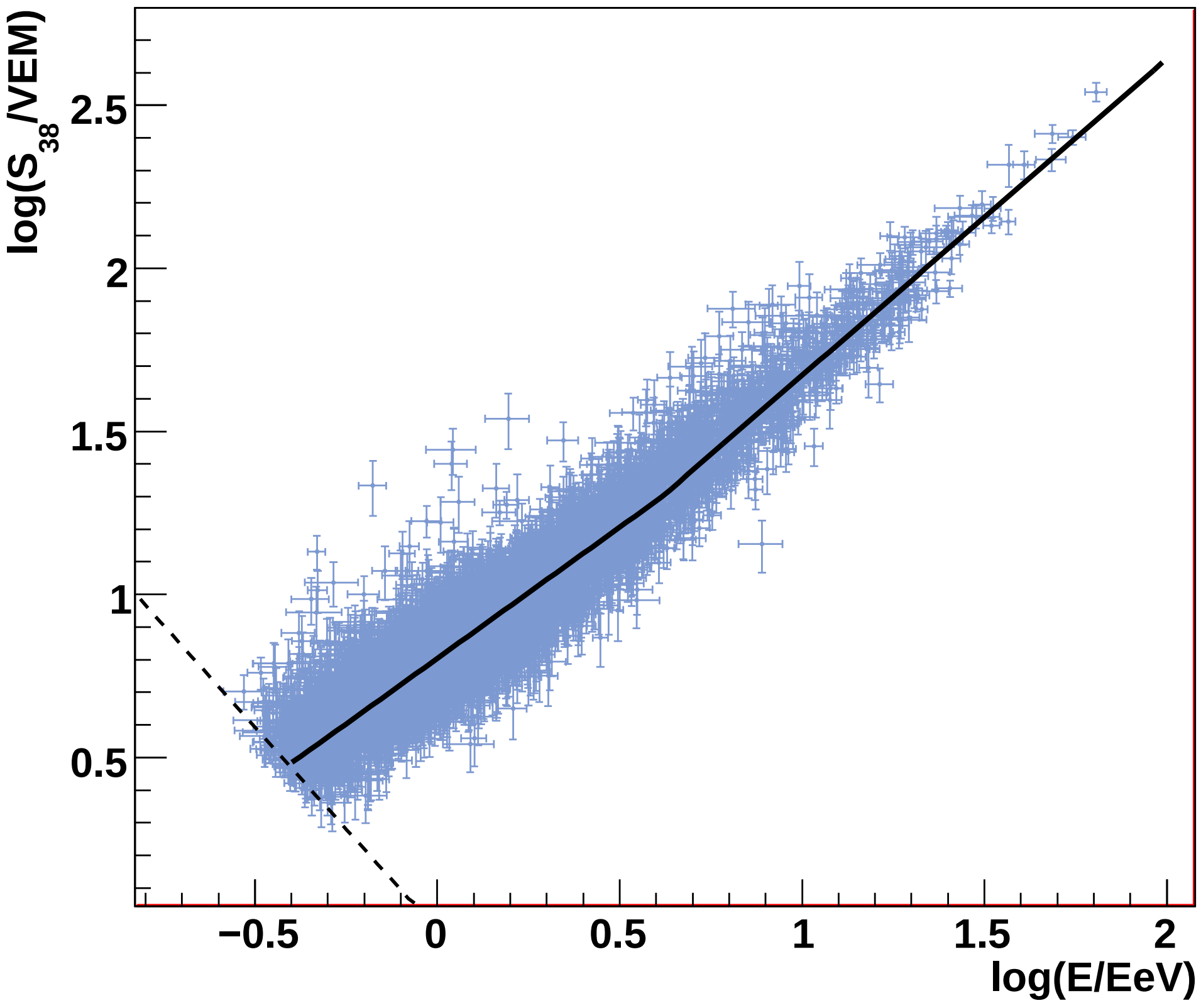

Fig. 1: $S_{38}$ (from the alt-SD method) vs. FD energy for high quality events. The solid line is the energy calibration equation

## V. Discussion and Conclusions

The alt-SD energy method expands the number of hybrid events with an energy estimate. The method is most useful for showers with energy below the 100% trigger efficiency of the SD. In the energy range $0.4 < E < 1$ EeV, the number of hybrid events that can be given an alt-SD energy estimate is approximately twice the number that can be given either a standard SD energy estimate or an FD energy estimate. This ratio increases as shower energy decreases.

Expanding the number of hybrid events with an energy estimate is particularly useful for point source studies. Generally, the hybrid geometry method returns a more accurate estimate of the shower axis direction compared to SD reconstruction. This is especially noticeable for low energy events. For example, for events where only 3 SD stations trigger, the angular resolution of SD reconstruction is approximately 1.75° [12]. However, for events in the same energy range, the angular resolution of the hybrid geometry is approximately 0.6° [13]. Expanding the number of hybrid events with an energy estimate allows for point source searches in narrow energy bands.

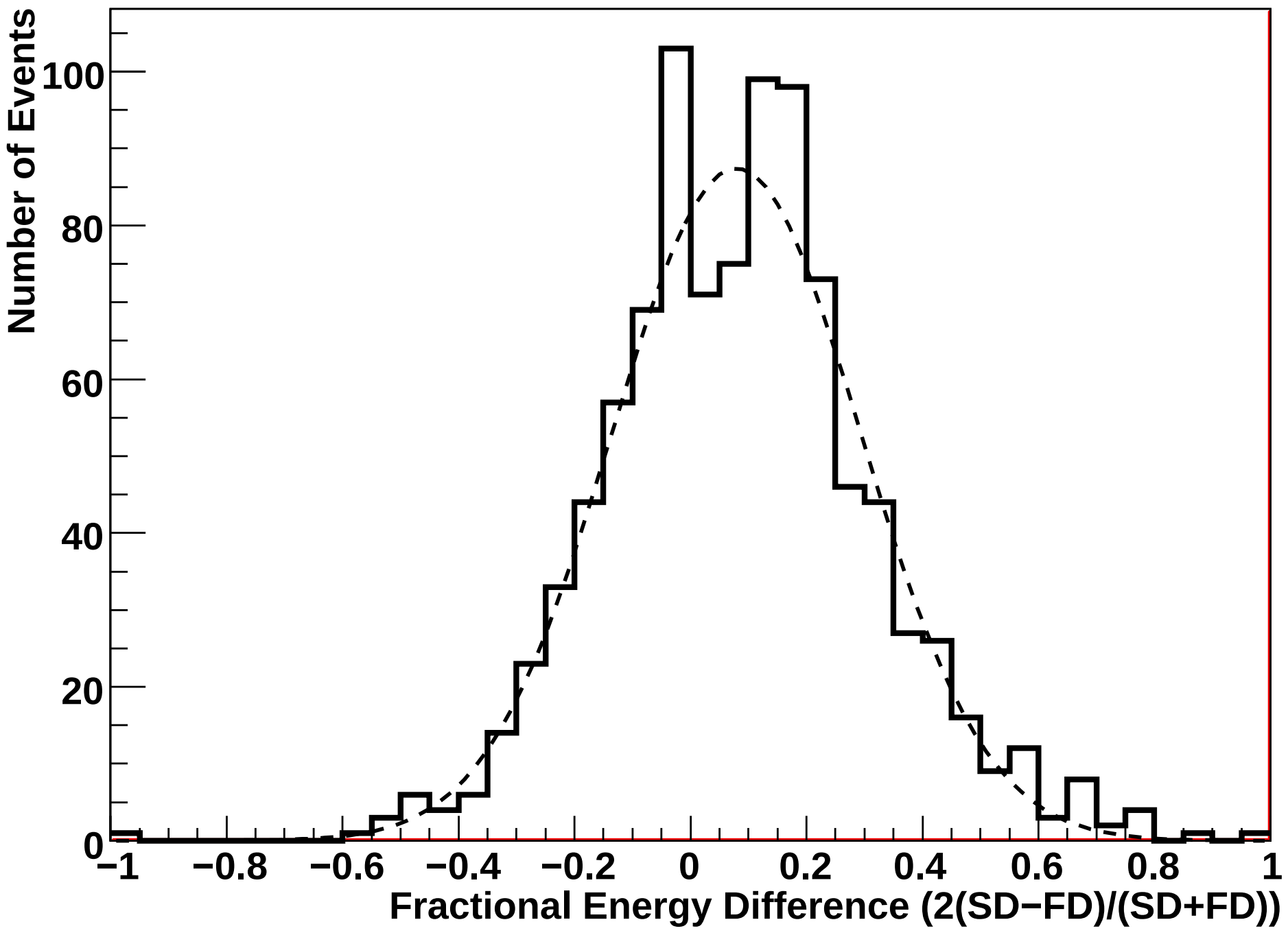

Fig. 2: Fractional difference between the FD and alt-SD energy for hybrid events above 3 EeV. The dotted line is a Gaussian with a standard deviation of 21%

# Acknowledgements

The successful installation and commissioning of the Pierre Auger Observatory would not have been possible without the strong commitment and effort from the technical and administrative staff in Malargüe.

We are very grateful to the following agencies and organizations for financial support:

Comisión Nacional de Energía Atómica, Fundación Antorchas, Gobierno De La Provincia de Mendoza, Municipalidad de Malargüe, NDM Holdings and Valle Las Leñas, in gratitude for their continuing cooperation over land access, Argentina; the Australian Research Council; Conselho Nacional de Desenvolvimento Científico e Tecnológico (CNPq), Financiadora de Estudos e Projetos (FINEP), Fundação de Amparo à Pesquisa do Estado de Rio de Janeiro (FAPERJ), Fundação de Amparo à Pesquisa do Estado de São Paulo (FAPESP), Ministério de Ciência e Tecnologia (MCT), Brazil; AVCR AV0Z10100502 and AV0Z10100522, GAAV KJB300100801 and KJB100100904, MSMT-CR LA08016, LC527, 1M06002, and MSM0021620859, Czech Republic; Centre de Calcul IN2P3/CNRS, Centre National de la Recherche Scientifique (CNRS), Conseil Régional Ile-de-France, Département Physique Nucléaire et Corpusculaire (PNC-IN2P3/CNRS), Département Sciences de l'Univers (SDU-INSU/CNRS), France; Bundesministerium für Bildung und Forschung (BMBF), Deutsche Forschungsgemeinschaft (DFG), Finanzministerium Baden-Württemberg, Helmholtz-Gemeinschaft Deutscher Forschungszentren (HGF), Ministerium für Wissenschaft und Forschung, Nordrhein-Westfalen, Ministerium für Wissenschaft, Forschung und Kunst, Baden-Württemberg, Germany; Istituto Nazionale di Fisica Nucleare (INFN), Ministero dell'Istruzione, dell'Università e della Ricerca (MIUR), Italy; Consejo Nacional de Ciencia y Tecnología (CONACYT), Mexico; Ministerie van Onderwijs, Cultuur en Wetenschap, Nederlandse Organisatie voor Wetenschappelijk Onderzoek (NWO), Stichting voor Fundamenteel Onderzoek der Materie (FOM), Netherlands; Ministry of Science and Higher Education, Grant Nos. 1 P03 D 014 30, N202 090 31/0623, and PAP/218/2006, Poland; Fundação para a Ciência e a Tecnologia, Portugal; Ministry for Higher Education, Science, and Technology, Slovenian Research Agency, Slovenia; Comunidad de Madrid, Consejería de Educación de la Comunidad de Castilla La Mancha, FEDER funds, Ministerio de Ciencia e Innovación, Xunta de Galicia, Spain; Science and Technology Facilities Council, United Kingdom; Department of Energy, Contract No. DE-AC02-07CH11359, National Science Foundation, Grant No. 0450696, The Grainger Foundation USA; ALFA-EC / HELEN, European Union 6th Framework Program, Grant No. MEIF-CT-2005-025057, European Union 7th Framework Program, Grant No. PIEF-GA-2008-220240, and UNESCO.